\documentclass[aps,prd,a4paper,nofootinbib,
preprintnumbers,superscriptaddress,showpacs,showkeys,twocolumn]{revtex4-2}

\usepackage{hyperref}
\usepackage{amsmath}
\usepackage{amssymb}
\usepackage{bm}
\usepackage{graphicx}
\usepackage{xcolor}
\usepackage{slashed}
\usepackage{comment}
\usepackage[normalem]{ulem}

\usepackage[sort,compress]{cleveref}

\crefname{section}{Sec.\!}{Secs.\!}
\crefname{equation}{Eq.\!}{Eqs.\!}
\crefname{figure}{Fig.\!}{Figs.\!}
\crefname{table}{Tab.\!}{Tabs.\!}
\crefname{appendix}{App.\!}{Apps.\!}
\crefname{chapter}{Chapter}{Chapters}

\crefname{section}{Sec.\!}{Secs.\!}
\crefname{equation}{Eq.\!}{Eqs.\!}
\crefname{figure}{Fig.\!}{Figs.\!}
\crefname{table}{Tab.\!}{Tabs.\!}
\crefname{appendix}{App.\!}{Apps.\!}
\newcommand{\be}{\begin{equation}}
\newcommand{\ee}{\end{equation}}
\newcommand{\ba}{\begin{eqnarray}}
\newcommand{\ea}{\end{eqnarray}}

\begin{document}

\title{Melting of $c \bar c$ and $b \bar b$ pairs
in the pre-equilibrium stage of proton-nucleus collisions at the Large Hadron Collider}

\author{Lucia Oliva} \email{lucia.oliva@dfa.unict.it}
\affiliation{Department of Physics and Astronomy "Ettore Majorana", University of Catania, Via Santa Sofia 64, I-95123 Catania, Italy}\affiliation{INFN-Sezione di Catania, Via Santa Sofia 64, I-95123 Catania, Italy}

 \author{Gabriele Parisi}\email{gabriele.parisi@dfa.unict.it}
\affiliation{Department of Physics and Astronomy "Ettore Majorana", University of Catania, Via Santa Sofia 64, I-95123 Catania, Italy}\affiliation{INFN-Laboratori Nazionali del Sud, Via Santa Sofia 62, I-95123 Catania, Italy}

 \author{Vincenzo Greco}\email{greco@lns.infn.it}
\affiliation{Department of Physics and Astronomy "Ettore Majorana", University of Catania, Via Santa Sofia 64, I-95123 Catania, Italy}\affiliation{INFN-Laboratori Nazionali del Sud, Via Santa Sofia 62, I-95123 Catania, Italy}

\author{Marco Ruggieri}\email{marco.ruggieri@dfa.unict.it}
\affiliation{Department of Physics and Astronomy "Ettore Majorana", University of Catania, Via Santa Sofia 64, I-95123 Catania, Italy}\affiliation{INFN-Sezione di Catania, Via Santa Sofia 64, I-95123 Catania, Italy}

%\vspace{10pt}
%\begin{indented}
%\item[]
%\end{indented}

\begin{abstract}
We study the melting of $c\bar c$ and $b\bar b$ pairs in the early stage
of high-energy proton-nucleus collisions. We describe the early stage in terms
of an evolving $SU(3)$ glasma stage, that is dominated by intense, out-of-equilibrium
gluon fields. On top of these fields, we liberate heavy quark-antiquark
pairs, whose constituents are let evolve according to 
relativistic kinetic theory coupled to the gluon fields.
We define a pair-by-pair probability that the pair melts during the evolution,
which we relate to the fluctuations of the color charges induced by 
the interaction of the quarks with the gluon fields.
We find that color decorrelation is the main melting mechanism within the
pre-equilibrium stage. Moreover, we estimate that within $\Delta \tau\approx0.4-0.5$ fm/c after the formation time of the pairs, 
about $50\%$ of $c\bar c$ and $b\bar b$ pairs are melted.
\end{abstract}



\maketitle

\section{Introduction}

Relativistic heavy-ion collisions, as performed at the Relativistic Heavy Ion Collider (RHIC) at Brookhaven National Laboratory and at the Large Hadron Collider (LHC) at CERN, provide a unique opportunity to investigate 
Quantum Chromodynamics (QCD)
at extreme temperatures and densities. 
Over the past few decades, many experimental results indicated the emergence of a new state of matter, referred to as the Quark-Gluon Plasma (QGP), which forms within $\sim 1$ fm/c after the collision and in which quarks and gluons behave as a hydrodynamic fluid of small $\eta/s\sim 0.1$~\cite{Heinz:2000bk, Ruggieri:2013ova, Ruggieri:2013bda, Plumari:2019gwq}.
Such facilities are able to perform measurements also in smaller collision systems such as 
proton-proton
(pp) and 
proton-nucleus
(pA)~\cite{Dusling:2015gta}, 
and these have shown very similar features as those found in heavy ion collisions. While calculations within the hydrodynamic framework have been quite successful in describing observables in these small collision systems, the applicability of hydrodynamics becomes increasingly doubtful as the system size decreases and gradients increase~\cite{Kurkela:2018qeb,Kurkela:2019kip}. It is also likely that both initial intrinsic momentum correlations and a hydro-like anisotropic expansion are necessary to understand the elliptic anisotropic emission in pA~\cite{Greif:2017bnr, Zhao:2022ugy, Oliva:2022rsv, Oliva:2023dti, Nugara:2024net, Grosse-Oetringhaus:2024bwr}.

For this reason, a thorough treatment of the initial stages is even more important for pA collisions, and the Classical Yang-Mills (CYM) theory offers a good framework for the study of this stage. 
In fact, CYM well describes the out-of-equilibrium 
evolution of the highly occupied gluonic system, named the  glasma~\cite{Lappi:2006fp}, 
that serves as the initial condition for the system produced in the collisions.
The glasma is based on the 
Color-Glass-Condensate  (CGC) effective theory, 
which stands as a valid description of high-energy nuclei~\cite{McLerran:1993ni,McLerran:1993ka,McLerran:1994vd}; see \cite{Gelis:2010nm,Iancu:2003xm,McLerran:2008es,Gelis:2012ri} for reviews.
The glasma initial condition has been widely used to 
establish the initial conditions for the subsequent hydrodynamic 
evolution~\cite{Schenke:2012hg,Schenke:2012wb}. 

Heavy Quarks (HQs), i.e. charm, $c$, and beauty, $b$,
represent a unique probe of the early stage of high-energy collisions as well as of the transport properties of the thermalized medium subsequently produced~\cite{Moore:2004tg, vanHees:2005wb, vanHees:2007me, Gossiaux:2008jv, He:2011qa, Alberico:2011zy, Lang:2012nqy, Das:2013kea, Song:2015sfa, Song:2015ykw, Cao:2015hia, Andronic:2015wma, Beraudo:2015wsd, Das:2015ana, Das:2015aga, Das:2016llg, Das:2016cwd, Cao:2016gvr, Aarts:2016hap, Prino:2016cni, Scardina:2017ipo, Das:2017dsh, Greco:2017rro, Xu:2017obm, Chatterjee:2017ahy, Chatterjee:2018lsx, Rapp:2018qla, Cao:2018ews, Xu:2018gux, Dong:2019unq, Song:2019cqz, Dong:2019byy, Song:2020tfm, Zhao:2020jqu, Oliva:2020doe, Oliva:2020mfr, Beraudo:2021ont, Ruggieri:2022kxv, Pooja:2023gqt, Sun:2023adv, Das:2024vac, Barata:2024xwy}. Indeed, due to the large masses, the HQs have a very short formation time of the order of $0.1$ fm/c. 
Therefore the HQ final states, the heavy-flavored hadrons, maintain imprint of the evolution dynamics during the whole collision time frame, with traces from both the initial stage and the subsequent QGP phase.
Besides studies on the diffusion of HQs in the QGP and on their hadronization
in relativistic nuclear collisions,
several works have investigated the impact of the early stage on heavy quarks~\cite{Mrowczynski:2017kso, Ruggieri:2018rzi, Sun:2019fud, Liu:2019lac, Liu:2020cpj, Boguslavski:2020tqz, Carrington:2020sww, Ipp:2020nfu, Khowal:2021zoo, Carrington:2021dvw, Pooja:2022ojj, Carrington:2022bnv, Das:2022lqh, Boguslavski:2023fdm, Avramescu:2023qvv, Pandey:2023dzz, Boguslavski:2023alu,Avramescu:2024poa,Avramescu:2024xts}.
These have shown that heavy quarks undergo significant diffusion in the 
early stage, suggesting an impact on experimental observables such as the nuclear modification factor and the elliptic flow, as analysed in~\cite{Sun:2019fud}.
Since for small systems the thermalized QGP stage is short lived or even absent, HQs could keep a cleaner trace of the early stage with respect to nucleus-nucleus reactions. Thus, for pA collisions the impact of the initial glasma state on final heavy-quark observables may be predominant in the whole evolution of the collision event.

A natural extension 
of the aforementioned works
is to look at the impact of the gluon-dominated initial state on the evolution and dissociation of heavy quarkonia, which are bound states of heavy quarks and antiquarks of the same flavour, namely charmonium $c\bar{c}$ and bottomonium $b\bar{b}$.
The production and suppression of heavy quarkonium states in hadronic collisions and in particular in QGP has been widely studied for almost four decades~\cite{Matsui:1986dk, Datta:2003ww, Burnier:2009yu, Brambilla:2011sg, Srivastava:2012pd, Casalderrey-Solana:2012yfo, Brambilla:2013dpa, Singh:2015eta, Andronic:2015wma, Brambilla:2016wgg, Song:2017phm, Akamatsu:2020ypb, Yao:2021lus, He:2021zej, Miura:2022arv, Villar:2022sbv, Brambilla:2022ynh, Song:2023zma, Song:2023ywt, Brambilla:2024tqg, Bai:2024xmm,Delorme:2024rdo, Du:2015wha, Du:2017qkv, Du:2018wsj, Yan:2006ve, Zhou:2014kka}. In particular,
the melting emerges from the interplay of several mechanisms, coming from the different phases of the collision, from initial HQ pair production through the QGP stage to final hadronization, see Ref.~\cite{Rothkopf:2019ipj,Andronic:2024oxz} for a review.
As in the QGP, 
$c\bar{c}$ and $b\bar{b}$ pairs can dissociate also in the evolving glasma~\cite{Pooja:2024rnn}, 
with important implications for their evolution in the ensuing QGP medium and for the determination of the heavy-flavoured final states produced at hadronization (both open heavy-flavour hadrons and regenerated quarkonia).
In the evolving glasma, it is the HQ interaction with the classical color fields that leads to a modification of their spatial and momentum coordinates as well as of their color charge, thus affecting the probability that they survive to the pre-equilibrium 
stage and enter into the possibly-formed QGP phase as a pair.

In this article, we investigate the melting of 
$c\bar c$ and $b\bar b$
pairs during the pre-equilibrium stage of high-energy pA collisions, the latter being modeled within the glasma framework.
The primary objective of this work is to determine the fraction of pairs that dissociate during the propagation in the evolving glasma, aiming also at disentangling the role of color-charge decorrelation of the HQ pair in the melting,
which was only partly addressed before.
Within our model the HQs in each pair, formed at
$\tau=\tau_\mathrm{form}$, couple to the strong gluon fields
via the Wong equations~\cite{Wong:1970fu,Heinz:1984yq}
as it has been done in~\cite{Pooja:2024rnn}.
With respect to the approach presented in~\cite{Pooja:2024rnn},
we implement a different criterion for the calculation of the number of melting pairs, based on a survival probability, $\mathcal{P}_\mathrm{survival}$,
that takes into account the decorrelation of the color charges
induced by the interactions of $Q$ and $\bar Q$ with the gluon fields
in the early stage. 
$\mathcal{P}_\mathrm{survival}$
is built similarly to the hadronization probability introduced in coalescence models~\cite{Greco:2003xt, Fries:2003vb, Greco:2003mm, Fries:2003kq, Greco:2003vf, Plumari:2017ntm}.
We are able to estimate an average melting time for both 
$c\bar c$ and $b\bar b$ pairs, that turns out to be around $0.4-0.5$ fm/c.
Differently from~\cite{Pooja:2024rnn}, we consider $N_c=3$ and
an initialization for pA collisions
that takes into account the event-by-event fluctuations of the color charges
in the proton and of the initial distribution of the 
heavy quarks in the transverse plane. Moreover,
within our model we allow for the expansion both along the longitudinal
direction and in the transverse plane, while in~\cite{Pooja:2024rnn}
only a static geometry was considered.

Throughout this article, we use natural units $k_B=c=\hbar=1$.
Moreover, we formulate the 
$SU(3)$
Yang-Mills equations using the Milne coordinates
\begin{equation}
    \tau=\sqrt{t^2-z^2},\quad \eta=\frac{1}{2}\ln{\left(\frac{t+z}{t-z}\right)},\label{eq:la7_11:13}
\end{equation}
where $\tau$ is the proper time and $\eta$ denotes the space-time rapidity.

\section{The evolving glasma in proton-nucleus collisions}

\subsection{Color charges for p and A}
The CGC effective theory \cite{McLerran:1993ni,McLerran:1993ka,McLerran:1994vd} is based on a separation of scales between the high Bjorken-$x$ degrees of freedom, i.e. the valence partons, and the low $x$ degrees of freedom which they generate, that is, soft gluons. To realize this separation,
the sources of the initial gluon fields 
are generated using the McLerran-Venugopalan (MV) model, 
in which the large-$x$ color sources 
are randomly distributed
on an infinitely thin color sheet.
The distribution of these charges is a Gaussian, characterized by
zero average 
\begin{equation}
    \langle \rho^a (\mathbf{x}_\perp)\rangle=0,
    \label{1}
\end{equation}
and variance given by
\begin{equation}
    \langle \rho^a (\mathbf{x}_\perp)\rho^b (\mathbf{y}_\perp)\rangle=g^2 \mu^2 \delta^{ab} \delta^{(2)}(\mathbf{x}_\perp-\mathbf{y}_\perp).
    \label{2}
\end{equation}
Here, $g$ is the coupling constant (we use $g=2$, corresponding to $\alpha_s=g^2/4\pi=0.3$), 
and $\mu$ is the so-called MV parameter, describing the number density of the color charges per unit of area in the transverse plane.
In our implementation, we relax the single-sheet hypothesis of the original MV model by generating a certain number $N_s$ of color sheets stacked on top of one another \cite{LAPPI:2010ykh}. 
Consequently, for each of the sheets, the numerical implementation of Eq.~\eqref{2}
we will use in this work is
\begin{equation}
    \langle \rho^a_{n,x}\rho^b_{m,y}\rangle=g^2 \mu^2 \frac{\delta_{n,m}}{N_s}\delta^{a,b} \frac{\delta_{x,y}}{a_\perp^2},
    \label{3}
\end{equation}
where $a,b=1,\dots, N_c^2-1$ are $SU(3)$ color indexes, $n,m=1,\dots,N_s$ are the color sheet indexes and $x,y$ span each point of a $N_\perp\times N_\perp$ lattice, whose transverse length is $L_\perp$ and lattice spacing is $a_\perp=L_\perp/N_\perp$. Operatively, conditions \eqref{1} and \eqref{3} are satisfied by generating random Gaussian numbers with mean zero and standard deviation equal to $\sqrt{(g^2\mu^2)/(N_s a_\perp^2)}$.
Below we explain, starting from these color charges, how to build up the gluon fields generated
by each sheet, and how to
combine all these fields to obtain the initial
gluon field in the glasma.
We use $N_s=50$ in Eq.~\eqref{3}, having checked that this number of sheets is enough to get numerical convergence.

These steps are shared by the charge generation of both a proton and a heavy ion. The original MV model assumed a source charge density that is homogeneous in the transverse plane. We adopt this assumption when dealing with large-A nuclei, whose details at the level of single nucleons can be neglected. On the other hand, a reasonable description of protons involves a varying nuclear density in the transverse plane to take into account the underlying quark structure. The main difference between those two cases lies in the choice of $\mu$: in the A-case $\mu$ is chosen as a constant equal to $\mu=0.5$ GeV, which translates into uniform fluctuations of color charge throughout the lattice. The p-case is instead more involved. In order to do so, let us introduce the thickness function of the proton $T_p$, normalized to $1$ and defined as
\begin{equation}
T_p(\mathbf{x}_\perp) = 
\frac{1}{3} \sum_{i=1}^3 \frac{1}{2\pi B_q} \exp\left(-\frac{(\mathbf{x}_\perp-\mathbf{x}_i)^2}{2B_q}\right),
\label{eq:bd1}
\end{equation}
where the $\mathbf{x}_i$ denote the positions of the constituent quarks, which are randomly extracted from the distribution
\begin{equation}
T_{cq}(\mathbf{x}_\perp) = \frac{1}{2\pi B_{cq}}
\exp\left(-\frac{\mathbf{x}_\perp^2}{2B_{cq}}\right).
\label{eq:bd2}
\end{equation}
Parameters have been chosen as $B_q=0.3$ GeV$^{-2}$, $B_{cq}=4$ GeV$^{-2}$, i.e. the widths of the two Gaussians differ by a factor around 4. After doing so, we evaluate the saturation scale $Q_s$ in this model as
\begin{equation}
Q_s^2(x,\mathbf{x}_\perp) = \frac{2\pi^2}{N_c}  \alpha_s\, 
xg(x,Q_0^2)\, T_p(\mathbf{x}_\perp).
\label{eq:qs_2_77}
\end{equation}
from which $\mu(x,\mathbf{x}_\perp)$ is given by \cite{Schenke:2020mbo}
\begin{equation}
g^2\mu(x,\mathbf{x}_\perp) = c Q_s(x,\mathbf{x}_\perp),
\label{eq:gmu_ajk}
\end{equation}
with $c=1.25$. Note that by virtue of Eq.~\eqref{eq:qs_2_77} not only $\mu$, but also $Q_s$ depend on the transverse plane coordinates.
Also, in principle $xg$ in \eqref{eq:qs_2_77} should be computed at the scale $Q_s$ by means of the DGLAP equation with a proper initialization. This was done in \cite{Schenke:2020mbo,Rezaeian:2012ji}, and we reserve this approach for future studies. For the sake of simplicity, in the present study we limit ourselves by assuming that $xg$ is given by the initial condition at $Q_0^2=1.51$ GeV$^2$ as in \cite{Schenke:2020mbo, Rezaeian:2012ji}, namely
\begin{equation}
xg(x,Q_0^2)=A_g x^{-\lambda_g}(1-x)^{f_g},
\label{eq:anna23}
\end{equation}
with $A_g=2.308$, $\lambda_g=0.058$ and $f_g=5.6$. This simplification is partly justified by the fact that the average saturation scale for the proton is of the order of $Q_s\sim 1$ GeV, hence we do not expect the DGLAP evolution of the $xg$ in Eq.~\eqref{eq:anna23} to lead to significant changes. For pA collisions at the LHC energy, the relevant values of $x$ are in the range $[10^{-4},10^{-3}]$. In our calculations, we will thus fix $x$ in the aforementioned range, then compute $xg$ by virtue of Eq.~\eqref{eq:anna23}. We will use the corresponding value obtained for $x=10^{-4}$, that is, $xg=3.94$.

\subsection{Gauge-link formulation of the evolving glasma}
\label{sec:gaugefields}
Once the charge has been generated, we know that the hard and the soft sectors are coupled via the Yang-Mills equations \cite{Yang:1954ek}
\begin{equation}
    D_\mu F^{\mu\nu}=J^\nu,
    \label{8.1}
\end{equation}
where $D_\mu=\partial_\mu-ig[A_\mu,\hspace{3pt} \cdot\hspace{3pt}]$ is the covariant derivative, $F^{\mu\nu}=\partial^\mu A^\nu-\partial^\nu A^\mu-ig[A^\mu,A^\nu]$ is the field strength tensor and $J^\mu$ the color current. Using the light-cone coordinates, the conservation of current implies $A^-=0$. Moreover, by working in the covariant gauge we can also impose $A^i=0$, where $i=x,y$. The only component left is therefore $A^+\equiv \alpha$, which can be seen to obey the following Poisson equation
\begin{equation}
    \Delta_\perp \alpha(\mathbf{x}_\perp)=-\rho(\mathbf{x}_\perp).
    \label{9}
\end{equation}
In order to solve this equation, we 
introduce a soft regulator, $m$,
then Fourier-transform both sides of~\eqref{9},
getting
\begin{equation}
\tilde{\alpha}_{n,k}^a=\frac{\tilde{\rho}_{n,k}^a}{\Tilde{k}_\perp^2+m^2},
    \label{9.4}
\end{equation}
where the discretized momentum $\Tilde{k}_\perp$ is given by
\begin{equation}
\Tilde{k}_\perp^2=\sum_{i=x,y}\left(\frac{2}{a_\perp}\right)^2\sin^2\left(\frac{k_ia_\perp}{2}\right),
    \label{9.2}
\end{equation}
and $m=0.2$ GeV $=1$ fm$^{-1}$, which acts as a screening mass
and effectively implements color confinement on lengths above $1/m$.
By then, Fourier-transforming back, we get $\alpha(\mathbf{x}_\perp)$ in coordinate space.

Once we obtained a solution for the gauge field in the covariant gauge, the Wilson line for each nucleus is evaluated as
\begin{equation}
    V_\mathbf{x}=\prod_{n=1}^{N_s}\exp\{ig\alpha_{n,\mathbf{x}}^a T^a\}
    \label{9.5}
\end{equation}
and then obtain the gauge links, $U$, for each of the two nuclei as
\begin{equation}
U_{\mathbf{x},i}^{A,B}=V_\mathbf{x}^{A,B}V_{\mathbf{x}+\hat{i}}^{\dagger,A,B}.
\label{9.6}
\end{equation}
In the above expressions, $T^a$ are the $SU(3)$ group generators in the fundamental representation, 
with normalization $\text{Tr}[T^a T^b]=\delta^{ab}/2$.
Periodic boundary conditions have been implemented on the transverse plane.

The gauge link resulting from the interaction of the two 
colored glasses 
in p and A, $U_{\bm x,i}$, at the position
$\bm x$ in the transverse plane
is determined by solving a  set of eight equations, which are
\begin{equation}
    \text{Tr}[T_a(U_{\mathbf{x},i}^A+U_{\mathbf{x},i}^B)(\mathbb{I}+U_{\mathbf{x},i})-\text{h.c.}]=0,
    \label{10}
\end{equation}
with $a=1,\dots,N_c^2 - 1$.
In the case of $N_c=2$ there is an exact solution of \eqref{10} for $U_{\mathbf{x},i}$ \cite{Krasnitz:1998ns}.
Instead, for $SU(3)$ we can solve \eqref{10} only numerically
via a standard iterative method. 
Using this procedure we calculate the gauge links along the $x$ and $y$ direction. 
For the longitudinal components we initialize simply $U_\eta=\mathbb{I}$
since the glasma has $A_\eta=0$ at $\tau=0^+$.
The initial color-electric fields are~\cite{Fukushima:2011nq}
\begin{widetext}
\begin{align}
    &E_x=E_y=0,\nonumber\\
   &E^\eta=-\frac{i}{4ga_\perp^2}\sum_{i=x,y}\left[(U_i(\mathbf{x}_\perp)-\mathbb{I})(U_i^{B,\dagger}(\mathbf{x}_\perp)-U_i^{A,\dagger}(\mathbf{x}_\perp))+(U_i^\dagger(\mathbf{x}_\perp-\hat{i})-\mathbb{I})(U_i^{B,\dagger}(\mathbf{x}_\perp-\hat{i})-U_i^{A,\dagger}(\mathbf{x}_\perp-\hat{i}))-h.c.\right].
   \label{11.2}
\end{align}

After the initial condition has been fixed,
The equations of motion for the gauge links 
are~\cite{Fukushima:2011nq}
\begin{align}
    \partial_\tau U_i(\mathbf{x})&=\frac{-iga_\perp}{\tau}E^i(\mathbf{x})U_i(\mathbf{x}),\label{eq:parisihatoltoilginger}\\
    \partial_\tau U_\eta(\mathbf{x})&=-iga_\eta\tau E^\eta(\mathbf{x}) U_\eta(\mathbf{x}),
    \label{11.3}
\end{align}
where $a_\eta$ denotes the discretization step in the $\eta$ direction.
In order to reduce the discretization error in time, we drive the evolution through a leapfrog algorithm, i.e. by letting the gauge links and the electric fields evolve in different steps alternatively.
Eqs.~\eqref{eq:parisihatoltoilginger} and~\eqref{11.3}
are discretized as
\begin{align}
U_i(\tau'')&=\exp\left[-2\Delta \tau\, iga_\perp E^i(\tau')/\tau'\right]U_i(\tau),\label{11.4bisAAA}\\
U_\eta(\tau'')&=\exp\left[-2\Delta \tau\, ig a_\eta \tau' E^\eta(\tau')\right]U_\eta(\tau),
\label{11.4}
\end{align}
where $\tau'=\tau+\Delta \tau/2$ and $\tau''=\tau+\Delta \tau$. Notice that the exponentiation of the electric field is important, in order to keep the up-to-date gauge links as unitary matrices. In the same fashion, the equations of motion of the electric field are discretized as:
\begin{align}
E^i(\tau')=&E^i(\tau-\Delta \tau)+2\Delta \tau \frac{i}{2ga_\eta^2 a_\perp\tau}[U_{\eta i}(\tau)+U_{-\eta i}(\tau)-(h.c.)]+2\Delta \tau \frac{i\tau}{2g a_\perp^3}\sum_{j\neq i}[U_{ji}(\tau)+U_{-ji}(\tau)-(h.c.)],\nonumber\\
E^\eta(\tau')=&E^\eta(\tau-\Delta \tau)+2\Delta \tau \frac{i}{2g a_\eta a_\perp^2 \tau} \sum_{j=x,y}[U_{j\eta}(\tau)+U_{-j\eta}(\tau)-(h.c.)],\label{12}
\end{align}
where
\begin{equation}
    U_{\mu\nu}(\mathbf{x})\equiv U_\mu(\mathbf{x})U_\nu (\mathbf{x}+\hat{\mu})U_\mu^\dagger(\mathbf{x}+\hat{\nu})U_\nu^\dagger(\mathbf{x})
    \label{13}
\end{equation}
are the plaquette variables. Notice that in this way the gauge links are initialized at time $\tau=0$, and evolve with integer steps: $\tau=\Delta \tau, 2\Delta \tau, \dots$ Instead, the electric field is initialized at $\tau=\Delta \tau/2$ and proceeds as $\tau=3\Delta \tau/2, 5 \Delta \tau/2,\dots$
Along with the electric field, using the gauge links we can derive the components of the color-magnetic field as \cite{Fukushima:2011nq}
\begin{align}
    B_x^a&=\frac{2}{iga_\eta \tau a_\perp}\text{Tr}[T^a(\mathbb{I}-U_{\eta y})]\label{eq:magnetic_field_x},\\
    B_y^a&=\frac{2}{iga_\eta \tau a_\perp}\text{Tr}[T^a(\mathbb{I}-U_{\eta x})]\label{eq:magnetic_field_y},\\
    B_z^a&=\frac{2}{iga_\perp^2}\text{Tr}[T^a(\mathbb{I}-U_{xy})].\label{eq:magnetic_field_z}
\end{align}
\end{widetext}

\section{Heavy quark dynamics in the pre-equilibrium stage\label{sec:devrijTOP}}

In this section, we discuss how we model the diffusion of the 
heavy quark pairs in the pre-equilibrium stage.
Following~\cite{Pooja:2024rnn},
we model 
the dynamics of the quarks in the $c\bar c$ and $b\bar b$ pairs
using semi-classical equations of motion,
namely the Wong equations. 
They describe the time evolution of the coordinate {\bf x}, 
momentum {\bf p}, and color charge $Q_a$ (with
$a=1,\dots,N_c^2-1$) of the quarks~\cite{Wong:1970fu, Heinz:1984yq}.
In the laboratory frame, the Wong equations read
\begin{eqnarray}
    \dfrac{\mathrm{d}x^i}{\mathrm{d}t} & =&\frac{p^i}{E},
    \label{eq:wong1}\\
    \dfrac{\mathrm{d}p^i}{\mathrm{d}t} & =&gQ^aF^{i\mu,a}\frac{p_\mu}{E}-\frac{\partial V}{\partial x_i},
    \label{eq:wong2}\\
    \dfrac{\mathrm{d}Q_a}{\mathrm{d}t} & =&-gf^{abc} A_\mu^bQ_c\frac{p^\mu}{E},
    \label{eq:wong3}
\end{eqnarray}
where $i=x,y,z$ and $\mu=0,\dots,3$. 
Moreover, $f^{abc}$ are the structure constants of the gauge group $SU(N_c)$ and $E=\sqrt{ \bm p^2+M^2}$ is the kinetic energy of the quark with mass $M$. 
The Wong equations have been recently used to study the dynamics of the heavy quarks in the pre-equilibrium stage of high-energy nuclear collisions; see \cite{Ruggieri:2018ies,Ruggieri:2018rzi,Sun:2019fud,Liu:2019lac,Jamal:2020fxo,Liu:2020cpj,Sun:2021req,Khowal:2021zoo, Pooja:2022ojj,Pooja:2024rnn, Avramescu:2024poa, Avramescu:2024xts} and references therein.

The gauge field $A_\mu$ 
in Eq.~\eqref{eq:wong3}
is determined by (the $A_\tau$ component is zero because of the gauge choice)
\begin{align}
    A_i(\mathbf{x})&=-\left(\frac{i}{g a_\perp}\right)\log U_i(\mathbf{x}), ~~i=x,y, \label{eq:gauge_field_xy}\\
    A_\eta(\mathbf{x})&=-\left(\frac{i}{g a_\eta}\right)\log U_\eta(\mathbf{x}),\label{eq:gauge_field_eta}
\end{align}
whereas the color-electric and color-magnetic fields are defined in terms of the field strength tensor through
\begin{eqnarray}
&&F_{\tau i}=E_i,~~~F_{\eta i}=\tau\epsilon_{ij}B_j,\\
&&F_{\tau\eta} = \tau E_\eta,~~~F_{xy}=-B_\eta.\label{eq:EBfields}
\end{eqnarray}
For $N_c=3$
the evolution of the color charges governed by Eq.~\eqref{eq:wong3} conserves the Casimir invariants
\begin{equation}
q_2 = Q_a Q_a,~~~q_3=d_{abc}Q_a Q_b Q_c,
\label{eq:mostrasumarte}
\end{equation} 
with
\begin{equation}
q_2=\frac{N_c^2-1}{2} ~~\text{and}~~q_3=\frac{(N_c^2-4)(N_c^2-1)}{4N_c}, \label{eq:casimir_values}
\end{equation}
and $d_{abc}$ are the symmetric structure constants \cite{Avramescu:2023qvv}. We solve Eqs.~\eqref{eq:wong1}-\eqref{eq:wong3} in the background of the evolving glasma fields. Within our model, 
we follow previous works and neglect the back-reaction of the quarks 
in the Yang-Mills equations: it has been shown~\cite{Liu:2020cpj} that 
such back-reaction 
does not substantially affect the diffusion of heavy quarks
within the lifetime of the pre-equilibrium stage.

In equation~\eqref{eq:wong2} $V$
denotes the interaction potential between 
a quark and its respective antiquark in the pair. 
In QCD,
the tree-level static potential between a quark and an 
antiquark can be written as
\begin{equation}
V_{\alpha\beta\gamma\delta}=
\frac{\alpha_s}{r_\text{rel}}T^a_{\alpha\beta}
\bar T^a_{\gamma\delta},
\label{HQ_potentialsing_buffet}
\end{equation}
which is related to the static limit of the amplitude of the 
one-gluon-exchange
interaction between two quarks
in the color states $\beta$ and $\delta$ that 
go out respectively in the color states
$\alpha$ and $\gamma$.
The $\bar T^a$ correspond to the 
generators in the $\bar{\bm 3}$
representation. In~\eqref{HQ_potentialsing_buffet} the quantity
$r_\mathrm{rel}=|\mathbf{x}_{Q}-\mathbf{x}_{\bar{Q}}|$,
is the relative distance between the quark and
the antiquark.

This potential depends on the
matrix element of the operator $T^a\otimes \bar T^a$,
where the first color generator acts on the Hilbert space of the
quark, and the second on that of the antiquark.
The potential~\eqref{HQ_potentialsing_buffet} can be
projected onto irreducible representations of 
$SU(3)$. In particular, 
if the quark-antiquark pair belongs to the 
irreducible representation $R$
we have
\begin{equation}
V_{R}=\lambda_R
\frac{\alpha_s}{r_\text{rel}},
\label{HQ_potentialsing_buffet_lambdaR}
\end{equation}
where
\begin{equation}
\lambda_R = \frac{1}{2}
\left[
C_2(R) - C_2(3) - C_2(\bar 3)\right].
\label{eq:giuliettamivuoisposare}
\end{equation}
Here, $C_2(R)$ denotes the eigenvalue of the quadratic Casimir
in the representation $R$, while
$C_2(3) = C_2(\bar 3) = 4/3$ are the Casimir for the quark
and the antiquark in the $\bm 3$ and $\bar{\bm 3}$ 
representations respectively.
If the quark-antiquark pair is in the
color-singlet representation then $C_2(1)=0$ and
\begin{equation}
V_1=-\frac{4}{3}\frac{\alpha_s}{r_\text{rel}}.
\label{HQ_potentialsing}
\end{equation}
Similarly, for the octet $C_2(8)=3$ and 
\begin{equation}
V_8=
+\frac{1}{6}\frac{\alpha_s}{r_\text{rel}}.
\label{HQ_potentialsingoct}
\end{equation}

Within our semi-classical approach, 
the most natural choice 
for $V$ in Eq.~\eqref{eq:wong2}
consists in replacing
$T^a\otimes T^a$ in Eq.~\eqref{HQ_potentialsing_buffet}
with $Q_a\bar Q_a/N_c$.
Hence, in Eq.~\eqref{eq:wong2} we consider
\begin{equation}
V=
\frac{ Q_a  \bar{Q}_a}{N_c}
\frac{\alpha_s}{r_\text{rel}}.
\label{HQ_potential}
\end{equation}
The overall $1/N_c$ in the right hand side of~\eqref{HQ_potential}
is included in order to reproduce $V=V_1$
when we initialize the pairs in a color-singlet
state, see below Eq.~\eqref{eq:giornicontati}.
For each pair,
$Q_a  \bar{Q}_a$ evolves with time,
therefore the coupling in $V$ 
follows the evolution of the color charges as a result
of the interaction with the background gluon fields.
For a justification of Eq.~\eqref{HQ_potential}
starting from the expectation value of the QCD potential
see section~\ref{sec:coequi}.

Before going ahead, we notice that 
in~\cite{Pooja:2024rnn} a similar problem was studied,
although for the simpler geometry of a static box
and for the $SU(2)$ gauge group.
The improvement that the potential~\eqref{HQ_potential}
brings with respect to the one used in~\cite{Pooja:2024rnn},
namely a pure singlet potential,
is that our $V$ dynamically evolves with the color charges
of the pair,
so it takes into account that even if the pairs are
initialized in color-singlet states, they can get
an octet component due to the interaction with the
background gluon fields. 

\section{Heavy quarks initialization}
\label{sec:HQ_init}

Both $c\bar c$ and $b\bar b$  quark pairs 
are produced by hard QCD scatterings among the nucleons of the colliding nuclei at the very early stage of high-energy nuclear collisions, within 0.1 fm/c from the overlap of the two initial nuclei.
In this work, we assume
the pairs to form at proper time 
$\tau_\mathrm{form} = 1/(2 M)$, with $M$ being the heavy-quark mass: $M=1.3$ GeV for $c$ and $M=4.2$ GeV for $b$. Therefore $\tau_\mathrm{form} \sim 0.08$ fm/c 
for $c$, and $\tau_\mathrm{form} \sim 0.02$ fm/c
for $b$.

\subsection{Initialization in the coordinate space and in momentum space}
In this work we consider the pairs produced at mid-rapidity, so that the longitudinal coordinate of the heavy quarks is kept $z=0$, if we indicate with positive $z$ and negative $z$ respectively the forward-going and backward-going directions of the colliding nuclei.
In the transverse plane, 
the initial positions of the heavy quarks
are extracted on an event-by-event basis.
For each event, after extracting the position of the three valence
quarks via the Gaussian profile~\eqref{eq:bd2}, we extract the position of the center of mass of each quark-antiquark pair via the profile~\eqref{eq:bd1}. In this way, the heavy quarks are distributed around the hotspots of the
initial energy density of the color fields. From that, we extract the relative distance between each quark and antiquark in the pair (in its rest frame) according to the distribution:
\begin{equation}
P_\mathrm{HQ}(r_\text{rel}) = r_\text{rel}\,\exp{(-r_\text{rel}^2/r_\text{0}^2)},
\label{eq:HQ_extraction}
\end{equation}
where $r_\text{0}=0.4$ fm for a $c\bar{c}$ pair and $r_\text{0}=0.2$ fm for a $b\bar{b}$ \cite{Zhao:2024vqp}. Using the $r_\text{rel}$ we extract for each pair, each quark is placed at a distance $r_\text{rel}/2$ from the center of mass of the pair, while the polar angle is extracted uniformly in the range $(0,2\pi)$. The corresponding antiquark is generated opposite to the quark, with respect to the center of mass. 
\begin{figure}[t!]
\centering
\includegraphics[trim={0 0 0 0},clip,width=\columnwidth]{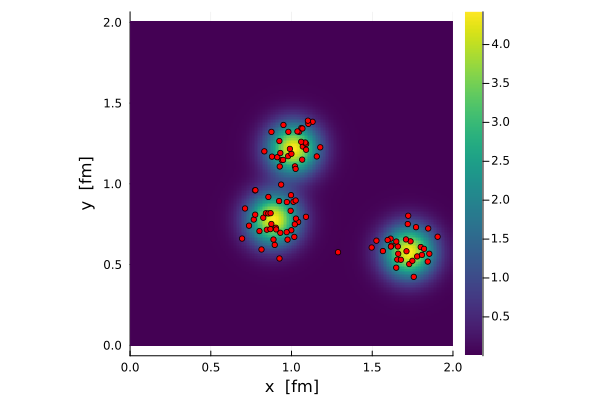}
\includegraphics[trim={0 0 0 0},clip,width=\columnwidth]{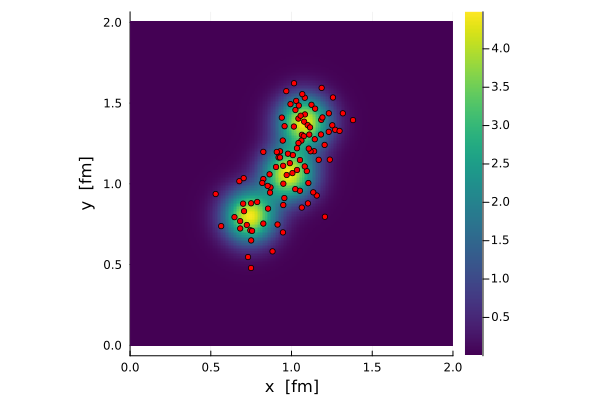}
\caption{Centers of mass of the production points of heavy-quark pairs (red dots) on the transverse plane for proton-nucleus collisions. 
The background image corresponds to the contour plot of the profile~\eqref{eq:bd1} in fm$^{-2}$, characterized by three hotspots that develop around the valence quarks in that event. 
We used $g\mu=1$ GeV for the nucleus. The two panels correspond to two different events.}
\label{fig:init_HQ_xT}
\end{figure}
For illustrative purposes, we show in Fig.~\ref{fig:init_HQ_xT} the centers of mass of 100 heavy-quark pairs in two events. 
We remark that this number of pairs is chosen only for a good visualization
and does not correspond to the real number of quarks produced in the considered collision system. 
In both panels of the figure, the three hotspots 
correspond to the profile~\eqref{eq:bd1} 
that develops around
the three valence quarks, whose location changes event-by-event; 
we checked that
these are the same locations where most of the energy density
gets deposited at the initial time.
The red dots represent the centers of mass of the production points of the heavy quark and antiquark in the pairs.

Moving on to the initialization in momentum space, we extract the modulus of the relative momentum in transverse plane of each quark and antiquark in the pair (in its rest frame) according to the distribution:
\begin{equation}
P_\mathrm{HQ}(p_\text{rel}) = p_\text{rel}\,\exp{(-p_\text{rel}^2 r_\text{0}^2)},
\label{eq:HQ_extraction_momentum}
\end{equation}
Once $p_\text{rel}$ is extracted for each pair, we set the modulus of the quark momentum as $p_\text{rel}/2$, while the polar angle is extracted uniformly in the range $(0,2\pi)$. From that, the components $p_x$ and $p_y$ of the heavy-quark momentum are determined and those of the companion antiquark are fixed opposite to them. 
We assume $p_z=0$ for quarks and their companion antiquarks,
as we simulate the mid-rapidity region and assume that the space-time rapidity of the produced particles $\eta$
is equal to the energy-momentum rapidity $y$.
With this initialization of momenta the quark and antiquark in a pair have momenta, ${\bm p}_q$ and ${\bm p}_{\bar q}$ respectively, such that the total momentum is zero, ${\bm p}_q + {\bm p}_{\bar q}=0$. Note also that the width of the gaussian \eqref{eq:HQ_extraction_momentum} is set as $1/r_\text{0}$, being $r_\text{0}$ the same parameter used in \eqref{eq:HQ_extraction} for the initialization in coordinate space.
The HQs in a $b\bar{b}$ pair will therefore be initialized with an average momentum which is larger than the HQs in a $c\bar{c}$ pair.

\subsection{Initialization in the color space\label{sec:colorspace}}
Let us deal with the initialization of the color charges appearing in the Wong equations~\eqref{eq:wong2} and~\eqref{eq:wong3}. 
The initial value of the color charges 
needs to satisfy the constraints~\eqref{eq:mostrasumarte}.
In order to get these, we firstly 
fix the charge of one heavy quark 
in the ensemble
to be~\cite{Avramescu:2023qvv}
\begin{equation}
Q_{0a}=-1.69469\,\delta_{a5}-1.06209\,\delta_{a8}
    \label{eq:HQcharge0}
\end{equation}
which explicitly satisfies 
the conditions~\eqref{eq:mostrasumarte}; 
then, we generate the charges of all the other heavy quarks as
\begin{equation}
Q_a=Q_{0b} U^{ab}, ~~~U^{ab}=2\text{Tr}[T^a U T^b U^\dagger],
    \label{eq:HQ_charge_all}
\end{equation}
where $U$ is an $SU(3)$ random matrix which is extracted for each heavy quark according to the Haar measure. 
By construction, all the HQs hereby obtained satisfy \eqref{eq:mostrasumarte}, and this condition will be satisfied throughout the time evolution, as already stated.

As anticipated in section~\ref{sec:devrijTOP},
we initialize the $c\bar c$  and $b\bar b$
pairs as an ensemble of color-singlet states. 
In a quantum-mechanical framework, this would amount to initialize the color part of the wave function of
each pair as
\begin{equation}
|S\rangle = \frac{1}{\sqrt{3}}(|r\bar r\rangle + |g\bar g\rangle + |b\bar b\rangle ).
\label{eq:parisi_1113}
\end{equation}
For the state~\eqref{eq:parisi_1113} 
we have
\begin{equation}
\langle S |\sum_a(T^a_1 + \bar T^a_2)^2 |S\rangle = 0,
\label{eq:visonobruttenotizie}
\end{equation}
where $T^a_{1,2}$ denotes the $SU(3)$ color generators
acting on the Hilbert spaces of the quark and the antiquark
respectively.
In the classical limit, we replace
operators with their expectation values.
Denoting by
$Q_a$ and $\bar{Q}_a$ the color charges of, respectively, the quark and the antiquark in a pair,
the classical version of Eq.~\eqref{eq:visonobruttenotizie} 
reads
\begin{equation}
\sum_a (Q_a + \bar Q_a)^2=0,
\label{eq:giocateallaguerra}
\end{equation}
which immediately implies
\begin{equation}
Q_a = - \bar Q_a,~~~a=1,\dots,N_c^2-1.
\label{eq:dottoresenzapantaloni}
\end{equation}
This also implies that, by taking into account the normalization
of the classical charges~\eqref{eq:mostrasumarte}, 
at initial time we have:
\begin{equation}
\sum_a Q_a \bar Q_a = -q_2.
\label{eq:giornicontati}
\end{equation}
Hence, we initialize the color charge of each antiquark by imposing the condition~\eqref{eq:dottoresenzapantaloni}.

\section{Results\label{sec:results}}

In this section, we show our results on the heavy-quark evolution and heavy quark-antiquark pair dissociation in proton-nucleus collisions at LHC energy.
The results shown in this section have been obtained for a square grid of transverse size $L=3$ fm, with $64\times64$ lattice points. In realistic collisions,
the pre-equilibrium stage lasts for a fraction of fm/c,
but for illustrative purposes
we performed simulations up to 1 fm/c with 1000 time steps
(we checked that this number of time steps is enough to have
numerical convergence). Moreover, in each simulation we considered 1000 HQ pairs, for a total of 2000 particles. In our numerical implementation,
following~\cite{Pooja:2024rnn}
we cure the numerical divergence for $r_\text{rel}=0$ in~\eqref{HQ_potential} 
by using  in Eq.~\eqref{eq:wong2}
the regularized potential
\begin{equation}
V=\frac{Q_a  \bar{Q}_a}{N_c}
\frac{\alpha_s}{r_\text{rel}}(1-e^{-A\, r_\text{rel}}),
    \label{HQ_potential_regularized}
\end{equation}
instead of~\eqref{HQ_potential}.
Here we present results obtained with
$A=(0.017\text{ fm})^{-1}$: 
we checked that taking $A$ larger than this value makes the screening of the divergence at $r_\text{rel}=0$ ineffective, not leading to the convergence of the numerical algorithm.

\subsection{Spreading and color decorrelation of the pairs\label{subse:rideremmo}}

\begin{figure}[t!]
\centering
\includegraphics[width=.9
\columnwidth]{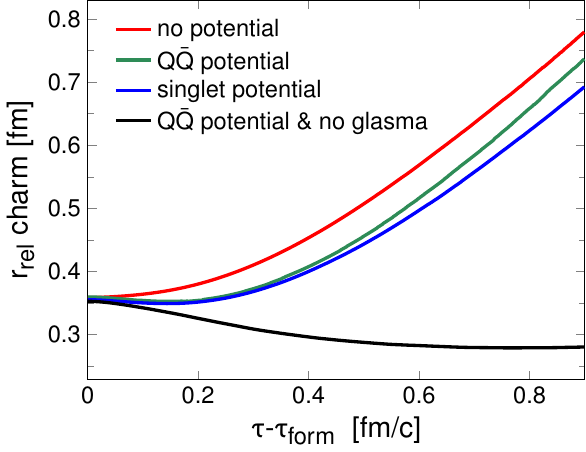}
\caption{Averaged relative distance for $c\bar c$ pairs,
computed in the pair rest frame.
The no potential data correspond to calculations without the
potential~\eqref{HQ_potential} in the Wong equation~\eqref{eq:wong2}.
$Q\bar Q$ potential denote the results obtained considering 
the potential as well as the evolving glasma fields. Blue line stands
for the results obtained assuming a pure color-singlet potential in
Eq.~\eqref{eq:wong2}. Finally, the $Q\bar Q$ potential \& no glasma curve denotes the results obtained by solving the equations of motion
switching off the interaction of the quarks with the fields.}
\label{Fig:rreldifferentpotentials}
\end{figure}

Let us focus on heavy quark-antiquark pairs and investigate the effect of the evolving glasma in dissociating the initially produced pairs.
The relative position of the quark and antiquark in the pairs is modified during their propagation in the fireball, both in the pre-equilibrium and in the thermalized QGP phases.
In particular, the interaction with the gluon fields 
leads to a diffusion in coordinate (as well as momentum) space,
resulting on average 
in the increase of the relative distance of the two particles in a pair.

In Fig.~\ref{Fig:rreldifferentpotentials} we plot the relative 
distance of $c$ and $\bar c$ in 
$c\bar{c}$ pairs, $r_{\mathrm{rel}}$, in the rest frame of the pairs, versus 
the proper time $\tau$ (measured with respect to the formation time
$\tau_\mathrm{form}$), averaged over the HQ pairs as well as over the gauge field configurations.  We show our results in four different cases: along with the result obtained with the dynamic potential~\eqref{HQ_potential} (solid green), we plot $r_{\mathrm{rel}}$ obtained by considering $V=0$ in Eq.~\eqref{eq:wong2}
(solid red),  by using $V=V_1$ 
defined in Eq.~\eqref{HQ_potentialsing} (solid blue), 
as well as
by turning off the gluon fields (solid black).
The latter is shown in order to illustrate the quantitative effect of the HQ potential and of the gluon fields
on $r_{\mathrm{rel}}$. 

In the scenario in which only the potential is considered, 
$r_{\mathrm{rel}}$ decreases over time due to the attractive interaction between the quark and the antiquark in the pair. Indeed, in this case the dynamical potential coincides with the singlet potential, since the color charges do not evolve under the action of the glasma.
On the other hand, we see that
the effect of the glasma is to increase the separation of the pair, 
due to the intense bombardment of gluons on both the quark and antiquark
of the pair. 
Notably, during the time interval where the glasma framework 
is phenomenologically relevant, i.e., up to approximately $0.3{-}0.4$ fm/c, 
$r_{\mathrm{rel}}$ remains nearly constant. This behavior results from the 
competition between the attractive potential and the interaction with the 
color fields of the glasma.

For completeness, in Fig.~\ref{Fig:rreldifferentpotentials} we also display the case where the potential 
is absent. 
By comparing this with the full calculation, we observe that, 
as expected, the presence of an attractive potential leads to an initial 
decrease in $r_{\mathrm{rel}}$.
Finally, when the potential is modeled as a pure singlet interaction,
$r_{\mathrm{rel}}$ remains
very close 
to the result of the full calculation within the time range where the glasma is phenomenologically relevant. 
This suggests that, during this period, the 
interaction between the quarks in the pair is not significantly different from that of a pure singlet. 
On the other hand, moving on at later times, $r_{\mathrm{rel}}$ in the pure singlet 
scenario is slightly lower if compared to the full calculation. 
This is a natural consequence of the fact that the singlet potential remains stronger than the full potential at larger times, leading to a tighter binding of the quark-antiquark pairs.

\begin{figure}[t!]
\centering
\includegraphics[trim={15 5 0 0},clip,width=\columnwidth]{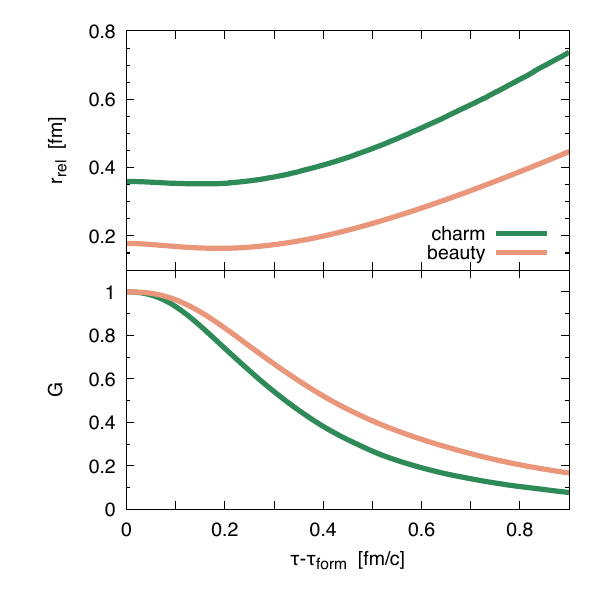}
\caption{Averaged relative coordinate in the pair rest frame (top panel) and color-charge correlator (bottom panel) of the heavy quark-antiquark pairs versus time $\tau-\tau_\mathrm{form}$, with $\tau_\mathrm{form}$ being the heavy-quark formation time. Green and brown curves correspond to charm and beauty heavy-quark pairs, respectively.}
\label{fig:r_rel_c_cor}
\end{figure}

By now limiting ourselves to the physical case of the potential in Eq.~\eqref{HQ_potential}, in the top panel of Fig.~\ref{fig:r_rel_c_cor} we show 
$r_{\mathrm{rel}}$ in the pair rest frame versus $\tau-\tau_\mathrm{form}$ for a $c\bar{c}$ pair (same solid green curve of Fig.~\ref{Fig:rreldifferentpotentials}) and for a $b\bar{b}$ pair (solid brown). Even when taking into account the different initialization in position for charm and beauty, we interestingly note that
the $c\bar c$ system
spreads slightly more quickly with respect to the $b\bar b$ one. This is expected since the higher masses of $b$ 
make them more static in the gluon fields.

Not only coordinates and momenta of heavy quarks are affected by
the background gluon fields, but also their color charges are, see Eq.~\eqref{eq:wong3}.
The quark-antiquark pairs can undergo transitions from singlet to octet (and viceversa) both in glasma and in QGP. In the latter stage, such transitions have been studied especially within the framework of open quantum systems \cite{Casalderrey-Solana:2012yfo, Brambilla:2016wgg, Miura:2022arv}. 
These singlet-octet transitions may also occur in the glasma stage due to the interaction of the quark and antiquark in the pair with the initial color fields. For this reason we study the decorrelation of the color charges of the quark and the antiquark in a pair, which serves as an estimate of the probability of singlet-to-octet transitions, and hence gives indications on the survival probability
of the pair in the gluon fields.

As explained in section~\ref{sec:colorspace},
we initialize the quark-antiquark pairs
in color-singlet states, then the interactions with the gluonic environment changes their color charges. 
Denoting by $Q_a$ and $\bar Q_a$ the color charges of a quark
and its antiquark respectively, let us consider for each pair the time-dependent quantity
\begin{equation}
\mathcal{G}(\tau) = -\frac{1}{q_2}\mathrm{Tr}(Q_a(\tau) \bar Q_b(\tau))
=-\frac{1}{q_2}\sum_a Q_a(\tau) \bar Q_a(\tau),
\label{eq:intascailtelefonino}
\end{equation}  
where $q_2$ is the quadratic Casimir invariant of
$SU(N_c)$ introduced in \eqref{eq:casimir_values}, and the trace is understood over the $(a,b)$ indices. 
For the color-singlet initialization we have $Q_a   = - \bar Q_a $ at
$\tau=\tau_\mathrm{form}$, therefore for each pair
\begin{equation}
\mathcal{G}=1~~~\text{ at }\tau=\tau_\mathrm{form},
\label{eq:mascheradoro}
\end{equation}
indicating that with our initialization the color charges of the two particles at $\tau=\tau_\mathrm{form}$ are maximally anti-correlated. We expect that on average the
magnitude of $\mathcal{G}$ decreases with time, as a result of the 
decorrelation of $Q_a$ and $\bar Q_b$
due to the random interactions with the gluon fields
of the evolving glasma. 
This decrease can be interpreted as
the raising of the weight of the color-octet component in the pair,
which eventually leads to the melting of the pair itself. 
We will use this idea to define a melting probability of the
$c\bar c$ and $b\bar b$ pairs, see section~\ref{subsec:melting_of_the_pairs}.

From the above definition \eqref{eq:intascailtelefonino} of $\mathcal{G}$, we define the gauge-invariant color-charge correlator as~\cite{Pooja:2024rnn}
\begin{equation}
    G(\tau) \equiv   \langle   \mathcal{G}(\tau)
    \rangle.
    \label{eq:c_cor}
\end{equation}
The brackets in~\eqref{eq:c_cor} 
indicate the average 
over the pairs as well as over the gauge field configurations. 
In the lower panel of Fig.~\ref{fig:r_rel_c_cor} we plot
$G(\tau)$ versus time $\tau-\tau_\mathrm{form}$,  
for charm (green line) and beauty (brown line) pairs.
We notice that $G$ decreases with time, 
as expected,
indicating a decorrelation of the color charges of the quark and of the companion antiquark in the pair. 

In order to be more quantitative
we define a decorrelation time,
$\tau_\mathrm{dec}$,
via the requirement that $G(\tau_\mathrm{dec})=1/2$.
We notice that
$\tau_\mathrm{dec}-\tau_\mathrm{form}$ stays in the range
$(0.3,0.4)$ fm/c for both charm and beauty.
We also find that $\tau_\mathrm{dec}-\tau_\mathrm{form}$ 
for $b$ quarks is larger than the corresponding quantity 
computed for the $c$ quarks: this is most likely due to the fact that
$b\bar b$ pairs are tighter, hence they spend more time within 
a single correlation domain
of the gluonic background. 

We could interpret $G(\tau)$ as the probability for pairs to stay
in the singlet channel during propagation in the 
pre-equilibrium stage, or equivalently,
$1-G(\tau)$ is the probability to have singlet-to-octet
fluctuations.
In fact, during the evolution,
the decrease of the correlator can be interpreted as the fluctuations
of the color charges becoming increasingly more important, making transitions to a color-octet state more probable.

One of the interesting aspects shown in 
Figs.~\ref{Fig:rreldifferentpotentials} and~\ref{fig:r_rel_c_cor} 
is that, while the interquark relative distance within the pairs remains nearly 
constant 
up to $\tau \approx 0.3$~fm/c for both $c$ and $b$ quark pairs, color decorrelation occurs rather quickly. 
In this time range, it is therefore color decorrelation that primarily 
drives 
the melting of the pair.

\subsection{Color equilibration\label{sec:coequi}}

The results shown in the lower panel of Fig.~\ref{fig:r_rel_c_cor} can be rephrased in terms of 
the expectation values of the color-projector operators,
that allow us to measure the probability of pairs 
to be in the color-singlet and the color-octet states.
To this end, we recall that in QCD one can introduce
the projectors onto the singlet, $\mathcal P_S$, and the octet, 
$\mathcal P_O$, spaces as
\begin{eqnarray}
\mathcal P_S &=& -\frac{2}{3}T^a \otimes \bar T^a   + \frac{ 1}{9}I,
\label{eq:proiettori_singoletto}\\
\mathcal P_O &=& \frac{2}{3}T^a\otimes \bar T^a + \frac{8 }{9}I,
\label{eq:proiettori_ottetto}
\end{eqnarray}
which can be easily obtained from the eigenvalues
of the operator $T^a \otimes \bar T^a$ in the color-singlet,
$\lambda_S$, and 
color-octet, $\lambda_O$, representations, namely
\begin{equation}
\lambda_S = -\frac{4}{3},~~~\lambda_O = + \frac{1}{6}.
\label{eq:autovalori}
\end{equation}
It is easy to verify that 
$\mathcal P_S^2 = \mathcal P_S$, 
$\mathcal P_O^2 = \mathcal P_O$,   
$\mathcal P_O \mathcal P_S = 
\mathcal P_S \mathcal P_O = 0$ and 
$\mathcal P_O + \mathcal P_S=1$.
The classical counterpart of
the projectors~\eqref{eq:proiettori_singoletto}
and~\eqref{eq:proiettori_ottetto} can be written as
\begin{eqnarray}
P_S &=& -\frac{2}{3}\frac{Q_a \bar Q_a}{N_c}
+ \frac{1}{9},\label{eq:pro_sin_cla}\\
P_O &=& \frac{2}{3}\frac{Q_a \bar Q_a}{N_c} 
+ \frac{8}{9},\label{eq:pro_oct_cla}
\end{eqnarray}
which can be obtained from
Eqs.~\eqref{eq:proiettori_singoletto}
and~\eqref{eq:proiettori_ottetto}
via the formal replacement
$T^a \otimes\bar T^a \rightarrow Q_a \bar Q_a/N_c$, 
similarly to what we have done in writing the classical 
counterpart of the potential~\eqref{HQ_potentialsing_buffet}
in Eq.~\eqref{HQ_potential}.
It is easy to check that
the singlet condition~\eqref{eq:giornicontati}
implies $P_S=1$ and $P_O=0$.

We now want to relate $P_S$ to the probability
that the quark-antiquark pair is in the
color-singlet state. In fact, 
in the quantum theory, the probability 
that the state $|\psi\rangle$ is a
color-singlet is 
\begin{equation}
p_S = 
|\langle\psi| \mathcal P_S|\psi \rangle |^2.
\label{eq:cb_ooo}
\end{equation}
If $|\psi\rangle$ represents a semi-classical
state, we can neglect
the quantum fluctuations and we can write
\begin{equation}
p_S
=
\langle\psi| \mathcal P_S^2|\psi \rangle,
\label{eq:cb_pbs}
\end{equation}
and taking into account that $\mathcal P_S^2 =
\mathcal P_S$, we get
\begin{equation}
p_S
=
\langle\psi| \mathcal P_S|\psi \rangle=P_S,
\label{eq:cb_f.it}
\end{equation}
where the last equality stands 
in the classical limit. 
Hence, we can interpret $P_S$ as the probability
that the semiclassical quark-antiquark
state is in the color-singlet state;
similarly, $P_O=1-P_S$ gives the probability that
the pair is in the color-octet state.

The identification of \( P_S \) and \( P_O \) as the probabilities for the state to be in the singlet and octet representations, respectively, allows us to justify \textit{a posteriori} the potential~\eqref{HQ_potential}, as it coincides with the classical limit of the QCD potential.
 In fact, 
let us consider
a linear combination of 
singlet, $|S\rangle$, and octet 
$|O_i\rangle$ with $i=1,\dots,8$,
states, namely
\begin{equation}
|\psi\rangle = c_S |S\rangle + \sum_{i=1}^8c_{i}|O_i\rangle,
\label{eq:psiso1}
\end{equation}
with $|c_S|^2 + \sum |c_i|^2=1$.
The QCD potential is $V=V_1\bm 1_S + V_8\bm 1_O$,
where $\bm 1_S$ and $\bm 1_O$ are identities in the singlet
and the octet representations, and
the potentials are given by Eqs.~\eqref{HQ_potentialsing} and~\eqref{HQ_potentialsingoct}. Hence, 
the projection of $V$ on the state $|\psi\rangle$ is
\begin{equation}
\langle \psi | V | \psi \rangle =
|c_S|^2 V_1 + (1-|c_S|^2) V_8.
\label{eq:projection1641}
\end{equation}
Within our semi-classical approach, we replace 
$|c_S|^2$  with $P_S$ 
in the
right hand side of 
Eq.~\eqref{eq:projection1641}, 
see Eq.~\eqref{eq:cb_f.it},
and 
the state $|\psi\rangle$ with a classical state
representing our quark-antiquark pairs. 
We then obtain
\begin{equation}
  V   = P_S V_1 + (1-P_S) V_8.
\label{eq:projection1643}
\end{equation}
Taking into account Eq.~\eqref{HQ_potentialsing}, \eqref{HQ_potentialsingoct} and~\eqref{eq:pro_sin_cla}, we 
finally get
Eq.~\eqref{HQ_potential}.
Therefore, we can identify the potential~\eqref{HQ_potential}
with the classical limit of the QCD potential~\eqref{eq:projection1641}.

\begin{figure}[t!]
\centering
\includegraphics[width=\columnwidth]{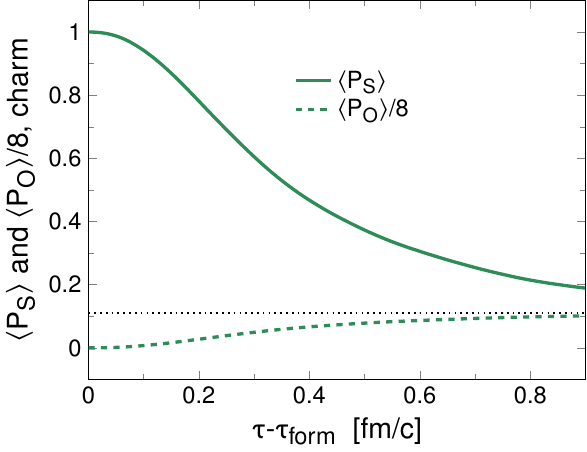}
\caption{Expectation values of the singlet and the (normalized)
octet projectors versus proper time for $c\bar c$ pairs.
The dashed horizontal black line stands for $1/9$.
The results show the almost-perfect equipartition
of probability 
among color-singlet and color-octet states at large times.}
\label{fig:Ps_Po}
\end{figure}

In Fig.~\ref{fig:Ps_Po} we plot $\langle P_S\rangle$ and 
$\langle P_O\rangle/8$
versus proper time, computed for $c\bar c$ pairs
(the results for $b\bar b$ pairs are similar).
The dashed horizontal black line stands for the value $1/9$.
In particular, we divided $\langle P_O\rangle $ 
by eight in order to
count the probability of being in one
of the eight states
of the octet.
The results shown in the figure have been obtained by
taking the ensemble average of Eqs.~\eqref{eq:pro_sin_cla}
and~\eqref{eq:pro_oct_cla}, similarly to the procedure we 
followed to produce the results 
for
$G(\tau)$ shown in Fig.~\ref{fig:r_rel_c_cor}.

We notice in Fig.~\ref{fig:Ps_Po}
that at the formation time the pairs 
are in 
the singlet state 
with probability $1$,
due to our initialization of the color
charges. As the pairs evolve and interact with the
gluon fields, there are transitions from the singlet
to the octet states as a result of color decorrelation.
At large times, when the colors of the quark and the antiquark
in the pair are uncorrelated, the averaged projectors
tend to the same value,
\begin{equation}
\langle P_S \rangle \approx \langle P_O \rangle/8
\approx 1/9.
\label{eq:equipartition_1656}
\end{equation}
This shows that asymptotically there is an equipartition
of probability for the pairs to be in the singlet
and in one of the eight octet states.

It is very interesting to notice that, from the
qualitative point of view, the color equilibration 
shown in Fig.~\ref{fig:Ps_Po} is in agreement
with the results of Ref.~\cite{Delorme:2024rdo}, in which 
the singlet-to-octet 
transitions have been studied for quarkonia states
in a thermalized QGP
within the framework of the quantum Brownian motion.
In that context, color equilibration takes place
on a longer timescale, likely because of
the different energy scales which are relevant in the
two problems. Nevertheless,
the interaction with the external environment (the QGP 
in~\cite{Delorme:2024rdo} and the background gluon fields
in the present work) in both cases leads to the equipartition
between singlet and octet states.

We finally comment that the
equilibration of color shown in Fig.~\ref{fig:Ps_Po}
does not immediately 
lead to the statement that 
$1/9$ of the pairs forms a bound state
at large times. In fact, in order to state anything about pair dissociation, the information on color equilibration
has to be supplemented with that on the relative distance
between the quark and the antiquark in the pair:
on average this distance increases with time (see the upper panel in Fig.~\ref{fig:r_rel_c_cor}),
therefore the pairs 
inevitably melt at large times
even if statistically the have probability
$1/9$ to be in a color-singlet state.
We analyze in more detail this problem in the next
subsection.

\subsection{Melting of the pairs}
\label{subsec:melting_of_the_pairs}

We make use of the results discussed in section~\ref{subse:rideremmo}
to define a survival probability of the pairs in the pre-equilibrium stage,
$\mathcal{P}_\mathrm{survival}$, and from this a melting probability
\begin{equation}
\mathcal{P}_\mathrm{melting}=1-\mathcal{P}_\mathrm{survival}.
\label{eq:presepi_a_napoli}
\end{equation}
In order to define $\mathcal{P}_\mathrm{survival}$, 
we notice in Fig.~\ref{fig:r_rel_c_cor}
that the relative distance is almost constant during the early evolution
of the system, while the color of the quarks in the pairs decorrelates
quickly. Therefore, 
it is reasonable to
assume that 
color rotation is the leading mechanism to dissolve the
pairs, hence $\mathcal{P}_\mathrm{survival}$ 
can be connected to the color-charge correlator $\mathcal{G}$ defined in
Eq.~\eqref{eq:intascailtelefonino}. 
We assume a Gaussian shape of $\mathcal{P}_\mathrm{survival}$ in color space, namely
\begin{equation}  \mathcal{P}_\mathrm{survival}=\exp{\left[-\kappa(\mathcal{G}-1)^2\right]},\label{eq:conte_non_contento}
\end{equation}
where $\kappa$ is a parameter that measures
the width of the fluctuations of the color charges
that are necessary to break the pair.
The form~\eqref{eq:conte_non_contento} of $\mathcal{P}_\mathrm{survival}$ is inspired by the Wigner function of coalescence models to form a vector meson from a quark and an antiquark of the same flavor \cite{Greco:2003vf}. 
In our model, at each time, the status of melted for each quark-antiquark pair is assigned with the probability given by Eq.~\eqref{eq:presepi_a_napoli}.

We have investigated the effect of 
the pre-equilibrium stage
on pair dissociation for several values of $\kappa$.
Here, for both charm and beauty
we show results for $\kappa=4$, 
that corresponds
to require that 
we have $\mathcal{P}_\mathrm{survival}=1/e$ when $\mathcal{G}=1/2$.

\begin{figure}[t!]
\centering
\includegraphics[trim={15 50 0 50},clip,width=1.05\columnwidth]{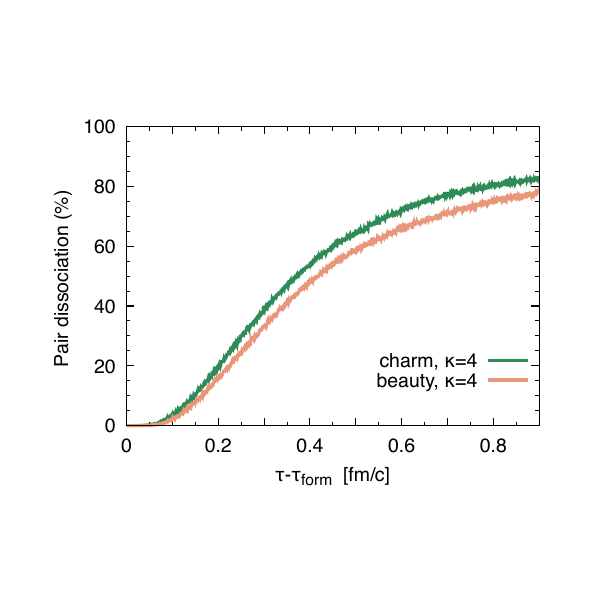}
\caption{Percentage of dissociated charm and bottom quark-antiquark pairs versus time $\tau-\tau_\mathrm{form}$, with $\tau_\mathrm{form}$ being the heavy quark formation time. The various curves correspond to different values of the dissociation parameters.}
\label{fig:dissoc}
\end{figure}

We quantify the impact of the early stage on the melting of the pairs by calculating the number of dissociated pairs with respect to their initial value. We show this quantity, in percentage, as a function of time $\tau-\tau_\mathrm{form}$ in Fig.~\ref{fig:dissoc} for both $c\bar{c}$ and $b\bar{b}$ pairs. 
By introducing the time to break half of the initial pairs, $\tau_\mathrm{break}$,
we find $\tau_\mathrm{break}-\tau_\mathrm{form}=0.37$ fm/c
for $c\bar c$ pairs, 
and  $\tau_\mathrm{break}-\tau_\mathrm{form}=0.42$ fm/c for $b\bar b$ pairs.
Similarly, taking $\kappa=9/4$ that amounts to require that 
$\mathcal{P}_\mathrm{survival}=1/e$ for $\mathcal{G}=1/3$, we find
$\tau_\mathrm{break}-\tau_\mathrm{form}=0.43$ fm/c
for $c\bar c$ pairs, 
and  $\tau_\mathrm{break}-\tau_\mathrm{form}=0.54$ fm/c for $b\bar b$ pairs.

We conclude that the effect of the interaction of the heavy quarks
with the gluon fields in the pre-equilibrium stage is to gradually
melt the pairs due to the color rotation of the quark and its companion
antiquark,
and the time needed to break half of the initial pairs
can be as small
as $0.4-0.5$ fm/c from the formation time. 
We finally notice 
that our results on the dissociation percentage and timescales are in the same 
ballpark as those shown in~\cite{Pooja:2024rnn},
in which the melting of pairs in the early stage was studied within a simpler model.

\section{Conclusions}
We studied the melting of $c\bar c$ and $b\bar b$ pairs in the 
pre-equilibrium stage of high-energy proton-nucleus collisions.
We modeled the early stage according to the color-glass-condensate
picture, choosing the glasma as the initial condition and solving 
the classical Yang-Mills equations to study its evolution.
The heavy quarks constituting the $c\bar c$ and $b\bar b$ pairs, 
which we assume to be produced in color-singlet states,
are then evolved on top of the classical gluon fields
via semi-classical equations of motion, known as the Wong equations.
Our main goal was the calculation of the percentage of pairs
that melt during their diffusion in the early stage.

We studied the evolution of the gauge-invariant correlator of the
color charges of the quark and its companion antiquark, $G(\tau)$.
At $\tau=\tau_\mathrm{form}$ we have $G(\tau_\mathrm{form})=1$
due to the color-singlet initialization of the pairs.
The interaction of the quarks with the dense
gluon environment leads to a degradation of the correlations.
Consequently, there are fluctuations to the color-octet state
at later times, and this process favors the melting of the pairs. 
The timescale of decorrelation,
$\tau_\mathrm{dec}$,
can be estimated, for example, 
by the requirement that $G(\tau_\mathrm{dec})=1/2$.
We found that $\tau_\mathrm{dec}-\tau_\mathrm{form}$ stays in the range
$(0.3,0.4)$ fm/c for both charm and beauty.
We also find that $\tau_\mathrm{dec}-\tau_\mathrm{form}$ 
for $b$ quarks is  larger than the similar quantity 
computed for the $c$ quarks: this is most likely due to the fact that
$b$ quarks are slower, so they spend more time within correlation domains
of the gluonic background. 

We reformulated the decorrelation of color charges 
in HQ pairs in terms of color-singlet and color-octet projectors, 
$P_S$ and $P_O$, 
which appear in our scheme as the classical counterparts of the quantum projection operators onto the singlet and octet representations of $SU(3)$. 
We identified $P_S$ and $P_O$ with the probabilities that the pair is in the singlet and octet states, respectively, 
thereby quantifying the singlet and octet components of the pair. Our findings indicate that the loss of color 
correlation leads to equilibration of color at large times,
similarly to the results of Ref.~\cite{Delorme:2024rdo}
for the QGP. Specifically, we found that at large times, 
$\langle P_S\rangle\approx \langle P_O\rangle/8\approx 1/9$, which we interpret as the pair having equal 
probability of being in the singlet state or in one of the 
eight possible octet states.

The diffusion of the heavy quarks in the evolving glasma fields leads
to the gradual melting of the pairs. Within our framework, the survival probability of each pair depends
on the fluctuations of the color charges that generate
a color-octet component for the pair and eventually break the pair itself.
These are related to the gauge field fluctuations
in the dense gluonic bath formed in the early stage.

For each pair at a given time, we accept its survival with the
probability $\mathcal{P}_\mathrm{survival}$
defined in Eq.~\eqref{eq:conte_non_contento}.
This probability depends on the color charges of the two particles that form the pair: these quantities are directly obtained from the solution of the Wong equations.
$\mathcal{P}_\mathrm{survival}$ depends on one parameter, $\kappa$, 
that measures
how large the color fluctuations need to be in order to melt the pair.
We introduced a breaking time, $\tau_\mathrm{break}$, defined as the value
of the proper time at which half of the pairs are melted.
The choice of not including the relative distance in 
the survival probability is motivated by noticing that $r_\mathrm{rel}$
remains almost constant during the whole pre-equilibrium stage, 
as a combined effect of the attractive potential and the broadening due
to the gluons in the bath. Therefore, it is natural to relate 
the melting probability solely to the color fluctuations.
We found that for both 
$c\bar c$ and $b\bar b$ pairs, 
$\tau_\mathrm{break}-\tau_\mathrm{form}$ 
is approximately in the range 
$(0.4,0.5)$ fm/c
Within the pre-equilibrium stage, it is color decorrelation that primarily 
drives the melting of the pair.

This work represents many improvements 
in comparison to Ref.~\cite{Pooja:2024rnn},
going beyond the $SU(2)$ glasma in a static box,
and more importantly, including for the first time a detailed study of
the color fluctuations in the pairs. Another improvement is the 
quark-antiquark potential for the pair,
with coupling proportional to $Q_a\bar Q_a$,
that evolves dynamically as a result of
the interaction of the pair with the background gluon field. 
This potential accounts for the fact that the pair does not remain in a pure singlet state during the evolution, allowing for the development of an octet component due to interactions with the background gluon field.
Finally,
we implemented the gauge-invariant formulation in terms of gauge links
and plaquettes, while in~\cite{Pooja:2024rnn} the authors used
the continuum formulation to solve the Yang-Mills equations.

The work presented here can be continued along several directions.
Firstly, it is desirable to link the final stage of our
pre-equilibrium evolution to relativistic transport,
in order to analyze the dynamics in the quark-gluon plasma stage. 
It is also possible to adopt a different approach, based on
a master equation for the density matrix of the pairs, to analyze
the color decorrelation and the melting of the pairs within a 
quantum-mechanical approach. Moreover, it will be interesting
to analyze the same problem within different scenarios in which
boost invariance is broken by longitudinal fluctuations of the background
gluon field. All these projects are currently under investigation and we
plan to report on them soon.

\subsection*{Acknowledgments}
The authors acknowledge 
discussions with Dana Avramescu and Raju Venugopalan.
Moreover, they acknowledge Pol Gossiaux,
who suggested them to expand the discussion on color projectors and color equilibration.
M. R. acknowledges Bruno Barbieri, Lautaro Martinez 
and John Petrucci for inspiration.
L. O. acknowledges funding from the European Union - Next Generation EU, Mission 4, Component 1, CUP E63C22002960006, for the project “Heavy flavour dynamics in the early stage of ultrarelativistic collisions” (HEFESTUS).
This work has been partly funded by 
PIACERI “Linea di intervento 1” (M@uRHIC) of the University of Catania, and 
the European Union – Next Generation EU through the research grants number P2022Z4P4B “SOPHYA - Sustainable Optimised PHYsics Algorithms: fundamental physics to build an advanced society” and 2022SM5YAS “Advanced probes of the Quark Gluon Plasma”, under the program PRIN 2022 PNRR of the Italian Ministero dell’Università e Ricerca (MUR).

\bibliography{References}

\begin{thebibliography}{131}%
\makeatletter
\providecommand \@ifxundefined [1]{%
 \@ifx{#1\undefined}
}%
\providecommand \@ifnum [1]{%
 \ifnum #1\expandafter \@firstoftwo
 \else \expandafter \@secondoftwo
 \fi
}%
\providecommand \@ifx [1]{%
 \ifx #1\expandafter \@firstoftwo
 \else \expandafter \@secondoftwo
 \fi
}%
\providecommand \natexlab [1]{#1}%
\providecommand \enquote  [1]{``#1''}%
\providecommand \bibnamefont  [1]{#1}%
\providecommand \bibfnamefont [1]{#1}%
\providecommand \citenamefont [1]{#1}%
\providecommand \href@noop [0]{\@secondoftwo}%
\providecommand \href [0]{\begingroup \@sanitize@url \@href}%
\providecommand \@href[1]{\@@startlink{#1}\@@href}%
\providecommand \@@href[1]{\endgroup#1\@@endlink}%
\providecommand \@sanitize@url [0]{\catcode `\\12\catcode `\$12\catcode `\&12\catcode `\#12\catcode `\^12\catcode `\_12\catcode `\%12\relax}%
\providecommand \@@startlink[1]{}%
\providecommand \@@endlink[0]{}%
\providecommand \url  [0]{\begingroup\@sanitize@url \@url }%
\providecommand \@url [1]{\endgroup\@href {#1}{\urlprefix }}%
\providecommand \urlprefix  [0]{URL }%
\providecommand \Eprint [0]{\href }%
\providecommand \doibase [0]{https://doi.org/}%
\providecommand \selectlanguage [0]{\@gobble}%
\providecommand \bibinfo  [0]{\@secondoftwo}%
\providecommand \bibfield  [0]{\@secondoftwo}%
\providecommand \translation [1]{[#1]}%
\providecommand \BibitemOpen [0]{}%
\providecommand \bibitemStop [0]{}%
\providecommand \bibitemNoStop [0]{.\EOS\space}%
\providecommand \EOS [0]{\spacefactor3000\relax}%
\providecommand \BibitemShut  [1]{\csname bibitem#1\endcsname}%
\let\auto@bib@innerbib\@empty
%</preamble>
\bibitem [{\citenamefont {Heinz}\ and\ \citenamefont {Jacob}(2000)}]{Heinz:2000bk}%
  \BibitemOpen
  \bibfield  {author} {\bibinfo {author} {\bibfnamefont {U.~W.}\ \bibnamefont {Heinz}}\ and\ \bibinfo {author} {\bibfnamefont {M.}~\bibnamefont {Jacob}},\ }\bibfield  {title} {\bibinfo {title} {{Evidence for a new state of matter: An Assessment of the results from the CERN lead beam program}},\ }\href@noop {} {\  (\bibinfo {year} {2000})},\ \Eprint {https://arxiv.org/abs/nucl-th/0002042} {arXiv:nucl-th/0002042} \BibitemShut {NoStop}%
\bibitem [{\citenamefont {Ruggieri}\ \emph {et~al.}(2014)\citenamefont {Ruggieri}, \citenamefont {Scardina}, \citenamefont {Plumari},\ and\ \citenamefont {Greco}}]{Ruggieri:2013ova}%
  \BibitemOpen
  \bibfield  {author} {\bibinfo {author} {\bibfnamefont {M.}~\bibnamefont {Ruggieri}}, \bibinfo {author} {\bibfnamefont {F.}~\bibnamefont {Scardina}}, \bibinfo {author} {\bibfnamefont {S.}~\bibnamefont {Plumari}},\ and\ \bibinfo {author} {\bibfnamefont {V.}~\bibnamefont {Greco}},\ }\bibfield  {title} {\bibinfo {title} {{Thermalization, Isotropization and Elliptic Flow from Nonequilibrium Initial Conditions with a Saturation Scale}},\ }\href {https://doi.org/10.1103/PhysRevC.89.054914} {\bibfield  {journal} {\bibinfo  {journal} {Phys. Rev. C}\ }\textbf {\bibinfo {volume} {89}},\ \bibinfo {pages} {054914} (\bibinfo {year} {2014})},\ \Eprint {https://arxiv.org/abs/1312.6060} {arXiv:1312.6060 [nucl-th]} \BibitemShut {NoStop}%
\bibitem [{\citenamefont {Ruggieri}\ \emph {et~al.}(2013)\citenamefont {Ruggieri}, \citenamefont {Scardina}, \citenamefont {Plumari},\ and\ \citenamefont {Greco}}]{Ruggieri:2013bda}%
  \BibitemOpen
  \bibfield  {author} {\bibinfo {author} {\bibfnamefont {M.}~\bibnamefont {Ruggieri}}, \bibinfo {author} {\bibfnamefont {F.}~\bibnamefont {Scardina}}, \bibinfo {author} {\bibfnamefont {S.}~\bibnamefont {Plumari}},\ and\ \bibinfo {author} {\bibfnamefont {V.}~\bibnamefont {Greco}},\ }\bibfield  {title} {\bibinfo {title} {{Elliptic Flow from Non-equilibrium Initial Condition with a Saturation Scale}},\ }\href {https://doi.org/10.1016/j.physletb.2013.10.014} {\bibfield  {journal} {\bibinfo  {journal} {Phys. Lett. B}\ }\textbf {\bibinfo {volume} {727}},\ \bibinfo {pages} {177} (\bibinfo {year} {2013})},\ \Eprint {https://arxiv.org/abs/1303.3178} {arXiv:1303.3178 [nucl-th]} \BibitemShut {NoStop}%
\bibitem [{\citenamefont {Plumari}(2019)}]{Plumari:2019gwq}%
  \BibitemOpen
  \bibfield  {author} {\bibinfo {author} {\bibfnamefont {S.}~\bibnamefont {Plumari}},\ }\bibfield  {title} {\bibinfo {title} {{Anisotropic flows and the shear viscosity of the QGP within an event-by-event massive parton transport approach}},\ }\href {https://doi.org/10.1140/epjc/s10052-018-6510-9} {\bibfield  {journal} {\bibinfo  {journal} {Eur. Phys. J. C}\ }\textbf {\bibinfo {volume} {79}},\ \bibinfo {pages} {2} (\bibinfo {year} {2019})}\BibitemShut {NoStop}%
\bibitem [{\citenamefont {Dusling}\ \emph {et~al.}(2016)\citenamefont {Dusling}, \citenamefont {Li},\ and\ \citenamefont {Schenke}}]{Dusling:2015gta}%
  \BibitemOpen
  \bibfield  {author} {\bibinfo {author} {\bibfnamefont {K.}~\bibnamefont {Dusling}}, \bibinfo {author} {\bibfnamefont {W.}~\bibnamefont {Li}},\ and\ \bibinfo {author} {\bibfnamefont {B.}~\bibnamefont {Schenke}},\ }\bibfield  {title} {\bibinfo {title} {{Novel collective phenomena in high-energy proton\textendash{}proton and proton\textendash{}nucleus collisions}},\ }\href {https://doi.org/10.1142/S0218301316300022} {\bibfield  {journal} {\bibinfo  {journal} {Int. J. Mod. Phys. E}\ }\textbf {\bibinfo {volume} {25}},\ \bibinfo {pages} {1630002} (\bibinfo {year} {2016})},\ \Eprint {https://arxiv.org/abs/1509.07939} {arXiv:1509.07939 [nucl-ex]} \BibitemShut {NoStop}%
\bibitem [{\citenamefont {Kurkela}\ \emph {et~al.}(2019{\natexlab{a}})\citenamefont {Kurkela}, \citenamefont {Wiedemann},\ and\ \citenamefont {Wu}}]{Kurkela:2018qeb}%
  \BibitemOpen
  \bibfield  {author} {\bibinfo {author} {\bibfnamefont {A.}~\bibnamefont {Kurkela}}, \bibinfo {author} {\bibfnamefont {U.~A.}\ \bibnamefont {Wiedemann}},\ and\ \bibinfo {author} {\bibfnamefont {B.}~\bibnamefont {Wu}},\ }\bibfield  {title} {\bibinfo {title} {{Opacity dependence of elliptic flow in kinetic theory}},\ }\href {https://doi.org/10.1140/epjc/s10052-019-7262-x} {\bibfield  {journal} {\bibinfo  {journal} {Eur. Phys. J. C}\ }\textbf {\bibinfo {volume} {79}},\ \bibinfo {pages} {759} (\bibinfo {year} {2019}{\natexlab{a}})},\ \Eprint {https://arxiv.org/abs/1805.04081} {arXiv:1805.04081 [hep-ph]} \BibitemShut {NoStop}%
\bibitem [{\citenamefont {Kurkela}\ \emph {et~al.}(2019{\natexlab{b}})\citenamefont {Kurkela}, \citenamefont {Wiedemann},\ and\ \citenamefont {Wu}}]{Kurkela:2019kip}%
  \BibitemOpen
  \bibfield  {author} {\bibinfo {author} {\bibfnamefont {A.}~\bibnamefont {Kurkela}}, \bibinfo {author} {\bibfnamefont {U.~A.}\ \bibnamefont {Wiedemann}},\ and\ \bibinfo {author} {\bibfnamefont {B.}~\bibnamefont {Wu}},\ }\bibfield  {title} {\bibinfo {title} {{Flow in AA and pA as an interplay of fluid-like and non-fluid like excitations}},\ }\href {https://doi.org/10.1140/epjc/s10052-019-7428-6} {\bibfield  {journal} {\bibinfo  {journal} {Eur. Phys. J. C}\ }\textbf {\bibinfo {volume} {79}},\ \bibinfo {pages} {965} (\bibinfo {year} {2019}{\natexlab{b}})},\ \Eprint {https://arxiv.org/abs/1905.05139} {arXiv:1905.05139 [hep-ph]} \BibitemShut {NoStop}%
\bibitem [{\citenamefont {Greif}\ \emph {et~al.}(2017)\citenamefont {Greif}, \citenamefont {Greiner}, \citenamefont {Schenke}, \citenamefont {Schlichting},\ and\ \citenamefont {Xu}}]{Greif:2017bnr}%
  \BibitemOpen
  \bibfield  {author} {\bibinfo {author} {\bibfnamefont {M.}~\bibnamefont {Greif}}, \bibinfo {author} {\bibfnamefont {C.}~\bibnamefont {Greiner}}, \bibinfo {author} {\bibfnamefont {B.}~\bibnamefont {Schenke}}, \bibinfo {author} {\bibfnamefont {S.}~\bibnamefont {Schlichting}},\ and\ \bibinfo {author} {\bibfnamefont {Z.}~\bibnamefont {Xu}},\ }\bibfield  {title} {\bibinfo {title} {{Importance of initial and final state effects for azimuthal correlations in p+Pb collisions}},\ }\href {https://doi.org/10.1103/PhysRevD.96.091504} {\bibfield  {journal} {\bibinfo  {journal} {Phys. Rev. D}\ }\textbf {\bibinfo {volume} {96}},\ \bibinfo {pages} {091504} (\bibinfo {year} {2017})},\ \Eprint {https://arxiv.org/abs/1708.02076} {arXiv:1708.02076 [hep-ph]} \BibitemShut {NoStop}%
\bibitem [{\citenamefont {Zhao}\ \emph {et~al.}(2023)\citenamefont {Zhao}, \citenamefont {Ryu}, \citenamefont {Shen},\ and\ \citenamefont {Schenke}}]{Zhao:2022ugy}%
  \BibitemOpen
  \bibfield  {author} {\bibinfo {author} {\bibfnamefont {W.}~\bibnamefont {Zhao}}, \bibinfo {author} {\bibfnamefont {S.}~\bibnamefont {Ryu}}, \bibinfo {author} {\bibfnamefont {C.}~\bibnamefont {Shen}},\ and\ \bibinfo {author} {\bibfnamefont {B.}~\bibnamefont {Schenke}},\ }\bibfield  {title} {\bibinfo {title} {{3D structure of anisotropic flow in small collision systems at energies available at the BNL Relativistic Heavy Ion Collider}},\ }\href {https://doi.org/10.1103/PhysRevC.107.014904} {\bibfield  {journal} {\bibinfo  {journal} {Phys. Rev. C}\ }\textbf {\bibinfo {volume} {107}},\ \bibinfo {pages} {014904} (\bibinfo {year} {2023})},\ \Eprint {https://arxiv.org/abs/2211.16376} {arXiv:2211.16376 [nucl-th]} \BibitemShut {NoStop}%
\bibitem [{\citenamefont {Oliva}\ \emph {et~al.}(2022)\citenamefont {Oliva}, \citenamefont {Fan}, \citenamefont {Moreau}, \citenamefont {Bass},\ and\ \citenamefont {Bratkovskaya}}]{Oliva:2022rsv}%
  \BibitemOpen
  \bibfield  {author} {\bibinfo {author} {\bibfnamefont {L.}~\bibnamefont {Oliva}}, \bibinfo {author} {\bibfnamefont {W.}~\bibnamefont {Fan}}, \bibinfo {author} {\bibfnamefont {P.}~\bibnamefont {Moreau}}, \bibinfo {author} {\bibfnamefont {S.~A.}\ \bibnamefont {Bass}},\ and\ \bibinfo {author} {\bibfnamefont {E.}~\bibnamefont {Bratkovskaya}},\ }\bibfield  {title} {\bibinfo {title} {{Nonequilibrium effects and transverse spherocity in ultrarelativistic proton-nucleus collisions}},\ }\href {https://doi.org/10.1103/PhysRevC.106.044910} {\bibfield  {journal} {\bibinfo  {journal} {Phys. Rev. C}\ }\textbf {\bibinfo {volume} {106}},\ \bibinfo {pages} {044910} (\bibinfo {year} {2022})},\ \Eprint {https://arxiv.org/abs/2204.04194} {arXiv:2204.04194 [nucl-th]} \BibitemShut {NoStop}%
\bibitem [{\citenamefont {Oliva}\ \emph {et~al.}(2023)\citenamefont {Oliva}, \citenamefont {Fan}, \citenamefont {Moreau}, \citenamefont {Bass},\ and\ \citenamefont {Bratkovskaya}}]{Oliva:2023dti}%
  \BibitemOpen
  \bibfield  {author} {\bibinfo {author} {\bibfnamefont {L.}~\bibnamefont {Oliva}}, \bibinfo {author} {\bibfnamefont {W.}~\bibnamefont {Fan}}, \bibinfo {author} {\bibfnamefont {P.}~\bibnamefont {Moreau}}, \bibinfo {author} {\bibfnamefont {S.~A.}\ \bibnamefont {Bass}},\ and\ \bibinfo {author} {\bibfnamefont {E.}~\bibnamefont {Bratkovskaya}},\ }\bibfield  {title} {\bibinfo {title} {{Non-equilibrium Dynamics and Collectivity in Ultra-relativistic Proton\textendash{}Nucleus Collisions}},\ }\href {https://doi.org/10.5506/APhysPolBSupp.16.1-A68} {\bibfield  {journal} {\bibinfo  {journal} {Acta Phys. Polon. Supp.}\ }\textbf {\bibinfo {volume} {16}},\ \bibinfo {pages} {1} (\bibinfo {year} {2023})}\BibitemShut {NoStop}%
\bibitem [{\citenamefont {Nugara}\ \emph {et~al.}(2025)\citenamefont {Nugara}, \citenamefont {Greco},\ and\ \citenamefont {Plumari}}]{Nugara:2024net}%
  \BibitemOpen
  \bibfield  {author} {\bibinfo {author} {\bibfnamefont {V.}~\bibnamefont {Nugara}}, \bibinfo {author} {\bibfnamefont {V.}~\bibnamefont {Greco}},\ and\ \bibinfo {author} {\bibfnamefont {S.}~\bibnamefont {Plumari}},\ }\bibfield  {title} {\bibinfo {title} {{Far-from-equilibrium attractors with Full Relativistic Boltzmann approach in 3+1D: moments of distribution function and~anisotropic flows $v_n$}},\ }\href {https://doi.org/10.1140/epjc/s10052-025-14029-9} {\bibfield  {journal} {\bibinfo  {journal} {Eur. Phys. J. C}\ }\textbf {\bibinfo {volume} {85}},\ \bibinfo {pages} {311} (\bibinfo {year} {2025})},\ \Eprint {https://arxiv.org/abs/2409.12123} {arXiv:2409.12123 [hep-ph]} \BibitemShut {NoStop}%
\bibitem [{\citenamefont {Grosse-Oetringhaus}\ and\ \citenamefont {Wiedemann}(2024)}]{Grosse-Oetringhaus:2024bwr}%
  \BibitemOpen
  \bibfield  {author} {\bibinfo {author} {\bibfnamefont {J.~F.}\ \bibnamefont {Grosse-Oetringhaus}}\ and\ \bibinfo {author} {\bibfnamefont {U.~A.}\ \bibnamefont {Wiedemann}},\ }\bibfield  {title} {\bibinfo {title} {{A Decade of Collectivity in Small Systems}},\ }\href@noop {} {\  (\bibinfo {year} {2024})}\BibitemShut {NoStop}%
\bibitem [{\citenamefont {Lappi}\ and\ \citenamefont {McLerran}(2006)}]{Lappi:2006fp}%
  \BibitemOpen
  \bibfield  {author} {\bibinfo {author} {\bibfnamefont {T.}~\bibnamefont {Lappi}}\ and\ \bibinfo {author} {\bibfnamefont {L.}~\bibnamefont {McLerran}},\ }\bibfield  {title} {\bibinfo {title} {{Some features of the glasma}},\ }\href {https://doi.org/10.1016/j.nuclphysa.2006.04.001} {\bibfield  {journal} {\bibinfo  {journal} {Nucl. Phys. A}\ }\textbf {\bibinfo {volume} {772}},\ \bibinfo {pages} {200} (\bibinfo {year} {2006})},\ \Eprint {https://arxiv.org/abs/hep-ph/0602189} {arXiv:hep-ph/0602189} \BibitemShut {NoStop}%
\bibitem [{\citenamefont {McLerran}\ and\ \citenamefont {Venugopalan}(1994{\natexlab{a}})}]{McLerran:1993ni}%
  \BibitemOpen
  \bibfield  {author} {\bibinfo {author} {\bibfnamefont {L.~D.}\ \bibnamefont {McLerran}}\ and\ \bibinfo {author} {\bibfnamefont {R.}~\bibnamefont {Venugopalan}},\ }\bibfield  {title} {\bibinfo {title} {{Computing quark and gluon distribution functions for very large nuclei}},\ }\href {https://doi.org/10.1103/PhysRevD.49.2233} {\bibfield  {journal} {\bibinfo  {journal} {Phys. Rev. D}\ }\textbf {\bibinfo {volume} {49}},\ \bibinfo {pages} {2233} (\bibinfo {year} {1994}{\natexlab{a}})},\ \Eprint {https://arxiv.org/abs/hep-ph/9309289} {arXiv:hep-ph/9309289} \BibitemShut {NoStop}%
\bibitem [{\citenamefont {McLerran}\ and\ \citenamefont {Venugopalan}(1994{\natexlab{b}})}]{McLerran:1993ka}%
  \BibitemOpen
  \bibfield  {author} {\bibinfo {author} {\bibfnamefont {L.~D.}\ \bibnamefont {McLerran}}\ and\ \bibinfo {author} {\bibfnamefont {R.}~\bibnamefont {Venugopalan}},\ }\bibfield  {title} {\bibinfo {title} {{Gluon distribution functions for very large nuclei at small transverse momentum}},\ }\href {https://doi.org/10.1103/PhysRevD.49.3352} {\bibfield  {journal} {\bibinfo  {journal} {Phys. Rev. D}\ }\textbf {\bibinfo {volume} {49}},\ \bibinfo {pages} {3352} (\bibinfo {year} {1994}{\natexlab{b}})},\ \Eprint {https://arxiv.org/abs/hep-ph/9311205} {arXiv:hep-ph/9311205} \BibitemShut {NoStop}%
\bibitem [{\citenamefont {McLerran}\ and\ \citenamefont {Venugopalan}(1994{\natexlab{c}})}]{McLerran:1994vd}%
  \BibitemOpen
  \bibfield  {author} {\bibinfo {author} {\bibfnamefont {L.~D.}\ \bibnamefont {McLerran}}\ and\ \bibinfo {author} {\bibfnamefont {R.}~\bibnamefont {Venugopalan}},\ }\bibfield  {title} {\bibinfo {title} {{Green's functions in the color field of a large nucleus}},\ }\href {https://doi.org/10.1103/PhysRevD.50.2225} {\bibfield  {journal} {\bibinfo  {journal} {Phys. Rev. D}\ }\textbf {\bibinfo {volume} {50}},\ \bibinfo {pages} {2225} (\bibinfo {year} {1994}{\natexlab{c}})},\ \Eprint {https://arxiv.org/abs/hep-ph/9402335} {arXiv:hep-ph/9402335} \BibitemShut {NoStop}%
\bibitem [{\citenamefont {Gelis}\ \emph {et~al.}(2010)\citenamefont {Gelis}, \citenamefont {Iancu}, \citenamefont {Jalilian-Marian},\ and\ \citenamefont {Venugopalan}}]{Gelis:2010nm}%
  \BibitemOpen
  \bibfield  {author} {\bibinfo {author} {\bibfnamefont {F.}~\bibnamefont {Gelis}}, \bibinfo {author} {\bibfnamefont {E.}~\bibnamefont {Iancu}}, \bibinfo {author} {\bibfnamefont {J.}~\bibnamefont {Jalilian-Marian}},\ and\ \bibinfo {author} {\bibfnamefont {R.}~\bibnamefont {Venugopalan}},\ }\bibfield  {title} {\bibinfo {title} {{The Color Glass Condensate}},\ }\href {https://doi.org/10.1146/annurev.nucl.010909.083629} {\bibfield  {journal} {\bibinfo  {journal} {Ann. Rev. Nucl. Part. Sci.}\ }\textbf {\bibinfo {volume} {60}},\ \bibinfo {pages} {463} (\bibinfo {year} {2010})},\ \Eprint {https://arxiv.org/abs/1002.0333} {arXiv:1002.0333 [hep-ph]} \BibitemShut {NoStop}%
\bibitem [{\citenamefont {Iancu}\ and\ \citenamefont {Venugopalan}(2003)}]{Iancu:2003xm}%
  \BibitemOpen
  \bibfield  {author} {\bibinfo {author} {\bibfnamefont {E.}~\bibnamefont {Iancu}}\ and\ \bibinfo {author} {\bibfnamefont {R.}~\bibnamefont {Venugopalan}},\ }\bibinfo {title} {{The Color glass condensate and high-energy scattering in QCD}},\ in\ \href {https://doi.org/10.1142/9789812795533_0005} {\emph {\bibinfo {booktitle} {{Quark-gluon plasma 4}}}},\ \bibinfo {editor} {edited by\ \bibinfo {editor} {\bibfnamefont {R.~C.}\ \bibnamefont {Hwa}}\ and\ \bibinfo {editor} {\bibfnamefont {X.-N.}\ \bibnamefont {Wang}}}\ (\bibinfo {year} {2003})\ pp.\ \bibinfo {pages} {249--3363},\ \Eprint {https://arxiv.org/abs/hep-ph/0303204} {arXiv:hep-ph/0303204} \BibitemShut {NoStop}%
\bibitem [{\citenamefont {McLerran}(2009)}]{McLerran:2008es}%
  \BibitemOpen
  \bibfield  {author} {\bibinfo {author} {\bibfnamefont {L.}~\bibnamefont {McLerran}},\ }\bibfield  {title} {\bibinfo {title} {{A Brief Introduction to the Color Glass Condensate and the Glasma}},\ }in\ \href {https://doi.org/10.3204/DESY-PROC-2009-01/26} {\emph {\bibinfo {booktitle} {{38th International Symposium on Multiparticle Dynamics}}}}\ (\bibinfo {year} {2009})\ pp.\ \bibinfo {pages} {3--18},\ \Eprint {https://arxiv.org/abs/0812.4989} {arXiv:0812.4989 [hep-ph]} \BibitemShut {NoStop}%
\bibitem [{\citenamefont {Gelis}(2013)}]{Gelis:2012ri}%
  \BibitemOpen
  \bibfield  {author} {\bibinfo {author} {\bibfnamefont {F.}~\bibnamefont {Gelis}},\ }\bibfield  {title} {\bibinfo {title} {{Color Glass Condensate and Glasma}},\ }\href {https://doi.org/10.1142/S0217751X13300019} {\bibfield  {journal} {\bibinfo  {journal} {Int. J. Mod. Phys. A}\ }\textbf {\bibinfo {volume} {28}},\ \bibinfo {pages} {1330001} (\bibinfo {year} {2013})},\ \Eprint {https://arxiv.org/abs/1211.3327} {arXiv:1211.3327 [hep-ph]} \BibitemShut {NoStop}%
\bibitem [{\citenamefont {Schenke}\ \emph {et~al.}(2012{\natexlab{a}})\citenamefont {Schenke}, \citenamefont {Tribedy},\ and\ \citenamefont {Venugopalan}}]{Schenke:2012hg}%
  \BibitemOpen
  \bibfield  {author} {\bibinfo {author} {\bibfnamefont {B.}~\bibnamefont {Schenke}}, \bibinfo {author} {\bibfnamefont {P.}~\bibnamefont {Tribedy}},\ and\ \bibinfo {author} {\bibfnamefont {R.}~\bibnamefont {Venugopalan}},\ }\bibfield  {title} {\bibinfo {title} {{Event-by-event gluon multiplicity, energy density, and eccentricities in ultrarelativistic heavy-ion collisions}},\ }\href {https://doi.org/10.1103/PhysRevC.86.034908} {\bibfield  {journal} {\bibinfo  {journal} {Phys. Rev. C}\ }\textbf {\bibinfo {volume} {86}},\ \bibinfo {pages} {034908} (\bibinfo {year} {2012}{\natexlab{a}})},\ \Eprint {https://arxiv.org/abs/1206.6805} {arXiv:1206.6805 [hep-ph]} \BibitemShut {NoStop}%
\bibitem [{\citenamefont {Schenke}\ \emph {et~al.}(2012{\natexlab{b}})\citenamefont {Schenke}, \citenamefont {Tribedy},\ and\ \citenamefont {Venugopalan}}]{Schenke:2012wb}%
  \BibitemOpen
  \bibfield  {author} {\bibinfo {author} {\bibfnamefont {B.}~\bibnamefont {Schenke}}, \bibinfo {author} {\bibfnamefont {P.}~\bibnamefont {Tribedy}},\ and\ \bibinfo {author} {\bibfnamefont {R.}~\bibnamefont {Venugopalan}},\ }\bibfield  {title} {\bibinfo {title} {{Fluctuating Glasma initial conditions and flow in heavy ion collisions}},\ }\href {https://doi.org/10.1103/PhysRevLett.108.252301} {\bibfield  {journal} {\bibinfo  {journal} {Phys. Rev. Lett.}\ }\textbf {\bibinfo {volume} {108}},\ \bibinfo {pages} {252301} (\bibinfo {year} {2012}{\natexlab{b}})},\ \Eprint {https://arxiv.org/abs/1202.6646} {arXiv:1202.6646 [nucl-th]} \BibitemShut {NoStop}%
\bibitem [{\citenamefont {Moore}\ and\ \citenamefont {Teaney}(2005)}]{Moore:2004tg}%
  \BibitemOpen
  \bibfield  {author} {\bibinfo {author} {\bibfnamefont {G.~D.}\ \bibnamefont {Moore}}\ and\ \bibinfo {author} {\bibfnamefont {D.}~\bibnamefont {Teaney}},\ }\bibfield  {title} {\bibinfo {title} {{How much do heavy quarks thermalize in a heavy ion collision?}},\ }\href {https://doi.org/10.1103/PhysRevC.71.064904} {\bibfield  {journal} {\bibinfo  {journal} {Phys. Rev. C}\ }\textbf {\bibinfo {volume} {71}},\ \bibinfo {pages} {064904} (\bibinfo {year} {2005})},\ \Eprint {https://arxiv.org/abs/hep-ph/0412346} {arXiv:hep-ph/0412346} \BibitemShut {NoStop}%
\bibitem [{\citenamefont {van Hees}\ \emph {et~al.}(2006)\citenamefont {van Hees}, \citenamefont {Greco},\ and\ \citenamefont {Rapp}}]{vanHees:2005wb}%
  \BibitemOpen
  \bibfield  {author} {\bibinfo {author} {\bibfnamefont {H.}~\bibnamefont {van Hees}}, \bibinfo {author} {\bibfnamefont {V.}~\bibnamefont {Greco}},\ and\ \bibinfo {author} {\bibfnamefont {R.}~\bibnamefont {Rapp}},\ }\bibfield  {title} {\bibinfo {title} {{Heavy-quark probes of the quark-gluon plasma at RHIC}},\ }\href {https://doi.org/10.1103/PhysRevC.73.034913} {\bibfield  {journal} {\bibinfo  {journal} {Phys. Rev. C}\ }\textbf {\bibinfo {volume} {73}},\ \bibinfo {pages} {034913} (\bibinfo {year} {2006})},\ \Eprint {https://arxiv.org/abs/nucl-th/0508055} {arXiv:nucl-th/0508055} \BibitemShut {NoStop}%
\bibitem [{\citenamefont {van Hees}\ \emph {et~al.}(2008)\citenamefont {van Hees}, \citenamefont {Mannarelli}, \citenamefont {Greco},\ and\ \citenamefont {Rapp}}]{vanHees:2007me}%
  \BibitemOpen
  \bibfield  {author} {\bibinfo {author} {\bibfnamefont {H.}~\bibnamefont {van Hees}}, \bibinfo {author} {\bibfnamefont {M.}~\bibnamefont {Mannarelli}}, \bibinfo {author} {\bibfnamefont {V.}~\bibnamefont {Greco}},\ and\ \bibinfo {author} {\bibfnamefont {R.}~\bibnamefont {Rapp}},\ }\bibfield  {title} {\bibinfo {title} {{Nonperturbative heavy-quark diffusion in the quark-gluon plasma}},\ }\href {https://doi.org/10.1103/PhysRevLett.100.192301} {\bibfield  {journal} {\bibinfo  {journal} {Phys. Rev. Lett.}\ }\textbf {\bibinfo {volume} {100}},\ \bibinfo {pages} {192301} (\bibinfo {year} {2008})},\ \Eprint {https://arxiv.org/abs/0709.2884} {arXiv:0709.2884 [hep-ph]} \BibitemShut {NoStop}%
\bibitem [{\citenamefont {Gossiaux}\ and\ \citenamefont {Aichelin}(2008)}]{Gossiaux:2008jv}%
  \BibitemOpen
  \bibfield  {author} {\bibinfo {author} {\bibfnamefont {P.~B.}\ \bibnamefont {Gossiaux}}\ and\ \bibinfo {author} {\bibfnamefont {J.}~\bibnamefont {Aichelin}},\ }\bibfield  {title} {\bibinfo {title} {{Towards an understanding of the RHIC single electron data}},\ }\href {https://doi.org/10.1103/PhysRevC.78.014904} {\bibfield  {journal} {\bibinfo  {journal} {Phys. Rev. C}\ }\textbf {\bibinfo {volume} {78}},\ \bibinfo {pages} {014904} (\bibinfo {year} {2008})},\ \Eprint {https://arxiv.org/abs/0802.2525} {arXiv:0802.2525 [hep-ph]} \BibitemShut {NoStop}%
\bibitem [{\citenamefont {He}\ \emph {et~al.}(2012)\citenamefont {He}, \citenamefont {Fries},\ and\ \citenamefont {Rapp}}]{He:2011qa}%
  \BibitemOpen
  \bibfield  {author} {\bibinfo {author} {\bibfnamefont {M.}~\bibnamefont {He}}, \bibinfo {author} {\bibfnamefont {R.~J.}\ \bibnamefont {Fries}},\ and\ \bibinfo {author} {\bibfnamefont {R.}~\bibnamefont {Rapp}},\ }\bibfield  {title} {\bibinfo {title} {{Heavy-Quark Diffusion and Hadronization in Quark-Gluon Plasma}},\ }\href {https://doi.org/10.1103/PhysRevC.86.014903} {\bibfield  {journal} {\bibinfo  {journal} {Phys. Rev. C}\ }\textbf {\bibinfo {volume} {86}},\ \bibinfo {pages} {014903} (\bibinfo {year} {2012})},\ \Eprint {https://arxiv.org/abs/1106.6006} {arXiv:1106.6006 [nucl-th]} \BibitemShut {NoStop}%
\bibitem [{\citenamefont {Alberico}\ \emph {et~al.}(2011)\citenamefont {Alberico}, \citenamefont {Beraudo}, \citenamefont {De~Pace}, \citenamefont {Molinari}, \citenamefont {Monteno}, \citenamefont {Nardi},\ and\ \citenamefont {Prino}}]{Alberico:2011zy}%
  \BibitemOpen
  \bibfield  {author} {\bibinfo {author} {\bibfnamefont {W.~M.}\ \bibnamefont {Alberico}}, \bibinfo {author} {\bibfnamefont {A.}~\bibnamefont {Beraudo}}, \bibinfo {author} {\bibfnamefont {A.}~\bibnamefont {De~Pace}}, \bibinfo {author} {\bibfnamefont {A.}~\bibnamefont {Molinari}}, \bibinfo {author} {\bibfnamefont {M.}~\bibnamefont {Monteno}}, \bibinfo {author} {\bibfnamefont {M.}~\bibnamefont {Nardi}},\ and\ \bibinfo {author} {\bibfnamefont {F.}~\bibnamefont {Prino}},\ }\bibfield  {title} {\bibinfo {title} {{Heavy-flavour spectra in high energy nucleus-nucleus collisions}},\ }\href {https://doi.org/10.1140/epjc/s10052-011-1666-6} {\bibfield  {journal} {\bibinfo  {journal} {Eur. Phys. J. C}\ }\textbf {\bibinfo {volume} {71}},\ \bibinfo {pages} {1666} (\bibinfo {year} {2011})},\ \Eprint {https://arxiv.org/abs/1101.6008} {arXiv:1101.6008 [hep-ph]} \BibitemShut {NoStop}%
\bibitem [{\citenamefont {Lang}\ \emph {et~al.}(2016)\citenamefont {Lang}, \citenamefont {van Hees}, \citenamefont {Steinheimer}, \citenamefont {Inghirami},\ and\ \citenamefont {Bleicher}}]{Lang:2012nqy}%
  \BibitemOpen
  \bibfield  {author} {\bibinfo {author} {\bibfnamefont {T.}~\bibnamefont {Lang}}, \bibinfo {author} {\bibfnamefont {H.}~\bibnamefont {van Hees}}, \bibinfo {author} {\bibfnamefont {J.}~\bibnamefont {Steinheimer}}, \bibinfo {author} {\bibfnamefont {G.}~\bibnamefont {Inghirami}},\ and\ \bibinfo {author} {\bibfnamefont {M.}~\bibnamefont {Bleicher}},\ }\bibfield  {title} {\bibinfo {title} {{Heavy quark transport in heavy ion collisions at energies available at the BNL Relativistic Heavy Ion Collider and at the CERN Large Hadron Collider within the UrQMD hybrid model}},\ }\href {https://doi.org/10.1103/PhysRevC.93.014901} {\bibfield  {journal} {\bibinfo  {journal} {Phys. Rev. C}\ }\textbf {\bibinfo {volume} {93}},\ \bibinfo {pages} {014901} (\bibinfo {year} {2016})},\ \Eprint {https://arxiv.org/abs/1211.6912} {arXiv:1211.6912 [hep-ph]} \BibitemShut {NoStop}%
\bibitem [{\citenamefont {Das}\ \emph {et~al.}(2014)\citenamefont {Das}, \citenamefont {Scardina}, \citenamefont {Plumari},\ and\ \citenamefont {Greco}}]{Das:2013kea}%
  \BibitemOpen
  \bibfield  {author} {\bibinfo {author} {\bibfnamefont {S.~K.}\ \bibnamefont {Das}}, \bibinfo {author} {\bibfnamefont {F.}~\bibnamefont {Scardina}}, \bibinfo {author} {\bibfnamefont {S.}~\bibnamefont {Plumari}},\ and\ \bibinfo {author} {\bibfnamefont {V.}~\bibnamefont {Greco}},\ }\bibfield  {title} {\bibinfo {title} {{Heavy-flavor in-medium momentum evolution: Langevin versus Boltzmann approach}},\ }\href {https://doi.org/10.1103/PhysRevC.90.044901} {\bibfield  {journal} {\bibinfo  {journal} {Phys. Rev. C}\ }\textbf {\bibinfo {volume} {90}},\ \bibinfo {pages} {044901} (\bibinfo {year} {2014})},\ \Eprint {https://arxiv.org/abs/1312.6857} {arXiv:1312.6857 [nucl-th]} \BibitemShut {NoStop}%
\bibitem [{\citenamefont {Song}\ \emph {et~al.}(2015)\citenamefont {Song}, \citenamefont {Berrehrah}, \citenamefont {Cabrera}, \citenamefont {Torres-Rincon}, \citenamefont {Tolos}, \citenamefont {Cassing},\ and\ \citenamefont {Bratkovskaya}}]{Song:2015sfa}%
  \BibitemOpen
  \bibfield  {author} {\bibinfo {author} {\bibfnamefont {T.}~\bibnamefont {Song}}, \bibinfo {author} {\bibfnamefont {H.}~\bibnamefont {Berrehrah}}, \bibinfo {author} {\bibfnamefont {D.}~\bibnamefont {Cabrera}}, \bibinfo {author} {\bibfnamefont {J.~M.}\ \bibnamefont {Torres-Rincon}}, \bibinfo {author} {\bibfnamefont {L.}~\bibnamefont {Tolos}}, \bibinfo {author} {\bibfnamefont {W.}~\bibnamefont {Cassing}},\ and\ \bibinfo {author} {\bibfnamefont {E.}~\bibnamefont {Bratkovskaya}},\ }\bibfield  {title} {\bibinfo {title} {{Tomography of the Quark-Gluon-Plasma by Charm Quarks}},\ }\href {https://doi.org/10.1103/PhysRevC.92.014910} {\bibfield  {journal} {\bibinfo  {journal} {Phys. Rev. C}\ }\textbf {\bibinfo {volume} {92}},\ \bibinfo {pages} {014910} (\bibinfo {year} {2015})},\ \Eprint {https://arxiv.org/abs/1503.03039} {arXiv:1503.03039 [nucl-th]} \BibitemShut {NoStop}%
\bibitem [{\citenamefont {Song}\ \emph {et~al.}(2016)\citenamefont {Song}, \citenamefont {Berrehrah}, \citenamefont {Cabrera}, \citenamefont {Cassing},\ and\ \citenamefont {Bratkovskaya}}]{Song:2015ykw}%
  \BibitemOpen
  \bibfield  {author} {\bibinfo {author} {\bibfnamefont {T.}~\bibnamefont {Song}}, \bibinfo {author} {\bibfnamefont {H.}~\bibnamefont {Berrehrah}}, \bibinfo {author} {\bibfnamefont {D.}~\bibnamefont {Cabrera}}, \bibinfo {author} {\bibfnamefont {W.}~\bibnamefont {Cassing}},\ and\ \bibinfo {author} {\bibfnamefont {E.}~\bibnamefont {Bratkovskaya}},\ }\bibfield  {title} {\bibinfo {title} {{Charm production in Pb + Pb collisions at energies available at the CERN Large Hadron Collider}},\ }\href {https://doi.org/10.1103/PhysRevC.93.034906} {\bibfield  {journal} {\bibinfo  {journal} {Phys. Rev. C}\ }\textbf {\bibinfo {volume} {93}},\ \bibinfo {pages} {034906} (\bibinfo {year} {2016})},\ \Eprint {https://arxiv.org/abs/1512.00891} {arXiv:1512.00891 [nucl-th]} \BibitemShut {NoStop}%
\bibitem [{\citenamefont {Cao}\ \emph {et~al.}(2015)\citenamefont {Cao}, \citenamefont {Qin},\ and\ \citenamefont {Bass}}]{Cao:2015hia}%
  \BibitemOpen
  \bibfield  {author} {\bibinfo {author} {\bibfnamefont {S.}~\bibnamefont {Cao}}, \bibinfo {author} {\bibfnamefont {G.-Y.}\ \bibnamefont {Qin}},\ and\ \bibinfo {author} {\bibfnamefont {S.~A.}\ \bibnamefont {Bass}},\ }\bibfield  {title} {\bibinfo {title} {{Energy loss, hadronization and hadronic interactions of heavy flavors in relativistic heavy-ion collisions}},\ }\href {https://doi.org/10.1103/PhysRevC.92.024907} {\bibfield  {journal} {\bibinfo  {journal} {Phys. Rev. C}\ }\textbf {\bibinfo {volume} {92}},\ \bibinfo {pages} {024907} (\bibinfo {year} {2015})},\ \Eprint {https://arxiv.org/abs/1505.01413} {arXiv:1505.01413 [nucl-th]} \BibitemShut {NoStop}%
\bibitem [{\citenamefont {Andronic}\ \emph {et~al.}(2016)\citenamefont {Andronic} \emph {et~al.}}]{Andronic:2015wma}%
  \BibitemOpen
  \bibfield  {author} {\bibinfo {author} {\bibfnamefont {A.}~\bibnamefont {Andronic}} \emph {et~al.},\ }\bibfield  {title} {\bibinfo {title} {{Heavy-flavour and quarkonium production in the LHC era: from proton\textendash{}proton to heavy-ion collisions}},\ }\href {https://doi.org/10.1140/epjc/s10052-015-3819-5} {\bibfield  {journal} {\bibinfo  {journal} {Eur. Phys. J. C}\ }\textbf {\bibinfo {volume} {76}},\ \bibinfo {pages} {107} (\bibinfo {year} {2016})},\ \Eprint {https://arxiv.org/abs/1506.03981} {arXiv:1506.03981 [nucl-ex]} \BibitemShut {NoStop}%
\bibitem [{\citenamefont {Beraudo}\ \emph {et~al.}(2016)\citenamefont {Beraudo}, \citenamefont {De~Pace}, \citenamefont {Monteno}, \citenamefont {Nardi},\ and\ \citenamefont {Prino}}]{Beraudo:2015wsd}%
  \BibitemOpen
  \bibfield  {author} {\bibinfo {author} {\bibfnamefont {A.}~\bibnamefont {Beraudo}}, \bibinfo {author} {\bibfnamefont {A.}~\bibnamefont {De~Pace}}, \bibinfo {author} {\bibfnamefont {M.}~\bibnamefont {Monteno}}, \bibinfo {author} {\bibfnamefont {M.}~\bibnamefont {Nardi}},\ and\ \bibinfo {author} {\bibfnamefont {F.}~\bibnamefont {Prino}},\ }\bibfield  {title} {\bibinfo {title} {{Heavy-flavour production in high-energy d-Au and p-Pb collisions}},\ }\href {https://doi.org/10.1007/JHEP03(2016)123} {\bibfield  {journal} {\bibinfo  {journal} {JHEP}\ }\textbf {\bibinfo {volume} {03}},\ \bibinfo {pages} {123}},\ \Eprint {https://arxiv.org/abs/1512.05186} {arXiv:1512.05186 [hep-ph]} \BibitemShut {NoStop}%
\bibitem [{\citenamefont {Das}\ \emph {et~al.}(2015{\natexlab{a}})\citenamefont {Das}, \citenamefont {Scardina}, \citenamefont {Plumari},\ and\ \citenamefont {Greco}}]{Das:2015ana}%
  \BibitemOpen
  \bibfield  {author} {\bibinfo {author} {\bibfnamefont {S.~K.}\ \bibnamefont {Das}}, \bibinfo {author} {\bibfnamefont {F.}~\bibnamefont {Scardina}}, \bibinfo {author} {\bibfnamefont {S.}~\bibnamefont {Plumari}},\ and\ \bibinfo {author} {\bibfnamefont {V.}~\bibnamefont {Greco}},\ }\bibfield  {title} {\bibinfo {title} {{Toward a solution to the $R_{AA}$ and $v_2$ puzzle for heavy quarks}},\ }\href {https://doi.org/10.1016/j.physletb.2015.06.003} {\bibfield  {journal} {\bibinfo  {journal} {Phys. Lett. B}\ }\textbf {\bibinfo {volume} {747}},\ \bibinfo {pages} {260} (\bibinfo {year} {2015}{\natexlab{a}})},\ \Eprint {https://arxiv.org/abs/1502.03757} {arXiv:1502.03757 [nucl-th]} \BibitemShut {NoStop}%
\bibitem [{\citenamefont {Das}\ \emph {et~al.}(2015{\natexlab{b}})\citenamefont {Das}, \citenamefont {Ruggieri}, \citenamefont {Mazumder}, \citenamefont {Greco},\ and\ \citenamefont {Alam}}]{Das:2015aga}%
  \BibitemOpen
  \bibfield  {author} {\bibinfo {author} {\bibfnamefont {S.~K.}\ \bibnamefont {Das}}, \bibinfo {author} {\bibfnamefont {M.}~\bibnamefont {Ruggieri}}, \bibinfo {author} {\bibfnamefont {S.}~\bibnamefont {Mazumder}}, \bibinfo {author} {\bibfnamefont {V.}~\bibnamefont {Greco}},\ and\ \bibinfo {author} {\bibfnamefont {J.-e.}\ \bibnamefont {Alam}},\ }\bibfield  {title} {\bibinfo {title} {{Heavy quark diffusion in the pre-equilibrium stage of heavy ion collisions}},\ }\href {https://doi.org/10.1088/0954-3899/42/9/095108} {\bibfield  {journal} {\bibinfo  {journal} {J. Phys. G}\ }\textbf {\bibinfo {volume} {42}},\ \bibinfo {pages} {095108} (\bibinfo {year} {2015}{\natexlab{b}})},\ \Eprint {https://arxiv.org/abs/1501.07521} {arXiv:1501.07521 [nucl-th]} \BibitemShut {NoStop}%
\bibitem [{\citenamefont {Das}\ \emph {et~al.}(2016)\citenamefont {Das}, \citenamefont {Torres-Rincon}, \citenamefont {Tolos}, \citenamefont {Minissale}, \citenamefont {Scardina},\ and\ \citenamefont {Greco}}]{Das:2016llg}%
  \BibitemOpen
  \bibfield  {author} {\bibinfo {author} {\bibfnamefont {S.~K.}\ \bibnamefont {Das}}, \bibinfo {author} {\bibfnamefont {J.~M.}\ \bibnamefont {Torres-Rincon}}, \bibinfo {author} {\bibfnamefont {L.}~\bibnamefont {Tolos}}, \bibinfo {author} {\bibfnamefont {V.}~\bibnamefont {Minissale}}, \bibinfo {author} {\bibfnamefont {F.}~\bibnamefont {Scardina}},\ and\ \bibinfo {author} {\bibfnamefont {V.}~\bibnamefont {Greco}},\ }\bibfield  {title} {\bibinfo {title} {{Propagation of heavy baryons in heavy-ion collisions}},\ }\href {https://doi.org/10.1103/PhysRevD.94.114039} {\bibfield  {journal} {\bibinfo  {journal} {Phys. Rev. D}\ }\textbf {\bibinfo {volume} {94}},\ \bibinfo {pages} {114039} (\bibinfo {year} {2016})},\ \Eprint {https://arxiv.org/abs/1604.05666} {arXiv:1604.05666 [nucl-th]} \BibitemShut {NoStop}%
\bibitem [{\citenamefont {Das}\ \emph {et~al.}(2017{\natexlab{a}})\citenamefont {Das}, \citenamefont {Plumari}, \citenamefont {Chatterjee}, \citenamefont {Alam}, \citenamefont {Scardina},\ and\ \citenamefont {Greco}}]{Das:2016cwd}%
  \BibitemOpen
  \bibfield  {author} {\bibinfo {author} {\bibfnamefont {S.~K.}\ \bibnamefont {Das}}, \bibinfo {author} {\bibfnamefont {S.}~\bibnamefont {Plumari}}, \bibinfo {author} {\bibfnamefont {S.}~\bibnamefont {Chatterjee}}, \bibinfo {author} {\bibfnamefont {J.}~\bibnamefont {Alam}}, \bibinfo {author} {\bibfnamefont {F.}~\bibnamefont {Scardina}},\ and\ \bibinfo {author} {\bibfnamefont {V.}~\bibnamefont {Greco}},\ }\bibfield  {title} {\bibinfo {title} {{Directed Flow of Charm Quarks as a Witness of the Initial Strong Magnetic Field in Ultra-Relativistic Heavy Ion Collisions}},\ }\href {https://doi.org/10.1016/j.physletb.2017.02.046} {\bibfield  {journal} {\bibinfo  {journal} {Phys. Lett. B}\ }\textbf {\bibinfo {volume} {768}},\ \bibinfo {pages} {260} (\bibinfo {year} {2017}{\natexlab{a}})},\ \Eprint {https://arxiv.org/abs/1608.02231} {arXiv:1608.02231 [nucl-th]} \BibitemShut {NoStop}%
\bibitem [{\citenamefont {Cao}\ \emph {et~al.}(2016)\citenamefont {Cao}, \citenamefont {Luo}, \citenamefont {Qin},\ and\ \citenamefont {Wang}}]{Cao:2016gvr}%
  \BibitemOpen
  \bibfield  {author} {\bibinfo {author} {\bibfnamefont {S.}~\bibnamefont {Cao}}, \bibinfo {author} {\bibfnamefont {T.}~\bibnamefont {Luo}}, \bibinfo {author} {\bibfnamefont {G.-Y.}\ \bibnamefont {Qin}},\ and\ \bibinfo {author} {\bibfnamefont {X.-N.}\ \bibnamefont {Wang}},\ }\bibfield  {title} {\bibinfo {title} {{Linearized Boltzmann transport model for jet propagation in the quark-gluon plasma: Heavy quark evolution}},\ }\href {https://doi.org/10.1103/PhysRevC.94.014909} {\bibfield  {journal} {\bibinfo  {journal} {Phys. Rev. C}\ }\textbf {\bibinfo {volume} {94}},\ \bibinfo {pages} {014909} (\bibinfo {year} {2016})},\ \Eprint {https://arxiv.org/abs/1605.06447} {arXiv:1605.06447 [nucl-th]} \BibitemShut {NoStop}%
\bibitem [{\citenamefont {Aarts}\ \emph {et~al.}(2017)\citenamefont {Aarts} \emph {et~al.}}]{Aarts:2016hap}%
  \BibitemOpen
  \bibfield  {author} {\bibinfo {author} {\bibfnamefont {G.}~\bibnamefont {Aarts}} \emph {et~al.},\ }\bibfield  {title} {\bibinfo {title} {{Heavy-flavor production and medium properties in high-energy nuclear collisions - What next?}},\ }\href {https://doi.org/10.1140/epja/i2017-12282-9} {\bibfield  {journal} {\bibinfo  {journal} {Eur. Phys. J. A}\ }\textbf {\bibinfo {volume} {53}},\ \bibinfo {pages} {93} (\bibinfo {year} {2017})},\ \Eprint {https://arxiv.org/abs/1612.08032} {arXiv:1612.08032 [nucl-th]} \BibitemShut {NoStop}%
\bibitem [{\citenamefont {Prino}\ and\ \citenamefont {Rapp}(2016)}]{Prino:2016cni}%
  \BibitemOpen
  \bibfield  {author} {\bibinfo {author} {\bibfnamefont {F.}~\bibnamefont {Prino}}\ and\ \bibinfo {author} {\bibfnamefont {R.}~\bibnamefont {Rapp}},\ }\bibfield  {title} {\bibinfo {title} {{Open Heavy Flavor in QCD Matter and in Nuclear Collisions}},\ }\href {https://doi.org/10.1088/0954-3899/43/9/093002} {\bibfield  {journal} {\bibinfo  {journal} {J. Phys. G}\ }\textbf {\bibinfo {volume} {43}},\ \bibinfo {pages} {093002} (\bibinfo {year} {2016})},\ \Eprint {https://arxiv.org/abs/1603.00529} {arXiv:1603.00529 [nucl-ex]} \BibitemShut {NoStop}%
\bibitem [{\citenamefont {Scardina}\ \emph {et~al.}(2017)\citenamefont {Scardina}, \citenamefont {Das}, \citenamefont {Minissale}, \citenamefont {Plumari},\ and\ \citenamefont {Greco}}]{Scardina:2017ipo}%
  \BibitemOpen
  \bibfield  {author} {\bibinfo {author} {\bibfnamefont {F.}~\bibnamefont {Scardina}}, \bibinfo {author} {\bibfnamefont {S.~K.}\ \bibnamefont {Das}}, \bibinfo {author} {\bibfnamefont {V.}~\bibnamefont {Minissale}}, \bibinfo {author} {\bibfnamefont {S.}~\bibnamefont {Plumari}},\ and\ \bibinfo {author} {\bibfnamefont {V.}~\bibnamefont {Greco}},\ }\bibfield  {title} {\bibinfo {title} {{Estimating the charm quark diffusion coefficient and thermalization time from D meson spectra at energies available at the BNL Relativistic Heavy Ion Collider and the CERN Large Hadron Collider}},\ }\href {https://doi.org/10.1103/PhysRevC.96.044905} {\bibfield  {journal} {\bibinfo  {journal} {Phys. Rev. C}\ }\textbf {\bibinfo {volume} {96}},\ \bibinfo {pages} {044905} (\bibinfo {year} {2017})},\ \Eprint {https://arxiv.org/abs/1707.05452} {arXiv:1707.05452 [nucl-th]} \BibitemShut {NoStop}%
\bibitem [{\citenamefont {Das}\ \emph {et~al.}(2017{\natexlab{b}})\citenamefont {Das}, \citenamefont {Ruggieri}, \citenamefont {Scardina}, \citenamefont {Plumari},\ and\ \citenamefont {Greco}}]{Das:2017dsh}%
  \BibitemOpen
  \bibfield  {author} {\bibinfo {author} {\bibfnamefont {S.~K.}\ \bibnamefont {Das}}, \bibinfo {author} {\bibfnamefont {M.}~\bibnamefont {Ruggieri}}, \bibinfo {author} {\bibfnamefont {F.}~\bibnamefont {Scardina}}, \bibinfo {author} {\bibfnamefont {S.}~\bibnamefont {Plumari}},\ and\ \bibinfo {author} {\bibfnamefont {V.}~\bibnamefont {Greco}},\ }\bibfield  {title} {\bibinfo {title} {{Effect of pre-equilibrium phase on $R_{AA}$ and $v_2$ of heavy quarks in heavy ion collisions}},\ }\href {https://doi.org/10.1088/1361-6471/aa815a} {\bibfield  {journal} {\bibinfo  {journal} {J. Phys. G}\ }\textbf {\bibinfo {volume} {44}},\ \bibinfo {pages} {095102} (\bibinfo {year} {2017}{\natexlab{b}})},\ \Eprint {https://arxiv.org/abs/1701.05123} {arXiv:1701.05123 [nucl-th]} \BibitemShut {NoStop}%
\bibitem [{\citenamefont {Greco}(2017)}]{Greco:2017rro}%
  \BibitemOpen
  \bibfield  {author} {\bibinfo {author} {\bibfnamefont {V.}~\bibnamefont {Greco}},\ }\bibfield  {title} {\bibinfo {title} {{Heavy Flavor Production, Flow and Energy Loss}},\ }\href {https://doi.org/10.1016/j.nuclphysa.2017.06.044} {\bibfield  {journal} {\bibinfo  {journal} {Nucl. Phys. A}\ }\textbf {\bibinfo {volume} {967}},\ \bibinfo {pages} {200} (\bibinfo {year} {2017})}\BibitemShut {NoStop}%
\bibitem [{\citenamefont {Xu}\ \emph {et~al.}(2018)\citenamefont {Xu}, \citenamefont {Bernhard}, \citenamefont {Bass}, \citenamefont {Nahrgang},\ and\ \citenamefont {Cao}}]{Xu:2017obm}%
  \BibitemOpen
  \bibfield  {author} {\bibinfo {author} {\bibfnamefont {Y.}~\bibnamefont {Xu}}, \bibinfo {author} {\bibfnamefont {J.~E.}\ \bibnamefont {Bernhard}}, \bibinfo {author} {\bibfnamefont {S.~A.}\ \bibnamefont {Bass}}, \bibinfo {author} {\bibfnamefont {M.}~\bibnamefont {Nahrgang}},\ and\ \bibinfo {author} {\bibfnamefont {S.}~\bibnamefont {Cao}},\ }\bibfield  {title} {\bibinfo {title} {{Data-driven analysis for the temperature and momentum dependence of the heavy-quark diffusion coefficient in relativistic heavy-ion collisions}},\ }\href {https://doi.org/10.1103/PhysRevC.97.014907} {\bibfield  {journal} {\bibinfo  {journal} {Phys. Rev. C}\ }\textbf {\bibinfo {volume} {97}},\ \bibinfo {pages} {014907} (\bibinfo {year} {2018})},\ \Eprint {https://arxiv.org/abs/1710.00807} {arXiv:1710.00807 [nucl-th]} \BibitemShut {NoStop}%
\bibitem [{\citenamefont {Chatterjee}\ and\ \citenamefont {Bo\.zek}(2018)}]{Chatterjee:2017ahy}%
  \BibitemOpen
  \bibfield  {author} {\bibinfo {author} {\bibfnamefont {S.}~\bibnamefont {Chatterjee}}\ and\ \bibinfo {author} {\bibfnamefont {P.}~\bibnamefont {Bo\.zek}},\ }\bibfield  {title} {\bibinfo {title} {{Large directed flow of open charm mesons probes the three dimensional distribution of matter in heavy ion collisions}},\ }\href {https://doi.org/10.1103/PhysRevLett.120.192301} {\bibfield  {journal} {\bibinfo  {journal} {Phys. Rev. Lett.}\ }\textbf {\bibinfo {volume} {120}},\ \bibinfo {pages} {192301} (\bibinfo {year} {2018})},\ \Eprint {https://arxiv.org/abs/1712.01189} {arXiv:1712.01189 [nucl-th]} \BibitemShut {NoStop}%
\bibitem [{\citenamefont {Chatterjee}\ and\ \citenamefont {Bozek}(2019)}]{Chatterjee:2018lsx}%
  \BibitemOpen
  \bibfield  {author} {\bibinfo {author} {\bibfnamefont {S.}~\bibnamefont {Chatterjee}}\ and\ \bibinfo {author} {\bibfnamefont {P.}~\bibnamefont {Bozek}},\ }\bibfield  {title} {\bibinfo {title} {{Interplay of drag by hot matter and electromagnetic force on the directed flow of heavy quarks}},\ }\href {https://doi.org/10.1016/j.physletb.2019.134955} {\bibfield  {journal} {\bibinfo  {journal} {Phys. Lett. B}\ }\textbf {\bibinfo {volume} {798}},\ \bibinfo {pages} {134955} (\bibinfo {year} {2019})},\ \Eprint {https://arxiv.org/abs/1804.04893} {arXiv:1804.04893 [nucl-th]} \BibitemShut {NoStop}%
\bibitem [{\citenamefont {Beraudo}\ \emph {et~al.}(2018)\citenamefont {Beraudo} \emph {et~al.}}]{Rapp:2018qla}%
  \BibitemOpen
  \bibfield  {author} {\bibinfo {author} {\bibfnamefont {A.}~\bibnamefont {Beraudo}} \emph {et~al.},\ }\bibfield  {title} {\bibinfo {title} {{Extraction of Heavy-Flavor Transport Coefficients in QCD Matter}},\ }\href {https://doi.org/10.1016/j.nuclphysa.2018.09.002} {\bibfield  {journal} {\bibinfo  {journal} {Nucl. Phys. A}\ }\textbf {\bibinfo {volume} {979}},\ \bibinfo {pages} {21} (\bibinfo {year} {2018})},\ \Eprint {https://arxiv.org/abs/1803.03824} {arXiv:1803.03824 [nucl-th]} \BibitemShut {NoStop}%
\bibitem [{\citenamefont {Cao}\ \emph {et~al.}(2019)\citenamefont {Cao} \emph {et~al.}}]{Cao:2018ews}%
  \BibitemOpen
  \bibfield  {author} {\bibinfo {author} {\bibfnamefont {S.}~\bibnamefont {Cao}} \emph {et~al.},\ }\bibfield  {title} {\bibinfo {title} {{Toward the determination of heavy-quark transport coefficients in quark-gluon plasma}},\ }\href {https://doi.org/10.1103/PhysRevC.99.054907} {\bibfield  {journal} {\bibinfo  {journal} {Phys. Rev. C}\ }\textbf {\bibinfo {volume} {99}},\ \bibinfo {pages} {054907} (\bibinfo {year} {2019})},\ \Eprint {https://arxiv.org/abs/1809.07894} {arXiv:1809.07894 [nucl-th]} \BibitemShut {NoStop}%
\bibitem [{\citenamefont {Xu}\ \emph {et~al.}(2019)\citenamefont {Xu} \emph {et~al.}}]{Xu:2018gux}%
  \BibitemOpen
  \bibfield  {author} {\bibinfo {author} {\bibfnamefont {Y.}~\bibnamefont {Xu}} \emph {et~al.},\ }\bibfield  {title} {\bibinfo {title} {{Resolving discrepancies in the estimation of heavy quark transport coefficients in relativistic heavy-ion collisions}},\ }\href {https://doi.org/10.1103/PhysRevC.99.014902} {\bibfield  {journal} {\bibinfo  {journal} {Phys. Rev. C}\ }\textbf {\bibinfo {volume} {99}},\ \bibinfo {pages} {014902} (\bibinfo {year} {2019})},\ \Eprint {https://arxiv.org/abs/1809.10734} {arXiv:1809.10734 [nucl-th]} \BibitemShut {NoStop}%
\bibitem [{\citenamefont {Dong}\ and\ \citenamefont {Greco}(2019)}]{Dong:2019unq}%
  \BibitemOpen
  \bibfield  {author} {\bibinfo {author} {\bibfnamefont {X.}~\bibnamefont {Dong}}\ and\ \bibinfo {author} {\bibfnamefont {V.}~\bibnamefont {Greco}},\ }\bibfield  {title} {\bibinfo {title} {{Heavy quark production and properties of Quark\textendash{}Gluon Plasma}},\ }\href {https://doi.org/10.1016/j.ppnp.2018.08.001} {\bibfield  {journal} {\bibinfo  {journal} {Prog. Part. Nucl. Phys.}\ }\textbf {\bibinfo {volume} {104}},\ \bibinfo {pages} {97} (\bibinfo {year} {2019})}\BibitemShut {NoStop}%
\bibitem [{\citenamefont {Song}\ \emph {et~al.}(2020{\natexlab{a}})\citenamefont {Song}, \citenamefont {Moreau}, \citenamefont {Aichelin},\ and\ \citenamefont {Bratkovskaya}}]{Song:2019cqz}%
  \BibitemOpen
  \bibfield  {author} {\bibinfo {author} {\bibfnamefont {T.}~\bibnamefont {Song}}, \bibinfo {author} {\bibfnamefont {P.}~\bibnamefont {Moreau}}, \bibinfo {author} {\bibfnamefont {J.}~\bibnamefont {Aichelin}},\ and\ \bibinfo {author} {\bibfnamefont {E.}~\bibnamefont {Bratkovskaya}},\ }\bibfield  {title} {\bibinfo {title} {{Exploring non-equilibrium quark-gluon plasma effects on charm transport coefficients}},\ }\href {https://doi.org/10.1103/PhysRevC.101.044901} {\bibfield  {journal} {\bibinfo  {journal} {Phys. Rev. C}\ }\textbf {\bibinfo {volume} {101}},\ \bibinfo {pages} {044901} (\bibinfo {year} {2020}{\natexlab{a}})},\ \Eprint {https://arxiv.org/abs/1910.09889} {arXiv:1910.09889 [nucl-th]} \BibitemShut {NoStop}%
\bibitem [{\citenamefont {Dong}\ \emph {et~al.}(2019)\citenamefont {Dong}, \citenamefont {Lee},\ and\ \citenamefont {Rapp}}]{Dong:2019byy}%
  \BibitemOpen
  \bibfield  {author} {\bibinfo {author} {\bibfnamefont {X.}~\bibnamefont {Dong}}, \bibinfo {author} {\bibfnamefont {Y.-J.}\ \bibnamefont {Lee}},\ and\ \bibinfo {author} {\bibfnamefont {R.}~\bibnamefont {Rapp}},\ }\bibfield  {title} {\bibinfo {title} {{Open Heavy-Flavor Production in Heavy-Ion Collisions}},\ }\href {https://doi.org/10.1146/annurev-nucl-101918-023806} {\bibfield  {journal} {\bibinfo  {journal} {Ann. Rev. Nucl. Part. Sci.}\ }\textbf {\bibinfo {volume} {69}},\ \bibinfo {pages} {417} (\bibinfo {year} {2019})},\ \Eprint {https://arxiv.org/abs/1903.07709} {arXiv:1903.07709 [nucl-ex]} \BibitemShut {NoStop}%
\bibitem [{\citenamefont {Song}\ \emph {et~al.}(2020{\natexlab{b}})\citenamefont {Song}, \citenamefont {Moreau}, \citenamefont {Xu}, \citenamefont {Ozvenchuk}, \citenamefont {Bratkovskaya}, \citenamefont {Aichelin}, \citenamefont {Bass}, \citenamefont {Gossiaux},\ and\ \citenamefont {Nahrgang}}]{Song:2020tfm}%
  \BibitemOpen
  \bibfield  {author} {\bibinfo {author} {\bibfnamefont {T.}~\bibnamefont {Song}}, \bibinfo {author} {\bibfnamefont {P.}~\bibnamefont {Moreau}}, \bibinfo {author} {\bibfnamefont {Y.}~\bibnamefont {Xu}}, \bibinfo {author} {\bibfnamefont {V.}~\bibnamefont {Ozvenchuk}}, \bibinfo {author} {\bibfnamefont {E.}~\bibnamefont {Bratkovskaya}}, \bibinfo {author} {\bibfnamefont {J.}~\bibnamefont {Aichelin}}, \bibinfo {author} {\bibfnamefont {S.~A.}\ \bibnamefont {Bass}}, \bibinfo {author} {\bibfnamefont {P.~B.}\ \bibnamefont {Gossiaux}},\ and\ \bibinfo {author} {\bibfnamefont {M.}~\bibnamefont {Nahrgang}},\ }\bibfield  {title} {\bibinfo {title} {{Traces of nonequilibrium effects, initial condition, bulk dynamics, and elementary collisions in the charm observables}},\ }\href {https://doi.org/10.1103/PhysRevC.101.044903} {\bibfield  {journal} {\bibinfo  {journal} {Phys. Rev. C}\ }\textbf {\bibinfo {volume} {101}},\ \bibinfo {pages} {044903} (\bibinfo {year} {2020}{\natexlab{b}})},\ \Eprint
  {https://arxiv.org/abs/2001.07951} {arXiv:2001.07951 [nucl-th]} \BibitemShut {NoStop}%
\bibitem [{\citenamefont {Zhao}\ \emph {et~al.}(2020)\citenamefont {Zhao}, \citenamefont {Zhou}, \citenamefont {Chen},\ and\ \citenamefont {Zhuang}}]{Zhao:2020jqu}%
  \BibitemOpen
  \bibfield  {author} {\bibinfo {author} {\bibfnamefont {J.}~\bibnamefont {Zhao}}, \bibinfo {author} {\bibfnamefont {K.}~\bibnamefont {Zhou}}, \bibinfo {author} {\bibfnamefont {S.}~\bibnamefont {Chen}},\ and\ \bibinfo {author} {\bibfnamefont {P.}~\bibnamefont {Zhuang}},\ }\bibfield  {title} {\bibinfo {title} {{Heavy flavors under extreme conditions in high energy nuclear collisions}},\ }\href {https://doi.org/10.1016/j.ppnp.2020.103801} {\bibfield  {journal} {\bibinfo  {journal} {Prog. Part. Nucl. Phys.}\ }\textbf {\bibinfo {volume} {114}},\ \bibinfo {pages} {103801} (\bibinfo {year} {2020})},\ \Eprint {https://arxiv.org/abs/2005.08277} {arXiv:2005.08277 [nucl-th]} \BibitemShut {NoStop}%
\bibitem [{\citenamefont {Oliva}\ \emph {et~al.}(2021)\citenamefont {Oliva}, \citenamefont {Plumari},\ and\ \citenamefont {Greco}}]{Oliva:2020doe}%
  \BibitemOpen
  \bibfield  {author} {\bibinfo {author} {\bibfnamefont {L.}~\bibnamefont {Oliva}}, \bibinfo {author} {\bibfnamefont {S.}~\bibnamefont {Plumari}},\ and\ \bibinfo {author} {\bibfnamefont {V.}~\bibnamefont {Greco}},\ }\bibfield  {title} {\bibinfo {title} {{Directed flow of D mesons at RHIC and LHC: non-perturbative dynamics, longitudinal bulk matter asymmetry and electromagnetic fields}},\ }\href {https://doi.org/10.1007/JHEP05(2021)034} {\bibfield  {journal} {\bibinfo  {journal} {JHEP}\ }\textbf {\bibinfo {volume} {05}},\ \bibinfo {pages} {034}},\ \Eprint {https://arxiv.org/abs/2009.11066} {arXiv:2009.11066 [hep-ph]} \BibitemShut {NoStop}%
\bibitem [{\citenamefont {Oliva}(2020)}]{Oliva:2020mfr}%
  \BibitemOpen
  \bibfield  {author} {\bibinfo {author} {\bibfnamefont {L.}~\bibnamefont {Oliva}},\ }\bibfield  {title} {\bibinfo {title} {{Electromagnetic fields and directed flow in large and small colliding systems at ultrarelativistic energies}},\ }\href {https://doi.org/10.1140/epja/s10050-020-00260-3} {\bibfield  {journal} {\bibinfo  {journal} {Eur. Phys. J. A}\ }\textbf {\bibinfo {volume} {56}},\ \bibinfo {pages} {255} (\bibinfo {year} {2020})},\ \Eprint {https://arxiv.org/abs/2007.00560} {arXiv:2007.00560 [nucl-th]} \BibitemShut {NoStop}%
\bibitem [{\citenamefont {Beraudo}\ \emph {et~al.}(2021)\citenamefont {Beraudo}, \citenamefont {De~Pace}, \citenamefont {Monteno}, \citenamefont {Nardi},\ and\ \citenamefont {Prino}}]{Beraudo:2021ont}%
  \BibitemOpen
  \bibfield  {author} {\bibinfo {author} {\bibfnamefont {A.}~\bibnamefont {Beraudo}}, \bibinfo {author} {\bibfnamefont {A.}~\bibnamefont {De~Pace}}, \bibinfo {author} {\bibfnamefont {M.}~\bibnamefont {Monteno}}, \bibinfo {author} {\bibfnamefont {M.}~\bibnamefont {Nardi}},\ and\ \bibinfo {author} {\bibfnamefont {F.}~\bibnamefont {Prino}},\ }\bibfield  {title} {\bibinfo {title} {{Rapidity dependence of heavy-flavour production in heavy-ion collisions within a full 3+1 transport approach: quenching, elliptic and directed flow}},\ }\href {https://doi.org/10.1007/JHEP05(2021)279} {\bibfield  {journal} {\bibinfo  {journal} {JHEP}\ }\textbf {\bibinfo {volume} {05}},\ \bibinfo {pages} {279}},\ \Eprint {https://arxiv.org/abs/2102.08064} {arXiv:2102.08064 [hep-ph]} \BibitemShut {NoStop}%
\bibitem [{\citenamefont {Ruggieri}\ \emph {et~al.}(2022)\citenamefont {Ruggieri}, \citenamefont {Pooja}, \citenamefont {Prakash},\ and\ \citenamefont {Das}}]{Ruggieri:2022kxv}%
  \BibitemOpen
  \bibfield  {author} {\bibinfo {author} {\bibfnamefont {M.}~\bibnamefont {Ruggieri}}, \bibinfo {author} {\bibnamefont {Pooja}}, \bibinfo {author} {\bibfnamefont {J.}~\bibnamefont {Prakash}},\ and\ \bibinfo {author} {\bibfnamefont {S.~K.}\ \bibnamefont {Das}},\ }\bibfield  {title} {\bibinfo {title} {{Memory effects on energy loss and diffusion of heavy quarks in the quark-gluon plasma}},\ }\href {https://doi.org/10.1103/PhysRevD.106.034032} {\bibfield  {journal} {\bibinfo  {journal} {Phys. Rev. D}\ }\textbf {\bibinfo {volume} {106}},\ \bibinfo {pages} {034032} (\bibinfo {year} {2022})},\ \Eprint {https://arxiv.org/abs/2203.06712} {arXiv:2203.06712 [hep-ph]} \BibitemShut {NoStop}%
\bibitem [{\citenamefont {Pooja}\ \emph {et~al.}(2023{\natexlab{a}})\citenamefont {Pooja}, \citenamefont {Das}, \citenamefont {Greco},\ and\ \citenamefont {Ruggieri}}]{Pooja:2023gqt}%
  \BibitemOpen
  \bibfield  {author} {\bibinfo {author} {\bibnamefont {Pooja}}, \bibinfo {author} {\bibfnamefont {S.~K.}\ \bibnamefont {Das}}, \bibinfo {author} {\bibfnamefont {V.}~\bibnamefont {Greco}},\ and\ \bibinfo {author} {\bibfnamefont {M.}~\bibnamefont {Ruggieri}},\ }\bibfield  {title} {\bibinfo {title} {{Thermalization and isotropization of heavy quarks in a non-Markovian medium in high-energy nuclear collisions}},\ }\href {https://doi.org/10.1103/PhysRevD.108.054026} {\bibfield  {journal} {\bibinfo  {journal} {Phys. Rev. D}\ }\textbf {\bibinfo {volume} {108}},\ \bibinfo {pages} {054026} (\bibinfo {year} {2023}{\natexlab{a}})},\ \Eprint {https://arxiv.org/abs/2306.13749} {arXiv:2306.13749 [hep-ph]} \BibitemShut {NoStop}%
\bibitem [{\citenamefont {Sun}\ \emph {et~al.}(2023)\citenamefont {Sun}, \citenamefont {Plumari},\ and\ \citenamefont {Das}}]{Sun:2023adv}%
  \BibitemOpen
  \bibfield  {author} {\bibinfo {author} {\bibfnamefont {Y.}~\bibnamefont {Sun}}, \bibinfo {author} {\bibfnamefont {S.}~\bibnamefont {Plumari}},\ and\ \bibinfo {author} {\bibfnamefont {S.~K.}\ \bibnamefont {Das}},\ }\bibfield  {title} {\bibinfo {title} {{Exploring the effects of electromagnetic fields and tilted bulk distribution on directed flow of D mesons in small systems}},\ }\href {https://doi.org/10.1016/j.physletb.2023.138043} {\bibfield  {journal} {\bibinfo  {journal} {Phys. Lett. B}\ }\textbf {\bibinfo {volume} {843}},\ \bibinfo {pages} {138043} (\bibinfo {year} {2023})},\ \Eprint {https://arxiv.org/abs/2304.12792} {arXiv:2304.12792 [nucl-th]} \BibitemShut {NoStop}%
\bibitem [{\citenamefont {Das}\ \emph {et~al.}(2024)\citenamefont {Das}, \citenamefont {Torres-Rincon},\ and\ \citenamefont {Rapp}}]{Das:2024vac}%
  \BibitemOpen
  \bibfield  {author} {\bibinfo {author} {\bibfnamefont {S.~K.}\ \bibnamefont {Das}}, \bibinfo {author} {\bibfnamefont {J.~M.}\ \bibnamefont {Torres-Rincon}},\ and\ \bibinfo {author} {\bibfnamefont {R.}~\bibnamefont {Rapp}},\ }\bibfield  {title} {\bibinfo {title} {{Charm and Bottom Hadrons in Hot Hadronic Matter}},\ }\href@noop {} {\  (\bibinfo {year} {2024})},\ \Eprint {https://arxiv.org/abs/2406.13286} {arXiv:2406.13286 [hep-ph]} \BibitemShut {NoStop}%
\bibitem [{\citenamefont {Barata}\ \emph {et~al.}(2024)\citenamefont {Barata}, \citenamefont {Hauksson}, \citenamefont {Mayo~L\'opez},\ and\ \citenamefont {Sadofyev}}]{Barata:2024xwy}%
  \BibitemOpen
  \bibfield  {author} {\bibinfo {author} {\bibfnamefont {J.~a.}\ \bibnamefont {Barata}}, \bibinfo {author} {\bibfnamefont {S.}~\bibnamefont {Hauksson}}, \bibinfo {author} {\bibfnamefont {X.}~\bibnamefont {Mayo~L\'opez}},\ and\ \bibinfo {author} {\bibfnamefont {A.~V.}\ \bibnamefont {Sadofyev}},\ }\bibfield  {title} {\bibinfo {title} {{Jet quenching in the glasma phase: Medium-induced radiation}},\ }\href {https://doi.org/10.1103/PhysRevD.110.094055} {\bibfield  {journal} {\bibinfo  {journal} {Phys. Rev. D}\ }\textbf {\bibinfo {volume} {110}},\ \bibinfo {pages} {094055} (\bibinfo {year} {2024})},\ \Eprint {https://arxiv.org/abs/2406.07615} {arXiv:2406.07615 [hep-ph]} \BibitemShut {NoStop}%
\bibitem [{\citenamefont {Mrowczynski}(2018)}]{Mrowczynski:2017kso}%
  \BibitemOpen
  \bibfield  {author} {\bibinfo {author} {\bibfnamefont {S.}~\bibnamefont {Mrowczynski}},\ }\bibfield  {title} {\bibinfo {title} {{Heavy Quarks in Turbulent QCD Plasmas}},\ }\href {https://doi.org/10.1140/epja/i2018-12478-5} {\bibfield  {journal} {\bibinfo  {journal} {Eur. Phys. J. A}\ }\textbf {\bibinfo {volume} {54}},\ \bibinfo {pages} {43} (\bibinfo {year} {2018})},\ \Eprint {https://arxiv.org/abs/1706.03127} {arXiv:1706.03127 [nucl-th]} \BibitemShut {NoStop}%
\bibitem [{\citenamefont {Ruggieri}\ and\ \citenamefont {Das}(2018{\natexlab{a}})}]{Ruggieri:2018rzi}%
  \BibitemOpen
  \bibfield  {author} {\bibinfo {author} {\bibfnamefont {M.}~\bibnamefont {Ruggieri}}\ and\ \bibinfo {author} {\bibfnamefont {S.~K.}\ \bibnamefont {Das}},\ }\bibfield  {title} {\bibinfo {title} {{Cathode tube effect: Heavy quarks probing the glasma in p -Pb collisions}},\ }\href {https://doi.org/10.1103/PhysRevD.98.094024} {\bibfield  {journal} {\bibinfo  {journal} {Phys. Rev. D}\ }\textbf {\bibinfo {volume} {98}},\ \bibinfo {pages} {094024} (\bibinfo {year} {2018}{\natexlab{a}})},\ \Eprint {https://arxiv.org/abs/1805.09617} {arXiv:1805.09617 [nucl-th]} \BibitemShut {NoStop}%
\bibitem [{\citenamefont {Sun}\ \emph {et~al.}(2019)\citenamefont {Sun}, \citenamefont {Coci}, \citenamefont {Das}, \citenamefont {Plumari}, \citenamefont {Ruggieri},\ and\ \citenamefont {Greco}}]{Sun:2019fud}%
  \BibitemOpen
  \bibfield  {author} {\bibinfo {author} {\bibfnamefont {Y.}~\bibnamefont {Sun}}, \bibinfo {author} {\bibfnamefont {G.}~\bibnamefont {Coci}}, \bibinfo {author} {\bibfnamefont {S.~K.}\ \bibnamefont {Das}}, \bibinfo {author} {\bibfnamefont {S.}~\bibnamefont {Plumari}}, \bibinfo {author} {\bibfnamefont {M.}~\bibnamefont {Ruggieri}},\ and\ \bibinfo {author} {\bibfnamefont {V.}~\bibnamefont {Greco}},\ }\bibfield  {title} {\bibinfo {title} {{Impact of Glasma on heavy quark observables in nucleus-nucleus collisions at LHC}},\ }\href {https://doi.org/10.1016/j.physletb.2019.134933} {\bibfield  {journal} {\bibinfo  {journal} {Phys. Lett. B}\ }\textbf {\bibinfo {volume} {798}},\ \bibinfo {pages} {134933} (\bibinfo {year} {2019})},\ \Eprint {https://arxiv.org/abs/1902.06254} {arXiv:1902.06254 [nucl-th]} \BibitemShut {NoStop}%
\bibitem [{\citenamefont {Liu}\ \emph {et~al.}(2020)\citenamefont {Liu}, \citenamefont {Plumari}, \citenamefont {Das}, \citenamefont {Greco},\ and\ \citenamefont {Ruggieri}}]{Liu:2019lac}%
  \BibitemOpen
  \bibfield  {author} {\bibinfo {author} {\bibfnamefont {J.~H.}\ \bibnamefont {Liu}}, \bibinfo {author} {\bibfnamefont {S.}~\bibnamefont {Plumari}}, \bibinfo {author} {\bibfnamefont {S.~K.}\ \bibnamefont {Das}}, \bibinfo {author} {\bibfnamefont {V.}~\bibnamefont {Greco}},\ and\ \bibinfo {author} {\bibfnamefont {M.}~\bibnamefont {Ruggieri}},\ }\bibfield  {title} {\bibinfo {title} {{Diffusion of heavy quarks in the early stage of high-energy nuclear collisions at energies available at the BNL Relativistic Heavy Ion Collider and at the CERN Large Hadron Collider}},\ }\href {https://doi.org/10.1103/PhysRevC.102.044902} {\bibfield  {journal} {\bibinfo  {journal} {Phys. Rev. C}\ }\textbf {\bibinfo {volume} {102}},\ \bibinfo {pages} {044902} (\bibinfo {year} {2020})},\ \Eprint {https://arxiv.org/abs/1911.02480} {arXiv:1911.02480 [nucl-th]} \BibitemShut {NoStop}%
\bibitem [{\citenamefont {Liu}\ \emph {et~al.}(2021)\citenamefont {Liu}, \citenamefont {Das}, \citenamefont {Greco},\ and\ \citenamefont {Ruggieri}}]{Liu:2020cpj}%
  \BibitemOpen
  \bibfield  {author} {\bibinfo {author} {\bibfnamefont {J.-H.}\ \bibnamefont {Liu}}, \bibinfo {author} {\bibfnamefont {S.~K.}\ \bibnamefont {Das}}, \bibinfo {author} {\bibfnamefont {V.}~\bibnamefont {Greco}},\ and\ \bibinfo {author} {\bibfnamefont {M.}~\bibnamefont {Ruggieri}},\ }\bibfield  {title} {\bibinfo {title} {{Ballistic diffusion of heavy quarks in the early stage of relativistic heavy ion collisions at RHIC and the LHC}},\ }\href {https://doi.org/10.1103/PhysRevD.103.034029} {\bibfield  {journal} {\bibinfo  {journal} {Phys. Rev. D}\ }\textbf {\bibinfo {volume} {103}},\ \bibinfo {pages} {034029} (\bibinfo {year} {2021})},\ \Eprint {https://arxiv.org/abs/2011.05818} {arXiv:2011.05818 [hep-ph]} \BibitemShut {NoStop}%
\bibitem [{\citenamefont {Boguslavski}\ \emph {et~al.}(2020)\citenamefont {Boguslavski}, \citenamefont {Kurkela}, \citenamefont {Lappi},\ and\ \citenamefont {Peuron}}]{Boguslavski:2020tqz}%
  \BibitemOpen
  \bibfield  {author} {\bibinfo {author} {\bibfnamefont {K.}~\bibnamefont {Boguslavski}}, \bibinfo {author} {\bibfnamefont {A.}~\bibnamefont {Kurkela}}, \bibinfo {author} {\bibfnamefont {T.}~\bibnamefont {Lappi}},\ and\ \bibinfo {author} {\bibfnamefont {J.}~\bibnamefont {Peuron}},\ }\bibfield  {title} {\bibinfo {title} {{Heavy quark diffusion in an overoccupied gluon plasma}},\ }\href {https://doi.org/10.1007/JHEP09(2020)077} {\bibfield  {journal} {\bibinfo  {journal} {JHEP}\ }\textbf {\bibinfo {volume} {09}},\ \bibinfo {pages} {077}},\ \Eprint {https://arxiv.org/abs/2005.02418} {arXiv:2005.02418 [hep-ph]} \BibitemShut {NoStop}%
\bibitem [{\citenamefont {Carrington}\ \emph {et~al.}(2020)\citenamefont {Carrington}, \citenamefont {Czajka},\ and\ \citenamefont {Mrowczynski}}]{Carrington:2020sww}%
  \BibitemOpen
  \bibfield  {author} {\bibinfo {author} {\bibfnamefont {M.~E.}\ \bibnamefont {Carrington}}, \bibinfo {author} {\bibfnamefont {A.}~\bibnamefont {Czajka}},\ and\ \bibinfo {author} {\bibfnamefont {S.}~\bibnamefont {Mrowczynski}},\ }\bibfield  {title} {\bibinfo {title} {{Heavy Quarks Embedded in Glasma}},\ }\href {https://doi.org/10.1016/j.nuclphysa.2020.121914} {\bibfield  {journal} {\bibinfo  {journal} {Nucl. Phys. A}\ }\textbf {\bibinfo {volume} {1001}},\ \bibinfo {pages} {121914} (\bibinfo {year} {2020})},\ \Eprint {https://arxiv.org/abs/2001.05074} {arXiv:2001.05074 [nucl-th]} \BibitemShut {NoStop}%
\bibitem [{\citenamefont {Ipp}\ \emph {et~al.}(2020)\citenamefont {Ipp}, \citenamefont {M\"uller},\ and\ \citenamefont {Schuh}}]{Ipp:2020nfu}%
  \BibitemOpen
  \bibfield  {author} {\bibinfo {author} {\bibfnamefont {A.}~\bibnamefont {Ipp}}, \bibinfo {author} {\bibfnamefont {D.~I.}\ \bibnamefont {M\"uller}},\ and\ \bibinfo {author} {\bibfnamefont {D.}~\bibnamefont {Schuh}},\ }\bibfield  {title} {\bibinfo {title} {{Jet momentum broadening in the pre-equilibrium Glasma}},\ }\href {https://doi.org/10.1016/j.physletb.2020.135810} {\bibfield  {journal} {\bibinfo  {journal} {Phys. Lett. B}\ }\textbf {\bibinfo {volume} {810}},\ \bibinfo {pages} {135810} (\bibinfo {year} {2020})},\ \Eprint {https://arxiv.org/abs/2009.14206} {arXiv:2009.14206 [hep-ph]} \BibitemShut {NoStop}%
\bibitem [{\citenamefont {Khowal}\ \emph {et~al.}(2022)\citenamefont {Khowal}, \citenamefont {Das}, \citenamefont {Oliva},\ and\ \citenamefont {Ruggieri}}]{Khowal:2021zoo}%
  \BibitemOpen
  \bibfield  {author} {\bibinfo {author} {\bibfnamefont {P.}~\bibnamefont {Khowal}}, \bibinfo {author} {\bibfnamefont {S.~K.}\ \bibnamefont {Das}}, \bibinfo {author} {\bibfnamefont {L.}~\bibnamefont {Oliva}},\ and\ \bibinfo {author} {\bibfnamefont {M.}~\bibnamefont {Ruggieri}},\ }\bibfield  {title} {\bibinfo {title} {{Heavy quarks in the early stage of high energy nuclear collisions at RHIC and LHC: Brownian motion versus diffusion in the evolving Glasma}},\ }\href {https://doi.org/10.1140/epjp/s13360-022-02517-w} {\bibfield  {journal} {\bibinfo  {journal} {Eur. Phys. J. Plus}\ }\textbf {\bibinfo {volume} {137}},\ \bibinfo {pages} {307} (\bibinfo {year} {2022})},\ \Eprint {https://arxiv.org/abs/2110.14610} {arXiv:2110.14610 [hep-ph]} \BibitemShut {NoStop}%
\bibitem [{\citenamefont {Carrington}\ \emph {et~al.}(2022{\natexlab{a}})\citenamefont {Carrington}, \citenamefont {Czajka},\ and\ \citenamefont {Mrowczynski}}]{Carrington:2021dvw}%
  \BibitemOpen
  \bibfield  {author} {\bibinfo {author} {\bibfnamefont {M.~E.}\ \bibnamefont {Carrington}}, \bibinfo {author} {\bibfnamefont {A.}~\bibnamefont {Czajka}},\ and\ \bibinfo {author} {\bibfnamefont {S.}~\bibnamefont {Mrowczynski}},\ }\bibfield  {title} {\bibinfo {title} {{Jet quenching in glasma}},\ }\href {https://doi.org/10.1016/j.physletb.2022.137464} {\bibfield  {journal} {\bibinfo  {journal} {Phys. Lett. B}\ }\textbf {\bibinfo {volume} {834}},\ \bibinfo {pages} {137464} (\bibinfo {year} {2022}{\natexlab{a}})},\ \Eprint {https://arxiv.org/abs/2112.06812} {arXiv:2112.06812 [hep-ph]} \BibitemShut {NoStop}%
\bibitem [{\citenamefont {Pooja}\ \emph {et~al.}(2023{\natexlab{b}})\citenamefont {Pooja}, \citenamefont {Das}, \citenamefont {Greco},\ and\ \citenamefont {Ruggieri}}]{Pooja:2022ojj}%
  \BibitemOpen
  \bibfield  {author} {\bibinfo {author} {\bibnamefont {Pooja}}, \bibinfo {author} {\bibfnamefont {S.~K.}\ \bibnamefont {Das}}, \bibinfo {author} {\bibfnamefont {V.}~\bibnamefont {Greco}},\ and\ \bibinfo {author} {\bibfnamefont {M.}~\bibnamefont {Ruggieri}},\ }\bibfield  {title} {\bibinfo {title} {{Anisotropic fluctuations of angular momentum of heavy quarks in the Glasma}},\ }\href {https://doi.org/10.1140/epjp/s13360-023-03913-6} {\bibfield  {journal} {\bibinfo  {journal} {Eur. Phys. J. Plus}\ }\textbf {\bibinfo {volume} {138}},\ \bibinfo {pages} {313} (\bibinfo {year} {2023}{\natexlab{b}})},\ \Eprint {https://arxiv.org/abs/2212.09725} {arXiv:2212.09725 [hep-ph]} \BibitemShut {NoStop}%
\bibitem [{\citenamefont {Carrington}\ \emph {et~al.}(2022{\natexlab{b}})\citenamefont {Carrington}, \citenamefont {Czajka},\ and\ \citenamefont {Mrowczynski}}]{Carrington:2022bnv}%
  \BibitemOpen
  \bibfield  {author} {\bibinfo {author} {\bibfnamefont {M.~E.}\ \bibnamefont {Carrington}}, \bibinfo {author} {\bibfnamefont {A.}~\bibnamefont {Czajka}},\ and\ \bibinfo {author} {\bibfnamefont {S.}~\bibnamefont {Mrowczynski}},\ }\bibfield  {title} {\bibinfo {title} {{Transport of hard probes through glasma}},\ }\href {https://doi.org/10.1103/PhysRevC.105.064910} {\bibfield  {journal} {\bibinfo  {journal} {Phys. Rev. C}\ }\textbf {\bibinfo {volume} {105}},\ \bibinfo {pages} {064910} (\bibinfo {year} {2022}{\natexlab{b}})},\ \Eprint {https://arxiv.org/abs/2202.00357} {arXiv:2202.00357 [nucl-th]} \BibitemShut {NoStop}%
\bibitem [{\citenamefont {Das}\ \emph {et~al.}(2022)\citenamefont {Das} \emph {et~al.}}]{Das:2022lqh}%
  \BibitemOpen
  \bibfield  {author} {\bibinfo {author} {\bibfnamefont {S.~K.}\ \bibnamefont {Das}} \emph {et~al.},\ }\bibfield  {title} {\bibinfo {title} {{Dynamics of Hot QCD Matter -- Current Status and Developments}},\ }\href {https://doi.org/10.1142/S0218301322500975} {\bibfield  {journal} {\bibinfo  {journal} {Int. J. Mod. Phys. E}\ }\textbf {\bibinfo {volume} {31}},\ \bibinfo {pages} {12} (\bibinfo {year} {2022})},\ \Eprint {https://arxiv.org/abs/2208.13440} {arXiv:2208.13440 [nucl-th]} \BibitemShut {NoStop}%
\bibitem [{\citenamefont {Boguslavski}\ \emph {et~al.}(2024{\natexlab{a}})\citenamefont {Boguslavski}, \citenamefont {Kurkela}, \citenamefont {Lappi}, \citenamefont {Lindenbauer},\ and\ \citenamefont {Peuron}}]{Boguslavski:2023fdm}%
  \BibitemOpen
  \bibfield  {author} {\bibinfo {author} {\bibfnamefont {K.}~\bibnamefont {Boguslavski}}, \bibinfo {author} {\bibfnamefont {A.}~\bibnamefont {Kurkela}}, \bibinfo {author} {\bibfnamefont {T.}~\bibnamefont {Lappi}}, \bibinfo {author} {\bibfnamefont {F.}~\bibnamefont {Lindenbauer}},\ and\ \bibinfo {author} {\bibfnamefont {J.}~\bibnamefont {Peuron}},\ }\bibfield  {title} {\bibinfo {title} {{Heavy quark diffusion coefficient in heavy-ion collisions via kinetic theory}},\ }\href {https://doi.org/10.1103/PhysRevD.109.014025} {\bibfield  {journal} {\bibinfo  {journal} {Phys. Rev. D}\ }\textbf {\bibinfo {volume} {109}},\ \bibinfo {pages} {014025} (\bibinfo {year} {2024}{\natexlab{a}})},\ \Eprint {https://arxiv.org/abs/2303.12520} {arXiv:2303.12520 [hep-ph]} \BibitemShut {NoStop}%
\bibitem [{\citenamefont {Avramescu}\ \emph {et~al.}(2023)\citenamefont {Avramescu}, \citenamefont {B\u{a}ran}, \citenamefont {Greco}, \citenamefont {Ipp}, \citenamefont {M\"uller},\ and\ \citenamefont {Ruggieri}}]{Avramescu:2023qvv}%
  \BibitemOpen
  \bibfield  {author} {\bibinfo {author} {\bibfnamefont {D.}~\bibnamefont {Avramescu}}, \bibinfo {author} {\bibfnamefont {V.}~\bibnamefont {B\u{a}ran}}, \bibinfo {author} {\bibfnamefont {V.}~\bibnamefont {Greco}}, \bibinfo {author} {\bibfnamefont {A.}~\bibnamefont {Ipp}}, \bibinfo {author} {\bibfnamefont {D.~I.}\ \bibnamefont {M\"uller}},\ and\ \bibinfo {author} {\bibfnamefont {M.}~\bibnamefont {Ruggieri}},\ }\bibfield  {title} {\bibinfo {title} {{Simulating jets and heavy quarks in the glasma using the colored particle-in-cell method}},\ }\href {https://doi.org/10.1103/PhysRevD.107.114021} {\bibfield  {journal} {\bibinfo  {journal} {Phys. Rev. D}\ }\textbf {\bibinfo {volume} {107}},\ \bibinfo {pages} {114021} (\bibinfo {year} {2023})},\ \Eprint {https://arxiv.org/abs/2303.05599} {arXiv:2303.05599 [hep-ph]} \BibitemShut {NoStop}%
\bibitem [{\citenamefont {Pandey}\ \emph {et~al.}(2024)\citenamefont {Pandey}, \citenamefont {Schlichting},\ and\ \citenamefont {Sharma}}]{Pandey:2023dzz}%
  \BibitemOpen
  \bibfield  {author} {\bibinfo {author} {\bibfnamefont {H.}~\bibnamefont {Pandey}}, \bibinfo {author} {\bibfnamefont {S.}~\bibnamefont {Schlichting}},\ and\ \bibinfo {author} {\bibfnamefont {S.}~\bibnamefont {Sharma}},\ }\bibfield  {title} {\bibinfo {title} {{Heavy-Quark Momentum Broadening in a Non-Abelian Plasma away from Thermal Equilibrium}},\ }\href {https://doi.org/10.1103/PhysRevLett.132.222301} {\bibfield  {journal} {\bibinfo  {journal} {Phys. Rev. Lett.}\ }\textbf {\bibinfo {volume} {132}},\ \bibinfo {pages} {222301} (\bibinfo {year} {2024})},\ \Eprint {https://arxiv.org/abs/2312.12280} {arXiv:2312.12280 [hep-lat]} \BibitemShut {NoStop}%
\bibitem [{\citenamefont {Boguslavski}\ \emph {et~al.}(2024{\natexlab{b}})\citenamefont {Boguslavski}, \citenamefont {Kurkela}, \citenamefont {Lappi}, \citenamefont {Lindenbauer},\ and\ \citenamefont {Peuron}}]{Boguslavski:2023alu}%
  \BibitemOpen
  \bibfield  {author} {\bibinfo {author} {\bibfnamefont {K.}~\bibnamefont {Boguslavski}}, \bibinfo {author} {\bibfnamefont {A.}~\bibnamefont {Kurkela}}, \bibinfo {author} {\bibfnamefont {T.}~\bibnamefont {Lappi}}, \bibinfo {author} {\bibfnamefont {F.}~\bibnamefont {Lindenbauer}},\ and\ \bibinfo {author} {\bibfnamefont {J.}~\bibnamefont {Peuron}},\ }\bibfield  {title} {\bibinfo {title} {{Jet momentum broadening during initial stages in heavy-ion collisions}},\ }\href {https://doi.org/10.1016/j.physletb.2024.138525} {\bibfield  {journal} {\bibinfo  {journal} {Phys. Lett. B}\ }\textbf {\bibinfo {volume} {850}},\ \bibinfo {pages} {138525} (\bibinfo {year} {2024}{\natexlab{b}})},\ \Eprint {https://arxiv.org/abs/2303.12595} {arXiv:2303.12595 [hep-ph]} \BibitemShut {NoStop}%
\bibitem [{\citenamefont {Avramescu}\ \emph {et~al.}(2024{\natexlab{a}})\citenamefont {Avramescu}, \citenamefont {Greco}, \citenamefont {Lappi}, \citenamefont {M\"antysaari},\ and\ \citenamefont {M\"uller}}]{Avramescu:2024poa}%
  \BibitemOpen
  \bibfield  {author} {\bibinfo {author} {\bibfnamefont {D.}~\bibnamefont {Avramescu}}, \bibinfo {author} {\bibfnamefont {V.}~\bibnamefont {Greco}}, \bibinfo {author} {\bibfnamefont {T.}~\bibnamefont {Lappi}}, \bibinfo {author} {\bibfnamefont {H.}~\bibnamefont {M\"antysaari}},\ and\ \bibinfo {author} {\bibfnamefont {D.}~\bibnamefont {M\"uller}},\ }\bibfield  {title} {\bibinfo {title} {{The impact of glasma on heavy flavor azimuthal correlations and spectra}},\ }\href@noop {} {\  (\bibinfo {year} {2024}{\natexlab{a}})},\ \Eprint {https://arxiv.org/abs/2409.10564} {arXiv:2409.10564 [hep-ph]} \BibitemShut {NoStop}%
\bibitem [{\citenamefont {Avramescu}\ \emph {et~al.}(2024{\natexlab{b}})\citenamefont {Avramescu}, \citenamefont {Greco}, \citenamefont {Lappi}, \citenamefont {M\"antysaari},\ and\ \citenamefont {M\"uller}}]{Avramescu:2024xts}%
  \BibitemOpen
  \bibfield  {author} {\bibinfo {author} {\bibfnamefont {D.}~\bibnamefont {Avramescu}}, \bibinfo {author} {\bibfnamefont {V.}~\bibnamefont {Greco}}, \bibinfo {author} {\bibfnamefont {T.}~\bibnamefont {Lappi}}, \bibinfo {author} {\bibfnamefont {H.}~\bibnamefont {M\"antysaari}},\ and\ \bibinfo {author} {\bibfnamefont {D.}~\bibnamefont {M\"uller}},\ }\bibfield  {title} {\bibinfo {title} {{Heavy flavor angular correlations as a direct probe of the glasma}},\ }\href@noop {} {\  (\bibinfo {year} {2024}{\natexlab{b}})},\ \Eprint {https://arxiv.org/abs/2409.10565} {arXiv:2409.10565 [hep-ph]} \BibitemShut {NoStop}%
\bibitem [{\citenamefont {Matsui}\ and\ \citenamefont {Satz}(1986)}]{Matsui:1986dk}%
  \BibitemOpen
  \bibfield  {author} {\bibinfo {author} {\bibfnamefont {T.}~\bibnamefont {Matsui}}\ and\ \bibinfo {author} {\bibfnamefont {H.}~\bibnamefont {Satz}},\ }\bibfield  {title} {\bibinfo {title} {{$J/\psi$ Suppression by Quark-Gluon Plasma Formation}},\ }\href {https://doi.org/10.1016/0370-2693(86)91404-8} {\bibfield  {journal} {\bibinfo  {journal} {Phys. Lett. B}\ }\textbf {\bibinfo {volume} {178}},\ \bibinfo {pages} {416} (\bibinfo {year} {1986})}\BibitemShut {NoStop}%
\bibitem [{\citenamefont {Datta}\ \emph {et~al.}(2004)\citenamefont {Datta}, \citenamefont {Karsch}, \citenamefont {Petreczky},\ and\ \citenamefont {Wetzorke}}]{Datta:2003ww}%
  \BibitemOpen
  \bibfield  {author} {\bibinfo {author} {\bibfnamefont {S.}~\bibnamefont {Datta}}, \bibinfo {author} {\bibfnamefont {F.}~\bibnamefont {Karsch}}, \bibinfo {author} {\bibfnamefont {P.}~\bibnamefont {Petreczky}},\ and\ \bibinfo {author} {\bibfnamefont {I.}~\bibnamefont {Wetzorke}},\ }\bibfield  {title} {\bibinfo {title} {{Behavior of charmonium systems after deconfinement}},\ }\href {https://doi.org/10.1103/PhysRevD.69.094507} {\bibfield  {journal} {\bibinfo  {journal} {Phys. Rev. D}\ }\textbf {\bibinfo {volume} {69}},\ \bibinfo {pages} {094507} (\bibinfo {year} {2004})},\ \Eprint {https://arxiv.org/abs/hep-lat/0312037} {arXiv:hep-lat/0312037} \BibitemShut {NoStop}%
\bibitem [{\citenamefont {Burnier}\ \emph {et~al.}(2009)\citenamefont {Burnier}, \citenamefont {Laine},\ and\ \citenamefont {Vepsalainen}}]{Burnier:2009yu}%
  \BibitemOpen
  \bibfield  {author} {\bibinfo {author} {\bibfnamefont {Y.}~\bibnamefont {Burnier}}, \bibinfo {author} {\bibfnamefont {M.}~\bibnamefont {Laine}},\ and\ \bibinfo {author} {\bibfnamefont {M.}~\bibnamefont {Vepsalainen}},\ }\bibfield  {title} {\bibinfo {title} {{Quarkonium dissociation in the presence of a small momentum space anisotropy}},\ }\href {https://doi.org/10.1016/j.physletb.2009.05.067} {\bibfield  {journal} {\bibinfo  {journal} {Phys. Lett. B}\ }\textbf {\bibinfo {volume} {678}},\ \bibinfo {pages} {86} (\bibinfo {year} {2009})},\ \Eprint {https://arxiv.org/abs/0903.3467} {arXiv:0903.3467 [hep-ph]} \BibitemShut {NoStop}%
\bibitem [{\citenamefont {Brambilla}\ \emph {et~al.}(2011)\citenamefont {Brambilla}, \citenamefont {Escobedo}, \citenamefont {Ghiglieri},\ and\ \citenamefont {Vairo}}]{Brambilla:2011sg}%
  \BibitemOpen
  \bibfield  {author} {\bibinfo {author} {\bibfnamefont {N.}~\bibnamefont {Brambilla}}, \bibinfo {author} {\bibfnamefont {M.~A.}\ \bibnamefont {Escobedo}}, \bibinfo {author} {\bibfnamefont {J.}~\bibnamefont {Ghiglieri}},\ and\ \bibinfo {author} {\bibfnamefont {A.}~\bibnamefont {Vairo}},\ }\bibfield  {title} {\bibinfo {title} {{Thermal width and gluo-dissociation of quarkonium in pNRQCD}},\ }\href {https://doi.org/10.1007/JHEP12(2011)116} {\bibfield  {journal} {\bibinfo  {journal} {JHEP}\ }\textbf {\bibinfo {volume} {12}},\ \bibinfo {pages} {116}},\ \Eprint {https://arxiv.org/abs/1109.5826} {arXiv:1109.5826 [hep-ph]} \BibitemShut {NoStop}%
\bibitem [{\citenamefont {Srivastava}\ \emph {et~al.}(2013)\citenamefont {Srivastava}, \citenamefont {Mishra},\ and\ \citenamefont {Singh}}]{Srivastava:2012pd}%
  \BibitemOpen
  \bibfield  {author} {\bibinfo {author} {\bibfnamefont {P.~K.}\ \bibnamefont {Srivastava}}, \bibinfo {author} {\bibfnamefont {M.}~\bibnamefont {Mishra}},\ and\ \bibinfo {author} {\bibfnamefont {C.~P.}\ \bibnamefont {Singh}},\ }\bibfield  {title} {\bibinfo {title} {{Color screening scenario for quarkonia suppression in a quasiparticle model compared with data obtained from experiments at the CERN SPS, BNL RHIC, and CERN LHC}},\ }\href {https://doi.org/10.1103/PhysRevC.87.034903} {\bibfield  {journal} {\bibinfo  {journal} {Phys. Rev. C}\ }\textbf {\bibinfo {volume} {87}},\ \bibinfo {pages} {034903} (\bibinfo {year} {2013})},\ \Eprint {https://arxiv.org/abs/1208.4796} {arXiv:1208.4796 [hep-ph]} \BibitemShut {NoStop}%
\bibitem [{\citenamefont {Casalderrey-Solana}(2013)}]{Casalderrey-Solana:2012yfo}%
  \BibitemOpen
  \bibfield  {author} {\bibinfo {author} {\bibfnamefont {J.}~\bibnamefont {Casalderrey-Solana}},\ }\bibfield  {title} {\bibinfo {title} {{Dynamical Quarkonia Suppression in a QGP-Brick}},\ }\href {https://doi.org/10.1007/JHEP03(2013)091} {\bibfield  {journal} {\bibinfo  {journal} {JHEP}\ }\textbf {\bibinfo {volume} {03}},\ \bibinfo {pages} {091}},\ \Eprint {https://arxiv.org/abs/1208.2602} {arXiv:1208.2602 [hep-ph]} \BibitemShut {NoStop}%
\bibitem [{\citenamefont {Brambilla}\ \emph {et~al.}(2013)\citenamefont {Brambilla}, \citenamefont {Escobedo}, \citenamefont {Ghiglieri},\ and\ \citenamefont {Vairo}}]{Brambilla:2013dpa}%
  \BibitemOpen
  \bibfield  {author} {\bibinfo {author} {\bibfnamefont {N.}~\bibnamefont {Brambilla}}, \bibinfo {author} {\bibfnamefont {M.~A.}\ \bibnamefont {Escobedo}}, \bibinfo {author} {\bibfnamefont {J.}~\bibnamefont {Ghiglieri}},\ and\ \bibinfo {author} {\bibfnamefont {A.}~\bibnamefont {Vairo}},\ }\bibfield  {title} {\bibinfo {title} {{Thermal width and quarkonium dissociation by inelastic parton scattering}},\ }\href {https://doi.org/10.1007/JHEP05(2013)130} {\bibfield  {journal} {\bibinfo  {journal} {JHEP}\ }\textbf {\bibinfo {volume} {05}},\ \bibinfo {pages} {130}},\ \Eprint {https://arxiv.org/abs/1303.6097} {arXiv:1303.6097 [hep-ph]} \BibitemShut {NoStop}%
\bibitem [{\citenamefont {Singh}\ \emph {et~al.}(2015)\citenamefont {Singh}, \citenamefont {Srivastava}, \citenamefont {Ganesh},\ and\ \citenamefont {Mishra}}]{Singh:2015eta}%
  \BibitemOpen
  \bibfield  {author} {\bibinfo {author} {\bibfnamefont {C.~R.}\ \bibnamefont {Singh}}, \bibinfo {author} {\bibfnamefont {P.~K.}\ \bibnamefont {Srivastava}}, \bibinfo {author} {\bibfnamefont {S.}~\bibnamefont {Ganesh}},\ and\ \bibinfo {author} {\bibfnamefont {M.}~\bibnamefont {Mishra}},\ }\bibfield  {title} {\bibinfo {title} {{Unified description of charmonium suppression in a quark-gluon plasma medium at RHIC and LHC energies}},\ }\href {https://doi.org/10.1103/PhysRevC.92.034916} {\bibfield  {journal} {\bibinfo  {journal} {Phys. Rev. C}\ }\textbf {\bibinfo {volume} {92}},\ \bibinfo {pages} {034916} (\bibinfo {year} {2015})},\ \Eprint {https://arxiv.org/abs/1505.05674} {arXiv:1505.05674 [hep-ph]} \BibitemShut {NoStop}%
\bibitem [{\citenamefont {Brambilla}\ \emph {et~al.}(2017)\citenamefont {Brambilla}, \citenamefont {Escobedo}, \citenamefont {Soto},\ and\ \citenamefont {Vairo}}]{Brambilla:2016wgg}%
  \BibitemOpen
  \bibfield  {author} {\bibinfo {author} {\bibfnamefont {N.}~\bibnamefont {Brambilla}}, \bibinfo {author} {\bibfnamefont {M.~A.}\ \bibnamefont {Escobedo}}, \bibinfo {author} {\bibfnamefont {J.}~\bibnamefont {Soto}},\ and\ \bibinfo {author} {\bibfnamefont {A.}~\bibnamefont {Vairo}},\ }\bibfield  {title} {\bibinfo {title} {{Quarkonium suppression in heavy-ion collisions: an open quantum system approach}},\ }\href {https://doi.org/10.1103/PhysRevD.96.034021} {\bibfield  {journal} {\bibinfo  {journal} {Phys. Rev. D}\ }\textbf {\bibinfo {volume} {96}},\ \bibinfo {pages} {034021} (\bibinfo {year} {2017})},\ \Eprint {https://arxiv.org/abs/1612.07248} {arXiv:1612.07248 [hep-ph]} \BibitemShut {NoStop}%
\bibitem [{\citenamefont {Song}\ \emph {et~al.}(2017)\citenamefont {Song}, \citenamefont {Aichelin},\ and\ \citenamefont {Bratkovskaya}}]{Song:2017phm}%
  \BibitemOpen
  \bibfield  {author} {\bibinfo {author} {\bibfnamefont {T.}~\bibnamefont {Song}}, \bibinfo {author} {\bibfnamefont {J.}~\bibnamefont {Aichelin}},\ and\ \bibinfo {author} {\bibfnamefont {E.}~\bibnamefont {Bratkovskaya}},\ }\bibfield  {title} {\bibinfo {title} {{Production of primordial $J/\psi$ in relativistic $p+p$ and heavy-ion collisions}},\ }\href {https://doi.org/10.1103/PhysRevC.96.014907} {\bibfield  {journal} {\bibinfo  {journal} {Phys. Rev. C}\ }\textbf {\bibinfo {volume} {96}},\ \bibinfo {pages} {014907} (\bibinfo {year} {2017})},\ \Eprint {https://arxiv.org/abs/1705.00046} {arXiv:1705.00046 [nucl-th]} \BibitemShut {NoStop}%
\bibitem [{\citenamefont {Akamatsu}(2022)}]{Akamatsu:2020ypb}%
  \BibitemOpen
  \bibfield  {author} {\bibinfo {author} {\bibfnamefont {Y.}~\bibnamefont {Akamatsu}},\ }\bibfield  {title} {\bibinfo {title} {{Quarkonium in quark\textendash{}gluon plasma: Open quantum system approaches re-examined}},\ }\href {https://doi.org/10.1016/j.ppnp.2021.103932} {\bibfield  {journal} {\bibinfo  {journal} {Prog. Part. Nucl. Phys.}\ }\textbf {\bibinfo {volume} {123}},\ \bibinfo {pages} {103932} (\bibinfo {year} {2022})},\ \Eprint {https://arxiv.org/abs/2009.10559} {arXiv:2009.10559 [nucl-th]} \BibitemShut {NoStop}%
\bibitem [{\citenamefont {Yao}(2021)}]{Yao:2021lus}%
  \BibitemOpen
  \bibfield  {author} {\bibinfo {author} {\bibfnamefont {X.}~\bibnamefont {Yao}},\ }\bibfield  {title} {\bibinfo {title} {{Open quantum systems for quarkonia}},\ }\href {https://doi.org/10.1142/S0217751X21300106} {\bibfield  {journal} {\bibinfo  {journal} {Int. J. Mod. Phys. A}\ }\textbf {\bibinfo {volume} {36}},\ \bibinfo {pages} {2130010} (\bibinfo {year} {2021})},\ \Eprint {https://arxiv.org/abs/2102.01736} {arXiv:2102.01736 [hep-ph]} \BibitemShut {NoStop}%
\bibitem [{\citenamefont {He}\ \emph {et~al.}(2022)\citenamefont {He}, \citenamefont {Wu},\ and\ \citenamefont {Rapp}}]{He:2021zej}%
  \BibitemOpen
  \bibfield  {author} {\bibinfo {author} {\bibfnamefont {M.}~\bibnamefont {He}}, \bibinfo {author} {\bibfnamefont {B.}~\bibnamefont {Wu}},\ and\ \bibinfo {author} {\bibfnamefont {R.}~\bibnamefont {Rapp}},\ }\bibfield  {title} {\bibinfo {title} {{Collectivity of J/\ensuremath{\psi} Mesons in Heavy-Ion Collisions}},\ }\href {https://doi.org/10.1103/PhysRevLett.128.162301} {\bibfield  {journal} {\bibinfo  {journal} {Phys. Rev. Lett.}\ }\textbf {\bibinfo {volume} {128}},\ \bibinfo {pages} {162301} (\bibinfo {year} {2022})},\ \Eprint {https://arxiv.org/abs/2111.13528} {arXiv:2111.13528 [nucl-th]} \BibitemShut {NoStop}%
\bibitem [{\citenamefont {Miura}\ \emph {et~al.}(2022)\citenamefont {Miura}, \citenamefont {Akamatsu}, \citenamefont {Asakawa},\ and\ \citenamefont {Kaida}}]{Miura:2022arv}%
  \BibitemOpen
  \bibfield  {author} {\bibinfo {author} {\bibfnamefont {T.}~\bibnamefont {Miura}}, \bibinfo {author} {\bibfnamefont {Y.}~\bibnamefont {Akamatsu}}, \bibinfo {author} {\bibfnamefont {M.}~\bibnamefont {Asakawa}},\ and\ \bibinfo {author} {\bibfnamefont {Y.}~\bibnamefont {Kaida}},\ }\bibfield  {title} {\bibinfo {title} {{Simulation of Lindblad equations for quarkonium in the quark-gluon plasma}},\ }\href {https://doi.org/10.1103/PhysRevD.106.074001} {\bibfield  {journal} {\bibinfo  {journal} {Phys. Rev. D}\ }\textbf {\bibinfo {volume} {106}},\ \bibinfo {pages} {074001} (\bibinfo {year} {2022})},\ \Eprint {https://arxiv.org/abs/2205.15551} {arXiv:2205.15551 [nucl-th]} \BibitemShut {NoStop}%
\bibitem [{\citenamefont {Villar}\ \emph {et~al.}(2023)\citenamefont {Villar}, \citenamefont {Zhao}, \citenamefont {Aichelin},\ and\ \citenamefont {Gossiaux}}]{Villar:2022sbv}%
  \BibitemOpen
  \bibfield  {author} {\bibinfo {author} {\bibfnamefont {D.~Y.~A.}\ \bibnamefont {Villar}}, \bibinfo {author} {\bibfnamefont {J.}~\bibnamefont {Zhao}}, \bibinfo {author} {\bibfnamefont {J.}~\bibnamefont {Aichelin}},\ and\ \bibinfo {author} {\bibfnamefont {P.~B.}\ \bibnamefont {Gossiaux}},\ }\bibfield  {title} {\bibinfo {title} {{New microscopic model for J/\ensuremath{\psi} production in heavy ion collisions}},\ }\href {https://doi.org/10.1103/PhysRevC.107.054913} {\bibfield  {journal} {\bibinfo  {journal} {Phys. Rev. C}\ }\textbf {\bibinfo {volume} {107}},\ \bibinfo {pages} {054913} (\bibinfo {year} {2023})},\ \Eprint {https://arxiv.org/abs/2206.01308} {arXiv:2206.01308 [nucl-th]} \BibitemShut {NoStop}%
\bibitem [{\citenamefont {Brambilla}\ \emph {et~al.}(2022)\citenamefont {Brambilla}, \citenamefont {Escobedo}, \citenamefont {Islam}, \citenamefont {Strickland}, \citenamefont {Tiwari}, \citenamefont {Vairo},\ and\ \citenamefont {Vander~Griend}}]{Brambilla:2022ynh}%
  \BibitemOpen
  \bibfield  {author} {\bibinfo {author} {\bibfnamefont {N.}~\bibnamefont {Brambilla}}, \bibinfo {author} {\bibfnamefont {M.~A.}\ \bibnamefont {Escobedo}}, \bibinfo {author} {\bibfnamefont {A.}~\bibnamefont {Islam}}, \bibinfo {author} {\bibfnamefont {M.}~\bibnamefont {Strickland}}, \bibinfo {author} {\bibfnamefont {A.}~\bibnamefont {Tiwari}}, \bibinfo {author} {\bibfnamefont {A.}~\bibnamefont {Vairo}},\ and\ \bibinfo {author} {\bibfnamefont {P.}~\bibnamefont {Vander~Griend}},\ }\bibfield  {title} {\bibinfo {title} {{Heavy quarkonium dynamics at next-to-leading order in the binding energy over temperature}},\ }\href {https://doi.org/10.1007/JHEP08(2022)303} {\bibfield  {journal} {\bibinfo  {journal} {JHEP}\ }\textbf {\bibinfo {volume} {08}},\ \bibinfo {pages} {303}},\ \Eprint {https://arxiv.org/abs/2205.10289} {arXiv:2205.10289 [hep-ph]} \BibitemShut {NoStop}%
\bibitem [{\citenamefont {Song}\ \emph {et~al.}(2023{\natexlab{a}})\citenamefont {Song}, \citenamefont {Aichelin}, \citenamefont {Zhao}, \citenamefont {Gossiaux},\ and\ \citenamefont {Bratkovskaya}}]{Song:2023zma}%
  \BibitemOpen
  \bibfield  {author} {\bibinfo {author} {\bibfnamefont {T.}~\bibnamefont {Song}}, \bibinfo {author} {\bibfnamefont {J.}~\bibnamefont {Aichelin}}, \bibinfo {author} {\bibfnamefont {J.}~\bibnamefont {Zhao}}, \bibinfo {author} {\bibfnamefont {P.~B.}\ \bibnamefont {Gossiaux}},\ and\ \bibinfo {author} {\bibfnamefont {E.}~\bibnamefont {Bratkovskaya}},\ }\bibfield  {title} {\bibinfo {title} {{Bottomonium production in pp and heavy-ion collisions}},\ }\href {https://doi.org/10.1103/PhysRevC.108.054908} {\bibfield  {journal} {\bibinfo  {journal} {Phys. Rev. C}\ }\textbf {\bibinfo {volume} {108}},\ \bibinfo {pages} {054908} (\bibinfo {year} {2023}{\natexlab{a}})},\ \Eprint {https://arxiv.org/abs/2305.10750} {arXiv:2305.10750 [nucl-th]} \BibitemShut {NoStop}%
\bibitem [{\citenamefont {Song}\ \emph {et~al.}(2023{\natexlab{b}})\citenamefont {Song}, \citenamefont {Aichelin},\ and\ \citenamefont {Bratkovskaya}}]{Song:2023ywt}%
  \BibitemOpen
  \bibfield  {author} {\bibinfo {author} {\bibfnamefont {T.}~\bibnamefont {Song}}, \bibinfo {author} {\bibfnamefont {J.}~\bibnamefont {Aichelin}},\ and\ \bibinfo {author} {\bibfnamefont {E.}~\bibnamefont {Bratkovskaya}},\ }\bibfield  {title} {\bibinfo {title} {{Charmonium production in a thermalizing heat bath}},\ }\href {https://doi.org/10.1103/PhysRevC.107.054906} {\bibfield  {journal} {\bibinfo  {journal} {Phys. Rev. C}\ }\textbf {\bibinfo {volume} {107}},\ \bibinfo {pages} {054906} (\bibinfo {year} {2023}{\natexlab{b}})},\ \Eprint {https://arxiv.org/abs/2302.14001} {arXiv:2302.14001 [hep-ph]} \BibitemShut {NoStop}%
\bibitem [{\citenamefont {Brambilla}\ \emph {et~al.}(2024)\citenamefont {Brambilla}, \citenamefont {Magorsch}, \citenamefont {Strickland}, \citenamefont {Vairo},\ and\ \citenamefont {Vander~Griend}}]{Brambilla:2024tqg}%
  \BibitemOpen
  \bibfield  {author} {\bibinfo {author} {\bibfnamefont {N.}~\bibnamefont {Brambilla}}, \bibinfo {author} {\bibfnamefont {T.}~\bibnamefont {Magorsch}}, \bibinfo {author} {\bibfnamefont {M.}~\bibnamefont {Strickland}}, \bibinfo {author} {\bibfnamefont {A.}~\bibnamefont {Vairo}},\ and\ \bibinfo {author} {\bibfnamefont {P.}~\bibnamefont {Vander~Griend}},\ }\bibfield  {title} {\bibinfo {title} {{Bottomonium suppression from the three-loop QCD potential}},\ }\href {https://doi.org/10.1103/PhysRevD.109.114016} {\bibfield  {journal} {\bibinfo  {journal} {Phys. Rev. D}\ }\textbf {\bibinfo {volume} {109}},\ \bibinfo {pages} {114016} (\bibinfo {year} {2024})},\ \Eprint {https://arxiv.org/abs/2403.15545} {arXiv:2403.15545 [hep-ph]} \BibitemShut {NoStop}%
\bibitem [{\citenamefont {Bai}\ and\ \citenamefont {Chen}(2024)}]{Bai:2024xmm}%
  \BibitemOpen
  \bibfield  {author} {\bibinfo {author} {\bibfnamefont {Y.}~\bibnamefont {Bai}}\ and\ \bibinfo {author} {\bibfnamefont {B.}~\bibnamefont {Chen}},\ }\bibfield  {title} {\bibinfo {title} {{Probing QGP droplets with charmonium in high-multiplicity proton\textendash{}proton collisions}},\ }\href {https://doi.org/10.1140/epjc/s10052-024-13599-4} {\bibfield  {journal} {\bibinfo  {journal} {Eur. Phys. J. C}\ }\textbf {\bibinfo {volume} {84}},\ \bibinfo {pages} {1193} (\bibinfo {year} {2024})},\ \Eprint {https://arxiv.org/abs/2407.10566} {arXiv:2407.10566 [nucl-th]} \BibitemShut {NoStop}%
\bibitem [{\citenamefont {Delorme}\ \emph {et~al.}(2024)\citenamefont {Delorme}, \citenamefont {Katz}, \citenamefont {Gousset}, \citenamefont {Gossiaux},\ and\ \citenamefont {Blaizot}}]{Delorme:2024rdo}%
  \BibitemOpen
  \bibfield  {author} {\bibinfo {author} {\bibfnamefont {S.}~\bibnamefont {Delorme}}, \bibinfo {author} {\bibfnamefont {R.}~\bibnamefont {Katz}}, \bibinfo {author} {\bibfnamefont {T.}~\bibnamefont {Gousset}}, \bibinfo {author} {\bibfnamefont {P.~B.}\ \bibnamefont {Gossiaux}},\ and\ \bibinfo {author} {\bibfnamefont {J.-P.}\ \bibnamefont {Blaizot}},\ }\bibfield  {title} {\bibinfo {title} {{Quarkonium dynamics in the quantum Brownian regime with non-abelian quantum master equations}},\ }\href {https://doi.org/10.1007/JHEP06(2024)060} {\bibfield  {journal} {\bibinfo  {journal} {JHEP}\ }\textbf {\bibinfo {volume} {06}},\ \bibinfo {pages} {060}},\ \Eprint {https://arxiv.org/abs/2402.04488} {arXiv:2402.04488 [hep-ph]} \BibitemShut {NoStop}%
\bibitem [{\citenamefont {Du}\ and\ \citenamefont {Rapp}(2015)}]{Du:2015wha}%
  \BibitemOpen
  \bibfield  {author} {\bibinfo {author} {\bibfnamefont {X.}~\bibnamefont {Du}}\ and\ \bibinfo {author} {\bibfnamefont {R.}~\bibnamefont {Rapp}},\ }\bibfield  {title} {\bibinfo {title} {{Sequential Regeneration of Charmonia in Heavy-Ion Collisions}},\ }\href {https://doi.org/10.1016/j.nuclphysa.2015.09.006} {\bibfield  {journal} {\bibinfo  {journal} {Nucl. Phys. A}\ }\textbf {\bibinfo {volume} {943}},\ \bibinfo {pages} {147} (\bibinfo {year} {2015})},\ \Eprint {https://arxiv.org/abs/1504.00670} {arXiv:1504.00670 [hep-ph]} \BibitemShut {NoStop}%
\bibitem [{\citenamefont {Du}\ \emph {et~al.}(2017)\citenamefont {Du}, \citenamefont {He},\ and\ \citenamefont {Rapp}}]{Du:2017qkv}%
  \BibitemOpen
  \bibfield  {author} {\bibinfo {author} {\bibfnamefont {X.}~\bibnamefont {Du}}, \bibinfo {author} {\bibfnamefont {M.}~\bibnamefont {He}},\ and\ \bibinfo {author} {\bibfnamefont {R.}~\bibnamefont {Rapp}},\ }\bibfield  {title} {\bibinfo {title} {{Color Screening and Regeneration of Bottomonia in High-Energy Heavy-Ion Collisions}},\ }\href {https://doi.org/10.1103/PhysRevC.96.054901} {\bibfield  {journal} {\bibinfo  {journal} {Phys. Rev. C}\ }\textbf {\bibinfo {volume} {96}},\ \bibinfo {pages} {054901} (\bibinfo {year} {2017})},\ \Eprint {https://arxiv.org/abs/1706.08670} {arXiv:1706.08670 [hep-ph]} \BibitemShut {NoStop}%
\bibitem [{\citenamefont {Du}\ and\ \citenamefont {Rapp}(2019)}]{Du:2018wsj}%
  \BibitemOpen
  \bibfield  {author} {\bibinfo {author} {\bibfnamefont {X.}~\bibnamefont {Du}}\ and\ \bibinfo {author} {\bibfnamefont {R.}~\bibnamefont {Rapp}},\ }\bibfield  {title} {\bibinfo {title} {{In-Medium Charmonium Production in Proton-Nucleus Collisions}},\ }\href {https://doi.org/10.1007/JHEP03(2019)015} {\bibfield  {journal} {\bibinfo  {journal} {JHEP}\ }\textbf {\bibinfo {volume} {03}},\ \bibinfo {pages} {015}},\ \Eprint {https://arxiv.org/abs/1808.10014} {arXiv:1808.10014 [nucl-th]} \BibitemShut {NoStop}%
\bibitem [{\citenamefont {Yan}\ \emph {et~al.}(2006)\citenamefont {Yan}, \citenamefont {Zhuang},\ and\ \citenamefont {Xu}}]{Yan:2006ve}%
  \BibitemOpen
  \bibfield  {author} {\bibinfo {author} {\bibfnamefont {L.}~\bibnamefont {Yan}}, \bibinfo {author} {\bibfnamefont {P.}~\bibnamefont {Zhuang}},\ and\ \bibinfo {author} {\bibfnamefont {N.}~\bibnamefont {Xu}},\ }\bibfield  {title} {\bibinfo {title} {{Competition between J / psi suppression and regeneration in quark-gluon plasma}},\ }\href {https://doi.org/10.1103/PhysRevLett.97.232301} {\bibfield  {journal} {\bibinfo  {journal} {Phys. Rev. Lett.}\ }\textbf {\bibinfo {volume} {97}},\ \bibinfo {pages} {232301} (\bibinfo {year} {2006})},\ \Eprint {https://arxiv.org/abs/nucl-th/0608010} {arXiv:nucl-th/0608010} \BibitemShut {NoStop}%
\bibitem [{\citenamefont {Zhou}\ \emph {et~al.}(2014)\citenamefont {Zhou}, \citenamefont {Xu}, \citenamefont {Xu},\ and\ \citenamefont {Zhuang}}]{Zhou:2014kka}%
  \BibitemOpen
  \bibfield  {author} {\bibinfo {author} {\bibfnamefont {K.}~\bibnamefont {Zhou}}, \bibinfo {author} {\bibfnamefont {N.}~\bibnamefont {Xu}}, \bibinfo {author} {\bibfnamefont {Z.}~\bibnamefont {Xu}},\ and\ \bibinfo {author} {\bibfnamefont {P.}~\bibnamefont {Zhuang}},\ }\bibfield  {title} {\bibinfo {title} {{Medium effects on charmonium production at ultrarelativistic energies available at the CERN Large Hadron Collider}},\ }\href {https://doi.org/10.1103/PhysRevC.89.054911} {\bibfield  {journal} {\bibinfo  {journal} {Phys. Rev. C}\ }\textbf {\bibinfo {volume} {89}},\ \bibinfo {pages} {054911} (\bibinfo {year} {2014})},\ \Eprint {https://arxiv.org/abs/1401.5845} {arXiv:1401.5845 [nucl-th]} \BibitemShut {NoStop}%
\bibitem [{\citenamefont {Rothkopf}(2020)}]{Rothkopf:2019ipj}%
  \BibitemOpen
  \bibfield  {author} {\bibinfo {author} {\bibfnamefont {A.}~\bibnamefont {Rothkopf}},\ }\bibfield  {title} {\bibinfo {title} {{Heavy Quarkonium in Extreme Conditions}},\ }\href {https://doi.org/10.1016/j.physrep.2020.02.006} {\bibfield  {journal} {\bibinfo  {journal} {Phys. Rept.}\ }\textbf {\bibinfo {volume} {858}},\ \bibinfo {pages} {1} (\bibinfo {year} {2020})},\ \Eprint {https://arxiv.org/abs/1912.02253} {arXiv:1912.02253 [hep-ph]} \BibitemShut {NoStop}%
\bibitem [{\citenamefont {Andronic}\ \emph {et~al.}(2024)\citenamefont {Andronic} \emph {et~al.}}]{Andronic:2024oxz}%
  \BibitemOpen
  \bibfield  {author} {\bibinfo {author} {\bibfnamefont {A.}~\bibnamefont {Andronic}} \emph {et~al.},\ }\bibfield  {title} {\bibinfo {title} {{Comparative study of quarkonium transport in hot QCD matter}},\ }\href {https://doi.org/10.1140/epja/s10050-024-01306-6} {\bibfield  {journal} {\bibinfo  {journal} {Eur. Phys. J. A}\ }\textbf {\bibinfo {volume} {60}},\ \bibinfo {pages} {88} (\bibinfo {year} {2024})},\ \Eprint {https://arxiv.org/abs/2402.04366} {arXiv:2402.04366 [nucl-th]} \BibitemShut {NoStop}%
\bibitem [{\citenamefont {Pooja}\ \emph {et~al.}(2024)\citenamefont {Pooja}, \citenamefont {Jamal}, \citenamefont {Bhaduri}, \citenamefont {Ruggieri},\ and\ \citenamefont {Das}}]{Pooja:2024rnn}%
  \BibitemOpen
  \bibfield  {author} {\bibinfo {author} {\bibnamefont {Pooja}}, \bibinfo {author} {\bibfnamefont {M.~Y.}\ \bibnamefont {Jamal}}, \bibinfo {author} {\bibfnamefont {P.~P.}\ \bibnamefont {Bhaduri}}, \bibinfo {author} {\bibfnamefont {M.}~\bibnamefont {Ruggieri}},\ and\ \bibinfo {author} {\bibfnamefont {S.~K.}\ \bibnamefont {Das}},\ }\bibfield  {title} {\bibinfo {title} {{$c\bar{c}$ and $b\bar{b}$ suppression in the glasma}},\ }\href {https://doi.org/10.1103/PhysRevD.110.094018} {\bibfield  {journal} {\bibinfo  {journal} {Phys. Rev. D}\ }\textbf {\bibinfo {volume} {110}},\ \bibinfo {pages} {094018} (\bibinfo {year} {2024})},\ \Eprint {https://arxiv.org/abs/2404.05315} {arXiv:2404.05315 [hep-ph]} \BibitemShut {NoStop}%
\bibitem [{\citenamefont {Wong}(1970)}]{Wong:1970fu}%
  \BibitemOpen
  \bibfield  {author} {\bibinfo {author} {\bibfnamefont {S.~K.}\ \bibnamefont {Wong}},\ }\bibfield  {title} {\bibinfo {title} {{Field and particle equations for the classical Yang-Mills field and particles with isotopic spin}},\ }\href {https://doi.org/10.1007/BF02892134} {\bibfield  {journal} {\bibinfo  {journal} {Nuovo Cim. A}\ }\textbf {\bibinfo {volume} {65}},\ \bibinfo {pages} {689} (\bibinfo {year} {1970})}\BibitemShut {NoStop}%
\bibitem [{\citenamefont {Heinz}(1985)}]{Heinz:1984yq}%
  \BibitemOpen
  \bibfield  {author} {\bibinfo {author} {\bibfnamefont {U.~W.}\ \bibnamefont {Heinz}},\ }\bibfield  {title} {\bibinfo {title} {{Quark - Gluon Transport Theory. Part 1. the Classical Theory}},\ }\href {https://doi.org/10.1016/0003-4916(85)90336-7} {\bibfield  {journal} {\bibinfo  {journal} {Annals Phys.}\ }\textbf {\bibinfo {volume} {161}},\ \bibinfo {pages} {48} (\bibinfo {year} {1985})}\BibitemShut {NoStop}%
\bibitem [{\citenamefont {Greco}\ \emph {et~al.}(2003{\natexlab{a}})\citenamefont {Greco}, \citenamefont {Ko},\ and\ \citenamefont {Levai}}]{Greco:2003xt}%
  \BibitemOpen
  \bibfield  {author} {\bibinfo {author} {\bibfnamefont {V.}~\bibnamefont {Greco}}, \bibinfo {author} {\bibfnamefont {C.~M.}\ \bibnamefont {Ko}},\ and\ \bibinfo {author} {\bibfnamefont {P.}~\bibnamefont {Levai}},\ }\bibfield  {title} {\bibinfo {title} {{Parton coalescence and anti-proton / pion anomaly at RHIC}},\ }\href {https://doi.org/10.1103/PhysRevLett.90.202302} {\bibfield  {journal} {\bibinfo  {journal} {Phys. Rev. Lett.}\ }\textbf {\bibinfo {volume} {90}},\ \bibinfo {pages} {202302} (\bibinfo {year} {2003}{\natexlab{a}})},\ \Eprint {https://arxiv.org/abs/nucl-th/0301093} {arXiv:nucl-th/0301093} \BibitemShut {NoStop}%
\bibitem [{\citenamefont {Fries}\ \emph {et~al.}(2003{\natexlab{a}})\citenamefont {Fries}, \citenamefont {Muller}, \citenamefont {Nonaka},\ and\ \citenamefont {Bass}}]{Fries:2003vb}%
  \BibitemOpen
  \bibfield  {author} {\bibinfo {author} {\bibfnamefont {R.~J.}\ \bibnamefont {Fries}}, \bibinfo {author} {\bibfnamefont {B.}~\bibnamefont {Muller}}, \bibinfo {author} {\bibfnamefont {C.}~\bibnamefont {Nonaka}},\ and\ \bibinfo {author} {\bibfnamefont {S.~A.}\ \bibnamefont {Bass}},\ }\bibfield  {title} {\bibinfo {title} {{Hadronization in heavy ion collisions: Recombination and fragmentation of partons}},\ }\href {https://doi.org/10.1103/PhysRevLett.90.202303} {\bibfield  {journal} {\bibinfo  {journal} {Phys. Rev. Lett.}\ }\textbf {\bibinfo {volume} {90}},\ \bibinfo {pages} {202303} (\bibinfo {year} {2003}{\natexlab{a}})},\ \Eprint {https://arxiv.org/abs/nucl-th/0301087} {arXiv:nucl-th/0301087} \BibitemShut {NoStop}%
\bibitem [{\citenamefont {Greco}\ \emph {et~al.}(2003{\natexlab{b}})\citenamefont {Greco}, \citenamefont {Ko},\ and\ \citenamefont {Levai}}]{Greco:2003mm}%
  \BibitemOpen
  \bibfield  {author} {\bibinfo {author} {\bibfnamefont {V.}~\bibnamefont {Greco}}, \bibinfo {author} {\bibfnamefont {C.~M.}\ \bibnamefont {Ko}},\ and\ \bibinfo {author} {\bibfnamefont {P.}~\bibnamefont {Levai}},\ }\bibfield  {title} {\bibinfo {title} {{Parton coalescence at RHIC}},\ }\href {https://doi.org/10.1103/PhysRevC.68.034904} {\bibfield  {journal} {\bibinfo  {journal} {Phys. Rev. C}\ }\textbf {\bibinfo {volume} {68}},\ \bibinfo {pages} {034904} (\bibinfo {year} {2003}{\natexlab{b}})},\ \Eprint {https://arxiv.org/abs/nucl-th/0305024} {arXiv:nucl-th/0305024} \BibitemShut {NoStop}%
\bibitem [{\citenamefont {Fries}\ \emph {et~al.}(2003{\natexlab{b}})\citenamefont {Fries}, \citenamefont {Muller}, \citenamefont {Nonaka},\ and\ \citenamefont {Bass}}]{Fries:2003kq}%
  \BibitemOpen
  \bibfield  {author} {\bibinfo {author} {\bibfnamefont {R.~J.}\ \bibnamefont {Fries}}, \bibinfo {author} {\bibfnamefont {B.}~\bibnamefont {Muller}}, \bibinfo {author} {\bibfnamefont {C.}~\bibnamefont {Nonaka}},\ and\ \bibinfo {author} {\bibfnamefont {S.~A.}\ \bibnamefont {Bass}},\ }\bibfield  {title} {\bibinfo {title} {{Hadron production in heavy ion collisions: Fragmentation and recombination from a dense parton phase}},\ }\href {https://doi.org/10.1103/PhysRevC.68.044902} {\bibfield  {journal} {\bibinfo  {journal} {Phys. Rev. C}\ }\textbf {\bibinfo {volume} {68}},\ \bibinfo {pages} {044902} (\bibinfo {year} {2003}{\natexlab{b}})},\ \Eprint {https://arxiv.org/abs/nucl-th/0306027} {arXiv:nucl-th/0306027} \BibitemShut {NoStop}%
\bibitem [{\citenamefont {Greco}\ \emph {et~al.}(2004)\citenamefont {Greco}, \citenamefont {Ko},\ and\ \citenamefont {Rapp}}]{Greco:2003vf}%
  \BibitemOpen
  \bibfield  {author} {\bibinfo {author} {\bibfnamefont {V.}~\bibnamefont {Greco}}, \bibinfo {author} {\bibfnamefont {C.~M.}\ \bibnamefont {Ko}},\ and\ \bibinfo {author} {\bibfnamefont {R.}~\bibnamefont {Rapp}},\ }\bibfield  {title} {\bibinfo {title} {{Quark coalescence for charmed mesons in ultrarelativistic heavy ion collisions}},\ }\href {https://doi.org/10.1016/j.physletb.2004.06.064} {\bibfield  {journal} {\bibinfo  {journal} {Phys. Lett. B}\ }\textbf {\bibinfo {volume} {595}},\ \bibinfo {pages} {202} (\bibinfo {year} {2004})},\ \Eprint {https://arxiv.org/abs/nucl-th/0312100} {arXiv:nucl-th/0312100} \BibitemShut {NoStop}%
\bibitem [{\citenamefont {Plumari}\ \emph {et~al.}(2018)\citenamefont {Plumari}, \citenamefont {Minissale}, \citenamefont {Das}, \citenamefont {Coci},\ and\ \citenamefont {Greco}}]{Plumari:2017ntm}%
  \BibitemOpen
  \bibfield  {author} {\bibinfo {author} {\bibfnamefont {S.}~\bibnamefont {Plumari}}, \bibinfo {author} {\bibfnamefont {V.}~\bibnamefont {Minissale}}, \bibinfo {author} {\bibfnamefont {S.~K.}\ \bibnamefont {Das}}, \bibinfo {author} {\bibfnamefont {G.}~\bibnamefont {Coci}},\ and\ \bibinfo {author} {\bibfnamefont {V.}~\bibnamefont {Greco}},\ }\bibfield  {title} {\bibinfo {title} {{Charmed Hadrons from Coalescence plus Fragmentation in relativistic nucleus-nucleus collisions at RHIC and LHC}},\ }\href {https://doi.org/10.1140/epjc/s10052-018-5828-7} {\bibfield  {journal} {\bibinfo  {journal} {Eur. Phys. J. C}\ }\textbf {\bibinfo {volume} {78}},\ \bibinfo {pages} {348} (\bibinfo {year} {2018})},\ \Eprint {https://arxiv.org/abs/1712.00730} {arXiv:1712.00730 [hep-ph]} \BibitemShut {NoStop}%
\bibitem [{\citenamefont {Lappi}(2011)}]{LAPPI:2010ykh}%
  \BibitemOpen
  \bibfield  {author} {\bibinfo {author} {\bibfnamefont {T.}~\bibnamefont {Lappi}},\ }\bibfield  {title} {\bibinfo {title} {{Small x physics and RHIC data}},\ }\href {https://doi.org/10.1142/S0218301311017302} {\bibfield  {journal} {\bibinfo  {journal} {Int. J. Mod. Phys. E}\ }\textbf {\bibinfo {volume} {20}},\ \bibinfo {pages} {1} (\bibinfo {year} {2011})},\ \Eprint {https://arxiv.org/abs/1003.1852} {arXiv:1003.1852 [hep-ph]} \BibitemShut {NoStop}%
\bibitem [{\citenamefont {Schenke}\ \emph {et~al.}(2020)\citenamefont {Schenke}, \citenamefont {Shen},\ and\ \citenamefont {Tribedy}}]{Schenke:2020mbo}%
  \BibitemOpen
  \bibfield  {author} {\bibinfo {author} {\bibfnamefont {B.}~\bibnamefont {Schenke}}, \bibinfo {author} {\bibfnamefont {C.}~\bibnamefont {Shen}},\ and\ \bibinfo {author} {\bibfnamefont {P.}~\bibnamefont {Tribedy}},\ }\bibfield  {title} {\bibinfo {title} {{Running the gamut of high energy nuclear collisions}},\ }\href {https://doi.org/10.1103/PhysRevC.102.044905} {\bibfield  {journal} {\bibinfo  {journal} {Phys. Rev. C}\ }\textbf {\bibinfo {volume} {102}},\ \bibinfo {pages} {044905} (\bibinfo {year} {2020})},\ \Eprint {https://arxiv.org/abs/2005.14682} {arXiv:2005.14682 [nucl-th]} \BibitemShut {NoStop}%
\bibitem [{\citenamefont {Rezaeian}\ \emph {et~al.}(2013)\citenamefont {Rezaeian}, \citenamefont {Siddikov}, \citenamefont {Van~de Klundert},\ and\ \citenamefont {Venugopalan}}]{Rezaeian:2012ji}%
  \BibitemOpen
  \bibfield  {author} {\bibinfo {author} {\bibfnamefont {A.~H.}\ \bibnamefont {Rezaeian}}, \bibinfo {author} {\bibfnamefont {M.}~\bibnamefont {Siddikov}}, \bibinfo {author} {\bibfnamefont {M.}~\bibnamefont {Van~de Klundert}},\ and\ \bibinfo {author} {\bibfnamefont {R.}~\bibnamefont {Venugopalan}},\ }\bibfield  {title} {\bibinfo {title} {{Analysis of combined HERA data in the Impact-Parameter dependent Saturation model}},\ }\href {https://doi.org/10.1103/PhysRevD.87.034002} {\bibfield  {journal} {\bibinfo  {journal} {Phys. Rev. D}\ }\textbf {\bibinfo {volume} {87}},\ \bibinfo {pages} {034002} (\bibinfo {year} {2013})},\ \Eprint {https://arxiv.org/abs/1212.2974} {arXiv:1212.2974 [hep-ph]} \BibitemShut {NoStop}%
\bibitem [{\citenamefont {Yang}\ and\ \citenamefont {Mills}(1954)}]{Yang:1954ek}%
  \BibitemOpen
  \bibfield  {author} {\bibinfo {author} {\bibfnamefont {C.-N.}\ \bibnamefont {Yang}}\ and\ \bibinfo {author} {\bibfnamefont {R.~L.}\ \bibnamefont {Mills}},\ }\bibfield  {title} {\bibinfo {title} {{Conservation of Isotopic Spin and Isotopic Gauge Invariance}},\ }\href {https://doi.org/10.1103/PhysRev.96.191} {\bibfield  {journal} {\bibinfo  {journal} {Phys. Rev.}\ }\textbf {\bibinfo {volume} {96}},\ \bibinfo {pages} {191} (\bibinfo {year} {1954})}\BibitemShut {NoStop}%
\bibitem [{\citenamefont {Krasnitz}\ and\ \citenamefont {Venugopalan}(1999)}]{Krasnitz:1998ns}%
  \BibitemOpen
  \bibfield  {author} {\bibinfo {author} {\bibfnamefont {A.}~\bibnamefont {Krasnitz}}\ and\ \bibinfo {author} {\bibfnamefont {R.}~\bibnamefont {Venugopalan}},\ }\bibfield  {title} {\bibinfo {title} {{Nonperturbative computation of gluon minijet production in nuclear collisions at very high-energies}},\ }\href {https://doi.org/10.1016/S0550-3213(99)00366-1} {\bibfield  {journal} {\bibinfo  {journal} {Nucl. Phys. B}\ }\textbf {\bibinfo {volume} {557}},\ \bibinfo {pages} {237} (\bibinfo {year} {1999})},\ \Eprint {https://arxiv.org/abs/hep-ph/9809433} {arXiv:hep-ph/9809433} \BibitemShut {NoStop}%
\bibitem [{\citenamefont {Fukushima}\ and\ \citenamefont {Gelis}(2012)}]{Fukushima:2011nq}%
  \BibitemOpen
  \bibfield  {author} {\bibinfo {author} {\bibfnamefont {K.}~\bibnamefont {Fukushima}}\ and\ \bibinfo {author} {\bibfnamefont {F.}~\bibnamefont {Gelis}},\ }\bibfield  {title} {\bibinfo {title} {{The evolving Glasma}},\ }\href {https://doi.org/10.1016/j.nuclphysa.2011.11.003} {\bibfield  {journal} {\bibinfo  {journal} {Nucl. Phys. A}\ }\textbf {\bibinfo {volume} {874}},\ \bibinfo {pages} {108} (\bibinfo {year} {2012})},\ \Eprint {https://arxiv.org/abs/1106.1396} {arXiv:1106.1396 [hep-ph]} \BibitemShut {NoStop}%
\bibitem [{\citenamefont {Ruggieri}\ and\ \citenamefont {Das}(2018{\natexlab{b}})}]{Ruggieri:2018ies}%
  \BibitemOpen
  \bibfield  {author} {\bibinfo {author} {\bibfnamefont {M.}~\bibnamefont {Ruggieri}}\ and\ \bibinfo {author} {\bibfnamefont {S.~K.}\ \bibnamefont {Das}},\ }\bibfield  {title} {\bibinfo {title} {{Diffusion of charm and beauty in the Glasma}},\ }\href {https://doi.org/10.1051/epjconf/201819200017} {\bibfield  {journal} {\bibinfo  {journal} {EPJ Web Conf.}\ }\textbf {\bibinfo {volume} {192}},\ \bibinfo {pages} {00017} (\bibinfo {year} {2018}{\natexlab{b}})},\ \Eprint {https://arxiv.org/abs/1809.07915} {arXiv:1809.07915 [nucl-th]} \BibitemShut {NoStop}%
\bibitem [{\citenamefont {Jamal}\ \emph {et~al.}(2021)\citenamefont {Jamal}, \citenamefont {Das},\ and\ \citenamefont {Ruggieri}}]{Jamal:2020fxo}%
  \BibitemOpen
  \bibfield  {author} {\bibinfo {author} {\bibfnamefont {M.~Y.}\ \bibnamefont {Jamal}}, \bibinfo {author} {\bibfnamefont {S.~K.}\ \bibnamefont {Das}},\ and\ \bibinfo {author} {\bibfnamefont {M.}~\bibnamefont {Ruggieri}},\ }\bibfield  {title} {\bibinfo {title} {{Energy Loss Versus Energy Gain of Heavy Quarks in a Hot Medium}},\ }\href {https://doi.org/10.1103/PhysRevD.103.054030} {\bibfield  {journal} {\bibinfo  {journal} {Phys. Rev. D}\ }\textbf {\bibinfo {volume} {103}},\ \bibinfo {pages} {054030} (\bibinfo {year} {2021})},\ \Eprint {https://arxiv.org/abs/2009.00561} {arXiv:2009.00561 [nucl-th]} \BibitemShut {NoStop}%
\bibitem [{\citenamefont {Sun}\ \emph {et~al.}(2021)\citenamefont {Sun}, \citenamefont {Coci}, \citenamefont {Das}, \citenamefont {Plumari}, \citenamefont {Ruggieri},\ and\ \citenamefont {Greco}}]{Sun:2021req}%
  \BibitemOpen
  \bibfield  {author} {\bibinfo {author} {\bibfnamefont {Y.}~\bibnamefont {Sun}}, \bibinfo {author} {\bibfnamefont {G.}~\bibnamefont {Coci}}, \bibinfo {author} {\bibfnamefont {S.~K.}\ \bibnamefont {Das}}, \bibinfo {author} {\bibfnamefont {S.}~\bibnamefont {Plumari}}, \bibinfo {author} {\bibfnamefont {M.}~\bibnamefont {Ruggieri}},\ and\ \bibinfo {author} {\bibfnamefont {V.}~\bibnamefont {Greco}},\ }\bibfield  {title} {\bibinfo {title} {{Impact of Glasma on heavy quark $R_{AA}$ and $v_2$ in nucleus-nucleus collisions at LHC}},\ }\href {https://doi.org/10.1016/j.nuclphysa.2020.121913} {\bibfield  {journal} {\bibinfo  {journal} {Nucl. Phys. A}\ }\textbf {\bibinfo {volume} {1005}},\ \bibinfo {pages} {121913} (\bibinfo {year} {2021})}\BibitemShut {NoStop}%
\bibitem [{\citenamefont {Zhao}\ \emph {et~al.}(2025)\citenamefont {Zhao}, \citenamefont {Gossiaux}, \citenamefont {Song}, \citenamefont {Bratkovskaya},\ and\ \citenamefont {Aichelin}}]{Zhao:2024vqp}%
  \BibitemOpen
  \bibfield  {author} {\bibinfo {author} {\bibfnamefont {J.}~\bibnamefont {Zhao}}, \bibinfo {author} {\bibfnamefont {P.~B.}\ \bibnamefont {Gossiaux}}, \bibinfo {author} {\bibfnamefont {T.}~\bibnamefont {Song}}, \bibinfo {author} {\bibfnamefont {E.}~\bibnamefont {Bratkovskaya}},\ and\ \bibinfo {author} {\bibfnamefont {J.}~\bibnamefont {Aichelin}},\ }\bibfield  {title} {\bibinfo {title} {{Wigner density approach to quarkonium production in high energy $pp$ collisions}},\ }\href {https://doi.org/10.1393/ncc/i2025-25005-6} {\bibfield  {journal} {\bibinfo  {journal} {Nuovo Cim. C}\ }\textbf {\bibinfo {volume} {48}},\ \bibinfo {pages} {5} (\bibinfo {year} {2025})}\BibitemShut {NoStop}%
\end{thebibliography}%

\end{document}